\documentclass{article}
\pdfoutput=1
\usepackage{jcappub}
\definecolor{AliceBlue}{rgb}{0.94,0.97,1.00}
\definecolor{AntiqueWhite1}{rgb}{1.00,0.94,0.86}
\definecolor{AntiqueWhite2}{rgb}{0.93,0.87,0.80}
\definecolor{AntiqueWhite3}{rgb}{0.80,0.75,0.69}
\definecolor{AntiqueWhite4}{rgb}{0.55,0.51,0.47}
\definecolor{AntiqueWhite}{rgb}{0.98,0.92,0.84}
\definecolor{BlanchedAlmond}{rgb}{1.00,0.92,0.80}
\definecolor{BlueViolet}{rgb}{0.54,0.17,0.89}
\definecolor{CadetBlue1}{rgb}{0.60,0.96,1.00}
\definecolor{CadetBlue2}{rgb}{0.56,0.90,0.93}
\definecolor{CadetBlue3}{rgb}{0.48,0.77,0.80}
\definecolor{CadetBlue4}{rgb}{0.33,0.53,0.55}
\definecolor{CadetBlue}{rgb}{0.37,0.62,0.63}
\definecolor{CornflowerBlue}{rgb}{0.39,0.58,0.93}
\definecolor{DarkBlue}{rgb}{0.00,0.00,0.55}
\definecolor{DarkCyan}{rgb}{0.00,0.55,0.55}
\definecolor{DarkGoldenrod1}{rgb}{1.00,0.73,0.06}
\definecolor{DarkGoldenrod2}{rgb}{0.93,0.68,0.05}
\definecolor{DarkGoldenrod3}{rgb}{0.80,0.58,0.05}
\definecolor{DarkGoldenrod4}{rgb}{0.55,0.40,0.03}
\definecolor{DarkGoldenrod}{rgb}{0.72,0.53,0.04}
\definecolor{DarkGray}{rgb}{0.66,0.66,0.66}
\definecolor{DarkGreen}{rgb}{0.00,0.39,0.00}
\definecolor{DarkGrey}{rgb}{0.66,0.66,0.66}
\definecolor{DarkKhaki}{rgb}{0.74,0.72,0.42}
\definecolor{DarkMagenta}{rgb}{0.55,0.00,0.55}
\definecolor{DarkOliveGreen1}{rgb}{0.79,1.00,0.44}
\definecolor{DarkOliveGreen2}{rgb}{0.74,0.93,0.41}
\definecolor{DarkOliveGreen3}{rgb}{0.64,0.80,0.35}
\definecolor{DarkOliveGreen4}{rgb}{0.43,0.55,0.24}
\definecolor{DarkOliveGreen}{rgb}{0.33,0.42,0.18}
\definecolor{DarkOrange1}{rgb}{1.00,0.50,0.00}
\definecolor{DarkOrange2}{rgb}{0.93,0.46,0.00}
\definecolor{DarkOrange3}{rgb}{0.80,0.40,0.00}
\definecolor{DarkOrange4}{rgb}{0.55,0.27,0.00}
\definecolor{DarkOrange}{rgb}{1.00,0.55,0.00}
\definecolor{DarkOrchid1}{rgb}{0.75,0.24,1.00}
\definecolor{DarkOrchid2}{rgb}{0.70,0.23,0.93}
\definecolor{DarkOrchid3}{rgb}{0.60,0.20,0.80}
\definecolor{DarkOrchid4}{rgb}{0.41,0.13,0.55}
\definecolor{DarkOrchid}{rgb}{0.60,0.20,0.80}
\definecolor{DarkRed}{rgb}{0.55,0.00,0.00}
\definecolor{DarkSalmon}{rgb}{0.91,0.59,0.48}
\definecolor{DarkSeaGreen1}{rgb}{0.76,1.00,0.76}
\definecolor{DarkSeaGreen2}{rgb}{0.71,0.93,0.71}
\definecolor{DarkSeaGreen3}{rgb}{0.61,0.80,0.61}
\definecolor{DarkSeaGreen4}{rgb}{0.41,0.55,0.41}
\definecolor{DarkSeaGreen}{rgb}{0.56,0.74,0.56}
\definecolor{DarkSlateBlue}{rgb}{0.28,0.24,0.55}
\definecolor{DarkSlateGray1}{rgb}{0.59,1.00,1.00}
\definecolor{DarkSlateGray2}{rgb}{0.55,0.93,0.93}
\definecolor{DarkSlateGray3}{rgb}{0.47,0.80,0.80}
\definecolor{DarkSlateGray4}{rgb}{0.32,0.55,0.55}
\definecolor{DarkSlateGray}{rgb}{0.18,0.31,0.31}
\definecolor{DarkSlateGrey}{rgb}{0.18,0.31,0.31}
\definecolor{DarkTurquoise}{rgb}{0.00,0.81,0.82}
\definecolor{DarkViolet}{rgb}{0.58,0.00,0.83}
\definecolor{DeepPink1}{rgb}{1.00,0.08,0.58}
\definecolor{DeepPink2}{rgb}{0.93,0.07,0.54}
\definecolor{DeepPink3}{rgb}{0.80,0.06,0.46}
\definecolor{DeepPink4}{rgb}{0.55,0.04,0.31}
\definecolor{DeepPink}{rgb}{1.00,0.08,0.58}
\definecolor{DeepSkyBlue1}{rgb}{0.00,0.75,1.00}
\definecolor{DeepSkyBlue2}{rgb}{0.00,0.70,0.93}
\definecolor{DeepSkyBlue3}{rgb}{0.00,0.60,0.80}
\definecolor{DeepSkyBlue4}{rgb}{0.00,0.41,0.55}
\definecolor{DeepSkyBlue}{rgb}{0.00,0.75,1.00}
\definecolor{DimGray}{rgb}{0.41,0.41,0.41}
\definecolor{DimGrey}{rgb}{0.41,0.41,0.41}
\definecolor{DodgerBlue1}{rgb}{0.12,0.56,1.00}
\definecolor{DodgerBlue2}{rgb}{0.11,0.53,0.93}
\definecolor{DodgerBlue3}{rgb}{0.09,0.45,0.80}
\definecolor{DodgerBlue4}{rgb}{0.06,0.31,0.55}
\definecolor{DodgerBlue}{rgb}{0.12,0.56,1.00}
\definecolor{FloralWhite}{rgb}{1.00,0.98,0.94}
\definecolor{ForestGreen}{rgb}{0.13,0.55,0.13}
\definecolor{GhostWhite}{rgb}{0.97,0.97,1.00}
\definecolor{GreenYellow}{rgb}{0.68,1.00,0.18}
\definecolor{HotPink1}{rgb}{1.00,0.43,0.71}
\definecolor{HotPink2}{rgb}{0.93,0.42,0.65}
\definecolor{HotPink3}{rgb}{0.80,0.38,0.56}
\definecolor{HotPink4}{rgb}{0.55,0.23,0.38}
\definecolor{HotPink}{rgb}{1.00,0.41,0.71}
\definecolor{IndianRed1}{rgb}{1.00,0.42,0.42}
\definecolor{IndianRed2}{rgb}{0.93,0.39,0.39}
\definecolor{IndianRed3}{rgb}{0.80,0.33,0.33}
\definecolor{IndianRed4}{rgb}{0.55,0.23,0.23}
\definecolor{IndianRed}{rgb}{0.80,0.36,0.36}
\definecolor{LavenderBlush1}{rgb}{1.00,0.94,0.96}
\definecolor{LavenderBlush2}{rgb}{0.93,0.88,0.90}
\definecolor{LavenderBlush3}{rgb}{0.80,0.76,0.77}
\definecolor{LavenderBlush4}{rgb}{0.55,0.51,0.53}
\definecolor{LavenderBlush}{rgb}{1.00,0.94,0.96}
\definecolor{LawnGreen}{rgb}{0.49,0.99,0.00}
\definecolor{LemonChiffon1}{rgb}{1.00,0.98,0.80}
\definecolor{LemonChiffon2}{rgb}{0.93,0.91,0.75}
\definecolor{LemonChiffon3}{rgb}{0.80,0.79,0.65}
\definecolor{LemonChiffon4}{rgb}{0.55,0.54,0.44}
\definecolor{LemonChiffon}{rgb}{1.00,0.98,0.80}
\definecolor{LightBlue1}{rgb}{0.75,0.94,1.00}
\definecolor{LightBlue2}{rgb}{0.70,0.87,0.93}
\definecolor{LightBlue3}{rgb}{0.60,0.75,0.80}
\definecolor{LightBlue4}{rgb}{0.41,0.51,0.55}
\definecolor{LightBlue}{rgb}{0.68,0.85,0.90}
\definecolor{LightCoral}{rgb}{0.94,0.50,0.50}
\definecolor{LightCyan1}{rgb}{0.88,1.00,1.00}
\definecolor{LightCyan2}{rgb}{0.82,0.93,0.93}
\definecolor{LightCyan3}{rgb}{0.71,0.80,0.80}
\definecolor{LightCyan4}{rgb}{0.48,0.55,0.55}
\definecolor{LightCyan}{rgb}{0.88,1.00,1.00}
\definecolor{LightGoldenrod1}{rgb}{1.00,0.93,0.55}
\definecolor{LightGoldenrod2}{rgb}{0.93,0.86,0.51}
\definecolor{LightGoldenrod3}{rgb}{0.80,0.75,0.44}
\definecolor{LightGoldenrod4}{rgb}{0.55,0.51,0.30}
\definecolor{LightGoldenrodYellow}{rgb}{0.98,0.98,0.82}
\definecolor{LightGoldenrod}{rgb}{0.93,0.87,0.51}
\definecolor{LightGray}{rgb}{0.83,0.83,0.83}
\definecolor{LightGreen}{rgb}{0.56,0.93,0.56}
\definecolor{LightGrey}{rgb}{0.83,0.83,0.83}
\definecolor{LightPink1}{rgb}{1.00,0.68,0.73}
\definecolor{LightPink2}{rgb}{0.93,0.64,0.68}
\definecolor{LightPink3}{rgb}{0.80,0.55,0.58}
\definecolor{LightPink4}{rgb}{0.55,0.37,0.40}
\definecolor{LightPink}{rgb}{1.00,0.71,0.76}
\definecolor{LightSalmon1}{rgb}{1.00,0.63,0.48}
\definecolor{LightSalmon2}{rgb}{0.93,0.58,0.45}
\definecolor{LightSalmon3}{rgb}{0.80,0.51,0.38}
\definecolor{LightSalmon4}{rgb}{0.55,0.34,0.26}
\definecolor{LightSalmon}{rgb}{1.00,0.63,0.48}
\definecolor{LightSeaGreen}{rgb}{0.13,0.70,0.67}
\definecolor{LightSkyBlue1}{rgb}{0.69,0.89,1.00}
\definecolor{LightSkyBlue2}{rgb}{0.64,0.83,0.93}
\definecolor{LightSkyBlue3}{rgb}{0.55,0.71,0.80}
\definecolor{LightSkyBlue4}{rgb}{0.38,0.48,0.55}
\definecolor{LightSkyBlue}{rgb}{0.53,0.81,0.98}
\definecolor{LightSlateBlue}{rgb}{0.52,0.44,1.00}
\definecolor{LightSlateGray}{rgb}{0.47,0.53,0.60}
\definecolor{LightSlateGrey}{rgb}{0.47,0.53,0.60}
\definecolor{LightSteelBlue1}{rgb}{0.79,0.88,1.00}
\definecolor{LightSteelBlue2}{rgb}{0.74,0.82,0.93}
\definecolor{LightSteelBlue3}{rgb}{0.64,0.71,0.80}
\definecolor{LightSteelBlue4}{rgb}{0.43,0.48,0.55}
\definecolor{LightSteelBlue}{rgb}{0.69,0.77,0.87}
\definecolor{LightYellow1}{rgb}{1.00,1.00,0.88}
\definecolor{LightYellow2}{rgb}{0.93,0.93,0.82}
\definecolor{LightYellow3}{rgb}{0.80,0.80,0.71}
\definecolor{LightYellow4}{rgb}{0.55,0.55,0.48}
\definecolor{LightYellow}{rgb}{1.00,1.00,0.88}
\definecolor{LimeGreen}{rgb}{0.20,0.80,0.20}
\definecolor{MediumAquamarine}{rgb}{0.40,0.80,0.67}
\definecolor{MediumBlue}{rgb}{0.00,0.00,0.80}
\definecolor{MediumOrchid1}{rgb}{0.88,0.40,1.00}
\definecolor{MediumOrchid2}{rgb}{0.82,0.37,0.93}
\definecolor{MediumOrchid3}{rgb}{0.71,0.32,0.80}
\definecolor{MediumOrchid4}{rgb}{0.48,0.22,0.55}
\definecolor{MediumOrchid}{rgb}{0.73,0.33,0.83}
\definecolor{MediumPurple1}{rgb}{0.67,0.51,1.00}
\definecolor{MediumPurple2}{rgb}{0.62,0.47,0.93}
\definecolor{MediumPurple3}{rgb}{0.54,0.41,0.80}
\definecolor{MediumPurple4}{rgb}{0.36,0.28,0.55}
\definecolor{MediumPurple}{rgb}{0.58,0.44,0.86}
\definecolor{MediumSeaGreen}{rgb}{0.24,0.70,0.44}
\definecolor{MediumSlateBlue}{rgb}{0.48,0.41,0.93}
\definecolor{MediumSpringGreen}{rgb}{0.00,0.98,0.60}
\definecolor{MediumTurquoise}{rgb}{0.28,0.82,0.80}
\definecolor{MediumVioletRed}{rgb}{0.78,0.08,0.52}
\definecolor{MidnightBlue}{rgb}{0.10,0.10,0.44}
\definecolor{MintCream}{rgb}{0.96,1.00,0.98}
\definecolor{MistyRose1}{rgb}{1.00,0.89,0.88}
\definecolor{MistyRose2}{rgb}{0.93,0.84,0.82}
\definecolor{MistyRose3}{rgb}{0.80,0.72,0.71}
\definecolor{MistyRose4}{rgb}{0.55,0.49,0.48}
\definecolor{MistyRose}{rgb}{1.00,0.89,0.88}
\definecolor{NavajoWhite1}{rgb}{1.00,0.87,0.68}
\definecolor{NavajoWhite2}{rgb}{0.93,0.81,0.63}
\definecolor{NavajoWhite3}{rgb}{0.80,0.70,0.55}
\definecolor{NavajoWhite4}{rgb}{0.55,0.47,0.37}
\definecolor{NavajoWhite}{rgb}{1.00,0.87,0.68}
\definecolor{NavyBlue}{rgb}{0.00,0.00,0.50}
\definecolor{OldLace}{rgb}{0.99,0.96,0.90}
\definecolor{OliveDrab1}{rgb}{0.75,1.00,0.24}
\definecolor{OliveDrab2}{rgb}{0.70,0.93,0.23}
\definecolor{OliveDrab3}{rgb}{0.60,0.80,0.20}
\definecolor{OliveDrab4}{rgb}{0.41,0.55,0.13}
\definecolor{OliveDrab}{rgb}{0.42,0.56,0.14}
\definecolor{OrangeRed1}{rgb}{1.00,0.27,0.00}
\definecolor{OrangeRed2}{rgb}{0.93,0.25,0.00}
\definecolor{OrangeRed3}{rgb}{0.80,0.22,0.00}
\definecolor{OrangeRed4}{rgb}{0.55,0.15,0.00}
\definecolor{OrangeRed}{rgb}{1.00,0.27,0.00}
\definecolor{PaleGoldenrod}{rgb}{0.93,0.91,0.67}
\definecolor{PaleGreen1}{rgb}{0.60,1.00,0.60}
\definecolor{PaleGreen2}{rgb}{0.56,0.93,0.56}
\definecolor{PaleGreen3}{rgb}{0.49,0.80,0.49}
\definecolor{PaleGreen4}{rgb}{0.33,0.55,0.33}
\definecolor{PaleGreen}{rgb}{0.60,0.98,0.60}
\definecolor{PaleTurquoise1}{rgb}{0.73,1.00,1.00}
\definecolor{PaleTurquoise2}{rgb}{0.68,0.93,0.93}
\definecolor{PaleTurquoise3}{rgb}{0.59,0.80,0.80}
\definecolor{PaleTurquoise4}{rgb}{0.40,0.55,0.55}
\definecolor{PaleTurquoise}{rgb}{0.69,0.93,0.93}
\definecolor{PaleVioletRed1}{rgb}{1.00,0.51,0.67}
\definecolor{PaleVioletRed2}{rgb}{0.93,0.47,0.62}
\definecolor{PaleVioletRed3}{rgb}{0.80,0.41,0.54}
\definecolor{PaleVioletRed4}{rgb}{0.55,0.28,0.36}
\definecolor{PaleVioletRed}{rgb}{0.86,0.44,0.58}
\definecolor{PapayaWhip}{rgb}{1.00,0.94,0.84}
\definecolor{PeachPuff1}{rgb}{1.00,0.85,0.73}
\definecolor{PeachPuff2}{rgb}{0.93,0.80,0.68}
\definecolor{PeachPuff3}{rgb}{0.80,0.69,0.58}
\definecolor{PeachPuff4}{rgb}{0.55,0.47,0.40}
\definecolor{PeachPuff}{rgb}{1.00,0.85,0.73}
\definecolor{PowderBlue}{rgb}{0.69,0.88,0.90}
\definecolor{RosyBrown1}{rgb}{1.00,0.76,0.76}
\definecolor{RosyBrown2}{rgb}{0.93,0.71,0.71}
\definecolor{RosyBrown3}{rgb}{0.80,0.61,0.61}
\definecolor{RosyBrown4}{rgb}{0.55,0.41,0.41}
\definecolor{RosyBrown}{rgb}{0.74,0.56,0.56}
\definecolor{RoyalBlue1}{rgb}{0.28,0.46,1.00}
\definecolor{RoyalBlue2}{rgb}{0.26,0.43,0.93}
\definecolor{RoyalBlue3}{rgb}{0.23,0.37,0.80}
\definecolor{RoyalBlue4}{rgb}{0.15,0.25,0.55}
\definecolor{RoyalBlue}{rgb}{0.25,0.41,0.88}
\definecolor{SaddleBrown}{rgb}{0.55,0.27,0.07}
\definecolor{SandyBrown}{rgb}{0.96,0.64,0.38}
\definecolor{SeaGreen1}{rgb}{0.33,1.00,0.62}
\definecolor{SeaGreen2}{rgb}{0.31,0.93,0.58}
\definecolor{SeaGreen3}{rgb}{0.26,0.80,0.50}
\definecolor{SeaGreen4}{rgb}{0.18,0.55,0.34}
\definecolor{SeaGreen}{rgb}{0.18,0.55,0.34}
\definecolor{SkyBlue1}{rgb}{0.53,0.81,1.00}
\definecolor{SkyBlue2}{rgb}{0.49,0.75,0.93}
\definecolor{SkyBlue3}{rgb}{0.42,0.65,0.80}
\definecolor{SkyBlue4}{rgb}{0.29,0.44,0.55}
\definecolor{SkyBlue}{rgb}{0.53,0.81,0.92}
\definecolor{SlateBlue1}{rgb}{0.51,0.44,1.00}
\definecolor{SlateBlue2}{rgb}{0.48,0.40,0.93}
\definecolor{SlateBlue3}{rgb}{0.41,0.35,0.80}
\definecolor{SlateBlue4}{rgb}{0.28,0.24,0.55}
\definecolor{SlateBlue}{rgb}{0.42,0.35,0.80}
\definecolor{SlateGray1}{rgb}{0.78,0.89,1.00}
\definecolor{SlateGray2}{rgb}{0.73,0.83,0.93}
\definecolor{SlateGray3}{rgb}{0.62,0.71,0.80}
\definecolor{SlateGray4}{rgb}{0.42,0.48,0.55}
\definecolor{SlateGray}{rgb}{0.44,0.50,0.56}
\definecolor{SlateGrey}{rgb}{0.44,0.50,0.56}
\definecolor{SpringGreen1}{rgb}{0.00,1.00,0.50}
\definecolor{SpringGreen2}{rgb}{0.00,0.93,0.46}
\definecolor{SpringGreen3}{rgb}{0.00,0.80,0.40}
\definecolor{SpringGreen4}{rgb}{0.00,0.55,0.27}
\definecolor{SpringGreen}{rgb}{0.00,1.00,0.50}
\definecolor{SteelBlue1}{rgb}{0.39,0.72,1.00}
\definecolor{SteelBlue2}{rgb}{0.36,0.67,0.93}
\definecolor{SteelBlue3}{rgb}{0.31,0.58,0.80}
\definecolor{SteelBlue4}{rgb}{0.21,0.39,0.55}
\definecolor{SteelBlue}{rgb}{0.27,0.51,0.71}
\definecolor{VioletRed1}{rgb}{1.00,0.24,0.59}
\definecolor{VioletRed2}{rgb}{0.93,0.23,0.55}
\definecolor{VioletRed3}{rgb}{0.80,0.20,0.47}
\definecolor{VioletRed4}{rgb}{0.55,0.13,0.32}
\definecolor{VioletRed}{rgb}{0.82,0.13,0.56}
\definecolor{WhiteSmoke}{rgb}{0.96,0.96,0.96}
\definecolor{YellowGreen}{rgb}{0.60,0.80,0.20}
\definecolor{aliceblue}{rgb}{0.94,0.97,1.00}
\definecolor{antiquewhite}{rgb}{0.98,0.92,0.84}
\definecolor{aquamarine1}{rgb}{0.50,1.00,0.83}
\definecolor{aquamarine2}{rgb}{0.46,0.93,0.78}
\definecolor{aquamarine3}{rgb}{0.40,0.80,0.67}
\definecolor{aquamarine4}{rgb}{0.27,0.55,0.45}
\definecolor{aquamarine}{rgb}{0.50,1.00,0.83}
\definecolor{azure1}{rgb}{0.94,1.00,1.00}
\definecolor{azure2}{rgb}{0.88,0.93,0.93}
\definecolor{azure3}{rgb}{0.76,0.80,0.80}
\definecolor{azure4}{rgb}{0.51,0.55,0.55}
\definecolor{azure}{rgb}{0.94,1.00,1.00}
\definecolor{beige}{rgb}{0.96,0.96,0.86}
\definecolor{bisque1}{rgb}{1.00,0.89,0.77}
\definecolor{bisque2}{rgb}{0.93,0.84,0.72}
\definecolor{bisque3}{rgb}{0.80,0.72,0.62}
\definecolor{bisque4}{rgb}{0.55,0.49,0.42}
\definecolor{bisque}{rgb}{1.00,0.89,0.77}
\definecolor{black}{rgb}{0.00,0.00,0.00}
\definecolor{blanchedalmond}{rgb}{1.00,0.92,0.80}
\definecolor{blue1}{rgb}{0.00,0.00,1.00}
\definecolor{blue2}{rgb}{0.00,0.00,0.93}
\definecolor{blue3}{rgb}{0.00,0.00,0.80}
\definecolor{blue4}{rgb}{0.00,0.00,0.55}
\definecolor{blueviolet}{rgb}{0.54,0.17,0.89}
\definecolor{blue}{rgb}{0.00,0.00,1.00}
\definecolor{brown1}{rgb}{1.00,0.25,0.25}
\definecolor{brown2}{rgb}{0.93,0.23,0.23}
\definecolor{brown3}{rgb}{0.80,0.20,0.20}
\definecolor{brown4}{rgb}{0.55,0.14,0.14}
\definecolor{brown}{rgb}{0.65,0.16,0.16}
\definecolor{burlywood1}{rgb}{1.00,0.83,0.61}
\definecolor{burlywood2}{rgb}{0.93,0.77,0.57}
\definecolor{burlywood3}{rgb}{0.80,0.67,0.49}
\definecolor{burlywood4}{rgb}{0.55,0.45,0.33}
\definecolor{burlywood}{rgb}{0.87,0.72,0.53}
\definecolor{cadetblue}{rgb}{0.37,0.62,0.63}
\definecolor{chartreuse1}{rgb}{0.50,1.00,0.00}
\definecolor{chartreuse2}{rgb}{0.46,0.93,0.00}
\definecolor{chartreuse3}{rgb}{0.40,0.80,0.00}
\definecolor{chartreuse4}{rgb}{0.27,0.55,0.00}
\definecolor{chartreuse}{rgb}{0.50,1.00,0.00}
\definecolor{chocolate1}{rgb}{1.00,0.50,0.14}
\definecolor{chocolate2}{rgb}{0.93,0.46,0.13}
\definecolor{chocolate3}{rgb}{0.80,0.40,0.11}
\definecolor{chocolate4}{rgb}{0.55,0.27,0.07}
\definecolor{chocolate}{rgb}{0.82,0.41,0.12}
\definecolor{coral1}{rgb}{1.00,0.45,0.34}
\definecolor{coral2}{rgb}{0.93,0.42,0.31}
\definecolor{coral3}{rgb}{0.80,0.36,0.27}
\definecolor{coral4}{rgb}{0.55,0.24,0.18}
\definecolor{coral}{rgb}{1.00,0.50,0.31}
\definecolor{cornflowerblue}{rgb}{0.39,0.58,0.93}
\definecolor{cornsilk1}{rgb}{1.00,0.97,0.86}
\definecolor{cornsilk2}{rgb}{0.93,0.91,0.80}
\definecolor{cornsilk3}{rgb}{0.80,0.78,0.69}
\definecolor{cornsilk4}{rgb}{0.55,0.53,0.47}
\definecolor{cornsilk}{rgb}{1.00,0.97,0.86}
\definecolor{cyan1}{rgb}{0.00,1.00,1.00}
\definecolor{cyan2}{rgb}{0.00,0.93,0.93}
\definecolor{cyan3}{rgb}{0.00,0.80,0.80}
\definecolor{cyan4}{rgb}{0.00,0.55,0.55}
\definecolor{cyan}{rgb}{0.00,1.00,1.00}
\definecolor{darkblue}{rgb}{0.00,0.00,0.55}
\definecolor{darkcyan}{rgb}{0.00,0.55,0.55}
\definecolor{darkgoldenrod}{rgb}{0.72,0.53,0.04}
\definecolor{darkgray}{rgb}{0.66,0.66,0.66}
\definecolor{darkgreen}{rgb}{0.00,0.39,0.00}
\definecolor{darkgrey}{rgb}{0.66,0.66,0.66}
\definecolor{darkkhaki}{rgb}{0.74,0.72,0.42}
\definecolor{darkmagenta}{rgb}{0.55,0.00,0.55}
\definecolor{darkolive}{rgb}{0.33,0.42,0.18}
\definecolor{darkorange}{rgb}{1.00,0.55,0.00}
\definecolor{darkorchid}{rgb}{0.60,0.20,0.80}
\definecolor{darkred}{rgb}{0.55,0.00,0.00}
\definecolor{darksalmon}{rgb}{0.91,0.59,0.48}
\definecolor{darksea}{rgb}{0.56,0.74,0.56}
\definecolor{darkslate}{rgb}{0.18,0.31,0.31}
\definecolor{darkslate}{rgb}{0.18,0.31,0.31}
\definecolor{darkslate}{rgb}{0.28,0.24,0.55}
\definecolor{darkturquoise}{rgb}{0.00,0.81,0.82}
\definecolor{darkviolet}{rgb}{0.58,0.00,0.83}
\definecolor{deeppink}{rgb}{1.00,0.08,0.58}
\definecolor{deepsky}{rgb}{0.00,0.75,1.00}
\definecolor{dimgray}{rgb}{0.41,0.41,0.41}
\definecolor{dimgrey}{rgb}{0.41,0.41,0.41}
\definecolor{dodgerblue}{rgb}{0.12,0.56,1.00}
\definecolor{firebrick1}{rgb}{1.00,0.19,0.19}
\definecolor{firebrick2}{rgb}{0.93,0.17,0.17}
\definecolor{firebrick3}{rgb}{0.80,0.15,0.15}
\definecolor{firebrick4}{rgb}{0.55,0.10,0.10}
\definecolor{firebrick}{rgb}{0.70,0.13,0.13}
\definecolor{floralwhite}{rgb}{1.00,0.98,0.94}
\definecolor{forestgreen}{rgb}{0.13,0.55,0.13}
\definecolor{gainsboro}{rgb}{0.86,0.86,0.86}
\definecolor{ghostwhite}{rgb}{0.97,0.97,1.00}
\definecolor{gold1}{rgb}{1.00,0.84,0.00}
\definecolor{gold2}{rgb}{0.93,0.79,0.00}
\definecolor{gold3}{rgb}{0.80,0.68,0.00}
\definecolor{gold4}{rgb}{0.55,0.46,0.00}
\definecolor{goldenrod1}{rgb}{1.00,0.76,0.15}
\definecolor{goldenrod2}{rgb}{0.93,0.71,0.13}
\definecolor{goldenrod3}{rgb}{0.80,0.61,0.11}
\definecolor{goldenrod4}{rgb}{0.55,0.41,0.08}
\definecolor{goldenrod}{rgb}{0.85,0.65,0.13}
\definecolor{gold}{rgb}{1.00,0.84,0.00}
\definecolor{gray0}{rgb}{0.00,0.00,0.00}
\definecolor{gray100}{rgb}{1.00,1.00,1.00}
\definecolor{gray10}{rgb}{0.10,0.10,0.10}
\definecolor{gray11}{rgb}{0.11,0.11,0.11}
\definecolor{gray12}{rgb}{0.12,0.12,0.12}
\definecolor{gray13}{rgb}{0.13,0.13,0.13}
\definecolor{gray14}{rgb}{0.14,0.14,0.14}
\definecolor{gray15}{rgb}{0.15,0.15,0.15}
\definecolor{gray16}{rgb}{0.16,0.16,0.16}
\definecolor{gray17}{rgb}{0.17,0.17,0.17}
\definecolor{gray18}{rgb}{0.18,0.18,0.18}
\definecolor{gray19}{rgb}{0.19,0.19,0.19}
\definecolor{gray1}{rgb}{0.01,0.01,0.01}
\definecolor{gray20}{rgb}{0.20,0.20,0.20}
\definecolor{gray21}{rgb}{0.21,0.21,0.21}
\definecolor{gray22}{rgb}{0.22,0.22,0.22}
\definecolor{gray23}{rgb}{0.23,0.23,0.23}
\definecolor{gray24}{rgb}{0.24,0.24,0.24}
\definecolor{gray25}{rgb}{0.25,0.25,0.25}
\definecolor{gray26}{rgb}{0.26,0.26,0.26}
\definecolor{gray27}{rgb}{0.27,0.27,0.27}
\definecolor{gray28}{rgb}{0.28,0.28,0.28}
\definecolor{gray29}{rgb}{0.29,0.29,0.29}
\definecolor{gray2}{rgb}{0.02,0.02,0.02}
\definecolor{gray30}{rgb}{0.30,0.30,0.30}
\definecolor{gray31}{rgb}{0.31,0.31,0.31}
\definecolor{gray32}{rgb}{0.32,0.32,0.32}
\definecolor{gray33}{rgb}{0.33,0.33,0.33}
\definecolor{gray34}{rgb}{0.34,0.34,0.34}
\definecolor{gray35}{rgb}{0.35,0.35,0.35}
\definecolor{gray36}{rgb}{0.36,0.36,0.36}
\definecolor{gray37}{rgb}{0.37,0.37,0.37}
\definecolor{gray38}{rgb}{0.38,0.38,0.38}
\definecolor{gray39}{rgb}{0.39,0.39,0.39}
\definecolor{gray3}{rgb}{0.03,0.03,0.03}
\definecolor{gray40}{rgb}{0.40,0.40,0.40}
\definecolor{gray41}{rgb}{0.41,0.41,0.41}
\definecolor{gray42}{rgb}{0.42,0.42,0.42}
\definecolor{gray43}{rgb}{0.43,0.43,0.43}
\definecolor{gray44}{rgb}{0.44,0.44,0.44}
\definecolor{gray45}{rgb}{0.45,0.45,0.45}
\definecolor{gray46}{rgb}{0.46,0.46,0.46}
\definecolor{gray47}{rgb}{0.47,0.47,0.47}
\definecolor{gray48}{rgb}{0.48,0.48,0.48}
\definecolor{gray49}{rgb}{0.49,0.49,0.49}
\definecolor{gray4}{rgb}{0.04,0.04,0.04}
\definecolor{gray50}{rgb}{0.50,0.50,0.50}
\definecolor{gray51}{rgb}{0.51,0.51,0.51}
\definecolor{gray52}{rgb}{0.52,0.52,0.52}
\definecolor{gray53}{rgb}{0.53,0.53,0.53}
\definecolor{gray54}{rgb}{0.54,0.54,0.54}
\definecolor{gray55}{rgb}{0.55,0.55,0.55}
\definecolor{gray56}{rgb}{0.56,0.56,0.56}
\definecolor{gray57}{rgb}{0.57,0.57,0.57}
\definecolor{gray58}{rgb}{0.58,0.58,0.58}
\definecolor{gray59}{rgb}{0.59,0.59,0.59}
\definecolor{gray5}{rgb}{0.05,0.05,0.05}
\definecolor{gray60}{rgb}{0.60,0.60,0.60}
\definecolor{gray61}{rgb}{0.61,0.61,0.61}
\definecolor{gray62}{rgb}{0.62,0.62,0.62}
\definecolor{gray63}{rgb}{0.63,0.63,0.63}
\definecolor{gray64}{rgb}{0.64,0.64,0.64}
\definecolor{gray65}{rgb}{0.65,0.65,0.65}
\definecolor{gray66}{rgb}{0.66,0.66,0.66}
\definecolor{gray67}{rgb}{0.67,0.67,0.67}
\definecolor{gray68}{rgb}{0.68,0.68,0.68}
\definecolor{gray69}{rgb}{0.69,0.69,0.69}
\definecolor{gray6}{rgb}{0.06,0.06,0.06}
\definecolor{gray70}{rgb}{0.70,0.70,0.70}
\definecolor{gray71}{rgb}{0.71,0.71,0.71}
\definecolor{gray72}{rgb}{0.72,0.72,0.72}
\definecolor{gray73}{rgb}{0.73,0.73,0.73}
\definecolor{gray74}{rgb}{0.74,0.74,0.74}
\definecolor{gray75}{rgb}{0.75,0.75,0.75}
\definecolor{gray76}{rgb}{0.76,0.76,0.76}
\definecolor{gray77}{rgb}{0.77,0.77,0.77}
\definecolor{gray78}{rgb}{0.78,0.78,0.78}
\definecolor{gray79}{rgb}{0.79,0.79,0.79}
\definecolor{gray7}{rgb}{0.07,0.07,0.07}
\definecolor{gray80}{rgb}{0.80,0.80,0.80}
\definecolor{gray81}{rgb}{0.81,0.81,0.81}
\definecolor{gray82}{rgb}{0.82,0.82,0.82}
\definecolor{gray83}{rgb}{0.83,0.83,0.83}
\definecolor{gray84}{rgb}{0.84,0.84,0.84}
\definecolor{gray85}{rgb}{0.85,0.85,0.85}
\definecolor{gray86}{rgb}{0.86,0.86,0.86}
\definecolor{gray87}{rgb}{0.87,0.87,0.87}
\definecolor{gray88}{rgb}{0.88,0.88,0.88}
\definecolor{gray89}{rgb}{0.89,0.89,0.89}
\definecolor{gray8}{rgb}{0.08,0.08,0.08}
\definecolor{gray90}{rgb}{0.90,0.90,0.90}
\definecolor{gray91}{rgb}{0.91,0.91,0.91}
\definecolor{gray92}{rgb}{0.92,0.92,0.92}
\definecolor{gray93}{rgb}{0.93,0.93,0.93}
\definecolor{gray94}{rgb}{0.94,0.94,0.94}
\definecolor{gray95}{rgb}{0.95,0.95,0.95}
\definecolor{gray96}{rgb}{0.96,0.96,0.96}
\definecolor{gray97}{rgb}{0.97,0.97,0.97}
\definecolor{gray98}{rgb}{0.98,0.98,0.98}
\definecolor{gray99}{rgb}{0.99,0.99,0.99}
\definecolor{gray9}{rgb}{0.09,0.09,0.09}
\definecolor{gray}{rgb}{0.75,0.75,0.75}
\definecolor{green1}{rgb}{0.00,1.00,0.00}
\definecolor{green2}{rgb}{0.00,0.93,0.00}
\definecolor{green3}{rgb}{0.00,0.80,0.00}
\definecolor{green4}{rgb}{0.00,0.55,0.00}
\definecolor{greenyellow}{rgb}{0.68,1.00,0.18}
\definecolor{green}{rgb}{0.00,1.00,0.00}
\definecolor{grey0}{rgb}{0.00,0.00,0.00}
\definecolor{grey100}{rgb}{1.00,1.00,1.00}
\definecolor{grey10}{rgb}{0.10,0.10,0.10}
\definecolor{grey11}{rgb}{0.11,0.11,0.11}
\definecolor{grey12}{rgb}{0.12,0.12,0.12}
\definecolor{grey13}{rgb}{0.13,0.13,0.13}
\definecolor{grey14}{rgb}{0.14,0.14,0.14}
\definecolor{grey15}{rgb}{0.15,0.15,0.15}
\definecolor{grey16}{rgb}{0.16,0.16,0.16}
\definecolor{grey17}{rgb}{0.17,0.17,0.17}
\definecolor{grey18}{rgb}{0.18,0.18,0.18}
\definecolor{grey19}{rgb}{0.19,0.19,0.19}
\definecolor{grey1}{rgb}{0.01,0.01,0.01}
\definecolor{grey20}{rgb}{0.20,0.20,0.20}
\definecolor{grey21}{rgb}{0.21,0.21,0.21}
\definecolor{grey22}{rgb}{0.22,0.22,0.22}
\definecolor{grey23}{rgb}{0.23,0.23,0.23}
\definecolor{grey24}{rgb}{0.24,0.24,0.24}
\definecolor{grey25}{rgb}{0.25,0.25,0.25}
\definecolor{grey26}{rgb}{0.26,0.26,0.26}
\definecolor{grey27}{rgb}{0.27,0.27,0.27}
\definecolor{grey28}{rgb}{0.28,0.28,0.28}
\definecolor{grey29}{rgb}{0.29,0.29,0.29}
\definecolor{grey2}{rgb}{0.02,0.02,0.02}
\definecolor{grey30}{rgb}{0.30,0.30,0.30}
\definecolor{grey31}{rgb}{0.31,0.31,0.31}
\definecolor{grey32}{rgb}{0.32,0.32,0.32}
\definecolor{grey33}{rgb}{0.33,0.33,0.33}
\definecolor{grey34}{rgb}{0.34,0.34,0.34}
\definecolor{grey35}{rgb}{0.35,0.35,0.35}
\definecolor{grey36}{rgb}{0.36,0.36,0.36}
\definecolor{grey37}{rgb}{0.37,0.37,0.37}
\definecolor{grey38}{rgb}{0.38,0.38,0.38}
\definecolor{grey39}{rgb}{0.39,0.39,0.39}
\definecolor{grey3}{rgb}{0.03,0.03,0.03}
\definecolor{grey40}{rgb}{0.40,0.40,0.40}
\definecolor{grey41}{rgb}{0.41,0.41,0.41}
\definecolor{grey42}{rgb}{0.42,0.42,0.42}
\definecolor{grey43}{rgb}{0.43,0.43,0.43}
\definecolor{grey44}{rgb}{0.44,0.44,0.44}
\definecolor{grey45}{rgb}{0.45,0.45,0.45}
\definecolor{grey46}{rgb}{0.46,0.46,0.46}
\definecolor{grey47}{rgb}{0.47,0.47,0.47}
\definecolor{grey48}{rgb}{0.48,0.48,0.48}
\definecolor{grey49}{rgb}{0.49,0.49,0.49}
\definecolor{grey4}{rgb}{0.04,0.04,0.04}
\definecolor{grey50}{rgb}{0.50,0.50,0.50}
\definecolor{grey51}{rgb}{0.51,0.51,0.51}
\definecolor{grey52}{rgb}{0.52,0.52,0.52}
\definecolor{grey53}{rgb}{0.53,0.53,0.53}
\definecolor{grey54}{rgb}{0.54,0.54,0.54}
\definecolor{grey55}{rgb}{0.55,0.55,0.55}
\definecolor{grey56}{rgb}{0.56,0.56,0.56}
\definecolor{grey57}{rgb}{0.57,0.57,0.57}
\definecolor{grey58}{rgb}{0.58,0.58,0.58}
\definecolor{grey59}{rgb}{0.59,0.59,0.59}
\definecolor{grey5}{rgb}{0.05,0.05,0.05}
\definecolor{grey60}{rgb}{0.60,0.60,0.60}
\definecolor{grey61}{rgb}{0.61,0.61,0.61}
\definecolor{grey62}{rgb}{0.62,0.62,0.62}
\definecolor{grey63}{rgb}{0.63,0.63,0.63}
\definecolor{grey64}{rgb}{0.64,0.64,0.64}
\definecolor{grey65}{rgb}{0.65,0.65,0.65}
\definecolor{grey66}{rgb}{0.66,0.66,0.66}
\definecolor{grey67}{rgb}{0.67,0.67,0.67}
\definecolor{grey68}{rgb}{0.68,0.68,0.68}
\definecolor{grey69}{rgb}{0.69,0.69,0.69}
\definecolor{grey6}{rgb}{0.06,0.06,0.06}
\definecolor{grey70}{rgb}{0.70,0.70,0.70}
\definecolor{grey71}{rgb}{0.71,0.71,0.71}
\definecolor{grey72}{rgb}{0.72,0.72,0.72}
\definecolor{grey73}{rgb}{0.73,0.73,0.73}
\definecolor{grey74}{rgb}{0.74,0.74,0.74}
\definecolor{grey75}{rgb}{0.75,0.75,0.75}
\definecolor{grey76}{rgb}{0.76,0.76,0.76}
\definecolor{grey77}{rgb}{0.77,0.77,0.77}
\definecolor{grey78}{rgb}{0.78,0.78,0.78}
\definecolor{grey79}{rgb}{0.79,0.79,0.79}
\definecolor{grey7}{rgb}{0.07,0.07,0.07}
\definecolor{grey80}{rgb}{0.80,0.80,0.80}
\definecolor{grey81}{rgb}{0.81,0.81,0.81}
\definecolor{grey82}{rgb}{0.82,0.82,0.82}
\definecolor{grey83}{rgb}{0.83,0.83,0.83}
\definecolor{grey84}{rgb}{0.84,0.84,0.84}
\definecolor{grey85}{rgb}{0.85,0.85,0.85}
\definecolor{grey86}{rgb}{0.86,0.86,0.86}
\definecolor{grey87}{rgb}{0.87,0.87,0.87}
\definecolor{grey88}{rgb}{0.88,0.88,0.88}
\definecolor{grey89}{rgb}{0.89,0.89,0.89}
\definecolor{grey8}{rgb}{0.08,0.08,0.08}
\definecolor{grey90}{rgb}{0.90,0.90,0.90}
\definecolor{grey91}{rgb}{0.91,0.91,0.91}
\definecolor{grey92}{rgb}{0.92,0.92,0.92}
\definecolor{grey93}{rgb}{0.93,0.93,0.93}
\definecolor{grey94}{rgb}{0.94,0.94,0.94}
\definecolor{grey95}{rgb}{0.95,0.95,0.95}
\definecolor{grey96}{rgb}{0.96,0.96,0.96}
\definecolor{grey97}{rgb}{0.97,0.97,0.97}
\definecolor{grey98}{rgb}{0.98,0.98,0.98}
\definecolor{grey99}{rgb}{0.99,0.99,0.99}
\definecolor{grey9}{rgb}{0.09,0.09,0.09}
\definecolor{grey}{rgb}{0.75,0.75,0.75}
\definecolor{honeydew1}{rgb}{0.94,1.00,0.94}
\definecolor{honeydew2}{rgb}{0.88,0.93,0.88}
\definecolor{honeydew3}{rgb}{0.76,0.80,0.76}
\definecolor{honeydew4}{rgb}{0.51,0.55,0.51}
\definecolor{honeydew}{rgb}{0.94,1.00,0.94}
\definecolor{hotpink}{rgb}{1.00,0.41,0.71}
\definecolor{indianred}{rgb}{0.80,0.36,0.36}
\definecolor{ivory1}{rgb}{1.00,1.00,0.94}
\definecolor{ivory2}{rgb}{0.93,0.93,0.88}
\definecolor{ivory3}{rgb}{0.80,0.80,0.76}
\definecolor{ivory4}{rgb}{0.55,0.55,0.51}
\definecolor{ivory}{rgb}{1.00,1.00,0.94}
\definecolor{khaki1}{rgb}{1.00,0.96,0.56}
\definecolor{khaki2}{rgb}{0.93,0.90,0.52}
\definecolor{khaki3}{rgb}{0.80,0.78,0.45}
\definecolor{khaki4}{rgb}{0.55,0.53,0.31}
\definecolor{khaki}{rgb}{0.94,0.90,0.55}
\definecolor{lavenderblush}{rgb}{1.00,0.94,0.96}
\definecolor{lavender}{rgb}{0.90,0.90,0.98}
\definecolor{lawngreen}{rgb}{0.49,0.99,0.00}
\definecolor{lemonchiffon}{rgb}{1.00,0.98,0.80}
\definecolor{lightblue}{rgb}{0.68,0.85,0.90}
\definecolor{lightcoral}{rgb}{0.94,0.50,0.50}
\definecolor{lightcyan}{rgb}{0.88,1.00,1.00}
\definecolor{lightgoldenrod}{rgb}{0.93,0.87,0.51}
\definecolor{lightgoldenrod}{rgb}{0.98,0.98,0.82}
\definecolor{lightgray}{rgb}{0.83,0.83,0.83}
\definecolor{lightgreen}{rgb}{0.56,0.93,0.56}
\definecolor{lightgrey}{rgb}{0.83,0.83,0.83}
\definecolor{lightpink}{rgb}{1.00,0.71,0.76}
\definecolor{lightsalmon}{rgb}{1.00,0.63,0.48}
\definecolor{lightsea}{rgb}{0.13,0.70,0.67}
\definecolor{lightsky}{rgb}{0.53,0.81,0.98}
\definecolor{lightslate}{rgb}{0.47,0.53,0.60}
\definecolor{lightslate}{rgb}{0.47,0.53,0.60}
\definecolor{lightslate}{rgb}{0.52,0.44,1.00}
\definecolor{lightsteel}{rgb}{0.69,0.77,0.87}
\definecolor{lightyellow}{rgb}{1.00,1.00,0.88}
\definecolor{limegreen}{rgb}{0.20,0.80,0.20}
\definecolor{linen}{rgb}{0.98,0.94,0.90}
\definecolor{magenta1}{rgb}{1.00,0.00,1.00}
\definecolor{magenta2}{rgb}{0.93,0.00,0.93}
\definecolor{magenta3}{rgb}{0.80,0.00,0.80}
\definecolor{magenta4}{rgb}{0.55,0.00,0.55}
\definecolor{magenta}{rgb}{1.00,0.00,1.00}
\definecolor{maroon1}{rgb}{1.00,0.20,0.70}
\definecolor{maroon2}{rgb}{0.93,0.19,0.65}
\definecolor{maroon3}{rgb}{0.80,0.16,0.56}
\definecolor{maroon4}{rgb}{0.55,0.11,0.38}
\definecolor{maroon}{rgb}{0.69,0.19,0.38}
\definecolor{mediumaquamarine}{rgb}{0.40,0.80,0.67}
\definecolor{mediumblue}{rgb}{0.00,0.00,0.80}
\definecolor{mediumorchid}{rgb}{0.73,0.33,0.83}
\definecolor{mediumpurple}{rgb}{0.58,0.44,0.86}
\definecolor{mediumsea}{rgb}{0.24,0.70,0.44}
\definecolor{mediumslate}{rgb}{0.48,0.41,0.93}
\definecolor{mediumspring}{rgb}{0.00,0.98,0.60}
\definecolor{mediumturquoise}{rgb}{0.28,0.82,0.80}
\definecolor{mediumviolet}{rgb}{0.78,0.08,0.52}
\definecolor{midnightblue}{rgb}{0.10,0.10,0.44}
\definecolor{mintcream}{rgb}{0.96,1.00,0.98}
\definecolor{mistyrose}{rgb}{1.00,0.89,0.88}
\definecolor{moccasin}{rgb}{1.00,0.89,0.71}
\definecolor{navajowhite}{rgb}{1.00,0.87,0.68}
\definecolor{navyblue}{rgb}{0.00,0.00,0.50}
\definecolor{navy}{rgb}{0.00,0.00,0.50}
\definecolor{oldlace}{rgb}{0.99,0.96,0.90}
\definecolor{olivedrab}{rgb}{0.42,0.56,0.14}
\definecolor{orange1}{rgb}{1.00,0.65,0.00}
\definecolor{orange2}{rgb}{0.93,0.60,0.00}
\definecolor{orange3}{rgb}{0.80,0.52,0.00}
\definecolor{orange4}{rgb}{0.55,0.35,0.00}
\definecolor{orangered}{rgb}{1.00,0.27,0.00}
\definecolor{orange}{rgb}{1.00,0.65,0.00}
\definecolor{orchid1}{rgb}{1.00,0.51,0.98}
\definecolor{orchid2}{rgb}{0.93,0.48,0.91}
\definecolor{orchid3}{rgb}{0.80,0.41,0.79}
\definecolor{orchid4}{rgb}{0.55,0.28,0.54}
\definecolor{orchid}{rgb}{0.85,0.44,0.84}
\definecolor{palegoldenrod}{rgb}{0.93,0.91,0.67}
\definecolor{palegreen}{rgb}{0.60,0.98,0.60}
\definecolor{paleturquoise}{rgb}{0.69,0.93,0.93}
\definecolor{paleviolet}{rgb}{0.86,0.44,0.58}
\definecolor{papayawhip}{rgb}{1.00,0.94,0.84}
\definecolor{peachpuff}{rgb}{1.00,0.85,0.73}
\definecolor{peru}{rgb}{0.80,0.52,0.25}
\definecolor{pink1}{rgb}{1.00,0.71,0.77}
\definecolor{pink2}{rgb}{0.93,0.66,0.72}
\definecolor{pink3}{rgb}{0.80,0.57,0.62}
\definecolor{pink4}{rgb}{0.55,0.39,0.42}
\definecolor{pink}{rgb}{1.00,0.75,0.80}
\definecolor{plum1}{rgb}{1.00,0.73,1.00}
\definecolor{plum2}{rgb}{0.93,0.68,0.93}
\definecolor{plum3}{rgb}{0.80,0.59,0.80}
\definecolor{plum4}{rgb}{0.55,0.40,0.55}
\definecolor{plum}{rgb}{0.87,0.63,0.87}
\definecolor{powderblue}{rgb}{0.69,0.88,0.90}
\definecolor{purple1}{rgb}{0.61,0.19,1.00}
\definecolor{purple2}{rgb}{0.57,0.17,0.93}
\definecolor{purple3}{rgb}{0.49,0.15,0.80}
\definecolor{purple4}{rgb}{0.33,0.10,0.55}
\definecolor{purple}{rgb}{0.63,0.13,0.94}
\definecolor{red1}{rgb}{1.00,0.00,0.00}
\definecolor{red2}{rgb}{0.93,0.00,0.00}
\definecolor{red3}{rgb}{0.80,0.00,0.00}
\definecolor{red4}{rgb}{0.55,0.00,0.00}
\definecolor{red}{rgb}{1.00,0.00,0.00}
\definecolor{rosybrown}{rgb}{0.74,0.56,0.56}
\definecolor{royalblue}{rgb}{0.25,0.41,0.88}
\definecolor{saddlebrown}{rgb}{0.55,0.27,0.07}
\definecolor{salmon1}{rgb}{1.00,0.55,0.41}
\definecolor{salmon2}{rgb}{0.93,0.51,0.38}
\definecolor{salmon3}{rgb}{0.80,0.44,0.33}
\definecolor{salmon4}{rgb}{0.55,0.30,0.22}
\definecolor{salmon}{rgb}{0.98,0.50,0.45}
\definecolor{sandybrown}{rgb}{0.96,0.64,0.38}
\definecolor{seagreen}{rgb}{0.18,0.55,0.34}
\definecolor{seashell1}{rgb}{1.00,0.96,0.93}
\definecolor{seashell2}{rgb}{0.93,0.90,0.87}
\definecolor{seashell3}{rgb}{0.80,0.77,0.75}
\definecolor{seashell4}{rgb}{0.55,0.53,0.51}
\definecolor{seashell}{rgb}{1.00,0.96,0.93}
\definecolor{sienna1}{rgb}{1.00,0.51,0.28}
\definecolor{sienna2}{rgb}{0.93,0.47,0.26}
\definecolor{sienna3}{rgb}{0.80,0.41,0.22}
\definecolor{sienna4}{rgb}{0.55,0.28,0.15}
\definecolor{sienna}{rgb}{0.63,0.32,0.18}
\definecolor{skyblue}{rgb}{0.53,0.81,0.92}
\definecolor{slateblue}{rgb}{0.42,0.35,0.80}
\definecolor{slategray}{rgb}{0.44,0.50,0.56}
\definecolor{slategrey}{rgb}{0.44,0.50,0.56}
\definecolor{snow1}{rgb}{1.00,0.98,0.98}
\definecolor{snow2}{rgb}{0.93,0.91,0.91}
\definecolor{snow3}{rgb}{0.80,0.79,0.79}
\definecolor{snow4}{rgb}{0.55,0.54,0.54}
\definecolor{snow}{rgb}{1.00,0.98,0.98}
\definecolor{springgreen}{rgb}{0.00,1.00,0.50}
\definecolor{steelblue}{rgb}{0.27,0.51,0.71}
\definecolor{tan1}{rgb}{1.00,0.65,0.31}
\definecolor{tan2}{rgb}{0.93,0.60,0.29}
\definecolor{tan3}{rgb}{0.80,0.52,0.25}
\definecolor{tan4}{rgb}{0.55,0.35,0.17}
\definecolor{tan}{rgb}{0.82,0.71,0.55}
\definecolor{thistle1}{rgb}{1.00,0.88,1.00}
\definecolor{thistle2}{rgb}{0.93,0.82,0.93}
\definecolor{thistle3}{rgb}{0.80,0.71,0.80}
\definecolor{thistle4}{rgb}{0.55,0.48,0.55}
\definecolor{thistle}{rgb}{0.85,0.75,0.85}
\definecolor{tomato1}{rgb}{1.00,0.39,0.28}
\definecolor{tomato2}{rgb}{0.93,0.36,0.26}
\definecolor{tomato3}{rgb}{0.80,0.31,0.22}
\definecolor{tomato4}{rgb}{0.55,0.21,0.15}
\definecolor{tomato}{rgb}{1.00,0.39,0.28}
\definecolor{turquoise1}{rgb}{0.00,0.96,1.00}
\definecolor{turquoise2}{rgb}{0.00,0.90,0.93}
\definecolor{turquoise3}{rgb}{0.00,0.77,0.80}
\definecolor{turquoise4}{rgb}{0.00,0.53,0.55}
\definecolor{turquoise}{rgb}{0.25,0.88,0.82}
\definecolor{violetred}{rgb}{0.82,0.13,0.56}
\definecolor{violet}{rgb}{0.93,0.51,0.93}
\definecolor{wheat1}{rgb}{1.00,0.91,0.73}
\definecolor{wheat2}{rgb}{0.93,0.85,0.68}
\definecolor{wheat3}{rgb}{0.80,0.73,0.59}
\definecolor{wheat4}{rgb}{0.55,0.49,0.40}
\definecolor{wheat}{rgb}{0.96,0.87,0.70}
\definecolor{whitesmoke}{rgb}{0.96,0.96,0.96}
\definecolor{white}{rgb}{1.00,1.00,1.00}
\definecolor{yellow1}{rgb}{1.00,1.00,0.00}
\definecolor{yellow2}{rgb}{0.93,0.93,0.00}
\definecolor{yellow3}{rgb}{0.80,0.80,0.00}
\definecolor{yellow4}{rgb}{0.55,0.55,0.00}
\definecolor{yellowgreen}{rgb}{0.60,0.80,0.20}
\definecolor{yellow}{rgb}{1.00,1.00,0.00}

\usepackage{amsmath}

\def\fun#1#2{\lower3.6pt\vbox{\baselineskip0pt\lineskip.9pt
\ialign{$\mathsurround=0pt#1\hfil##\hfil$\crcr#2\crcr\sim\crcr}}}

\title{Multi-variate joint PDF for non-Gaussianities: exact formulation and generic approximations}

\author[a,b]{Licia Verde,}
\author[a,b]{Raul Jimenez,} 
\author[b]{Luis Alvarez-Gaume,}
\author[c]{Alan F. Heavens,}
\author[d,e]{Sabino Matarrese}

\affiliation[a]{ICREA \& ICC, Instituto de Ciencias del Cosmos, University of Barcelona (IEEC-UB), Marti i Franques 1, Barcelona 08028, Spain.}
\affiliation[b]{Theory Group, Physics Department, CERN, CH-1211, Geneva 23, Switzerland.}
\affiliation[c]{Imperial Centre for Inference and Cosmology, Imperial College, Blackett Laboratory, Prince Consort Road, London SW7 2AZ U.K.}
\affiliation[d]{Dipartimento di Fisica e Astronomia G. Galilei, Universita degli Studi di Padova, I-35131 Padova, Italy.}
\affiliation[e]{INFN, Sezione di Padova, I-35131 Padova, Italy.}

\emailAdd{liciaverde@icc.ub.edu}
\emailAdd{raul.jimenez@icc.ub.edu}
\emailAdd{luis.alvarez-gaume@cern.ch}
\emailAdd{a.heavens@imperial.ac.uk}
\emailAdd{sabino.matarrese@pd.infn.it}

\abstract{We provide an exact expression for the multi-variate joint probability distribution function of non-Gaussian fields primordially  arising from local transformations of a Gaussian field. This kind 
of non-Gaussianity is generated in many models of inflation. We apply our expression to the  non-Gaussianity estimation from Cosmic Microwave Background maps and  the  halo mass function
where we obtain  analytical expressions. We also provide  analytic  approximations  and their range of validity. For the Cosmic Microwave Background we give a fast way to compute the PDF which is valid up to  more than  $7\sigma$ for $f_{\rm NL}$ values (both true and sampled) not ruled out by current observations, which consists of expressing the PDF as a combination of bispectrum and trispectrum of the temperature maps. The resulting expression is valid  for any kind of non-Gaussianity and is not limited to the local type. The above results may serve as the basis for a fully Bayesian analysis of the non-Gaussianity parameter.}

\begin{document}

\maketitle

\section{Introduction}
\label{sec:intro}

One of the most interesting predictions of slow-roll single field inflation is that it should generate primordial perturbations that are nearly Gaussian distributed, with deviations from non-Gaussianity present at the level of the value of the slow-roll parameters \cite{Allenetal87,SalopekBond90,Falketal93,Ganguietal94,Verde:1999ij,WangKamionkowski2000, Komatsu:2001rj,Acquaviva:2002ud,Maldacena:2002vr,Bartolo:2004if}. Current constraints on the value of the slow-roll parameters e.g., \cite{Norena:2012rs} indicate that non-Gaussianity from single field slow-roll inflation will not be measured from the sky, even if one could access all modes in the sky with futuristic 21cm experiments \cite{Pillepich:2006fj, Cooray:2006km}. However, any deviation from the simplest slow roll (with  standard Bunch-Davis vacuum and  canonical kinetic term) implies non-Gaussianity, possibly at a detectable level. In particular  many multi-field models of inflation will produce detectable non-Gaussianity that current or future surveys could measure: see the review by Refs.\cite{Bartolo:2004if, Chen2010Adv} for a detailed explanation of the physics behind it. 
Although in principle  the primordial non-Gaussian signal should be present both in the Cosmic Microwave background and in the late large-scale structure of the  Universe, most of the effort has  so far been  devoted to the cosmic microwave background where the gravitationally-induced non-Gaussianity is  small  e.g.,\cite{Verde:1999ij,Komatsu10,Liguorietal10,YadavWandelt2010,planck2013-p09a} and refs therein.   
However, if  the non-Gaussian signal  is at a clearly detectable level, current estimators (which are exact in the Gaussian limit) become sub-optimal and in particular can underestimate the size of confidence regions \cite{LiguoriYadavetal2007, Smith:2011rm}. Therefore it is important to  understand exquisitely the distribution of errors of any estimator, not only its variance (see \cite{Kamionkowski:2010me}): the full  probability distribution function (PDF) is needed.

Further, a possible detection of non-Gaussianity from the sky would be a direct probe of the properties of de Sitter space, as a study of the different correlators of non-Gaussianity would amount to performing transition experiments on de Sitter; something that will never be accessible from high-energy experiments in the imaginable future. This could open a new window to understand the properties of gravity in de Sitter. One could, for example, answer the question if fluctuations could be treated semi-classically as we do now.

Work so far has concentrated in finding the  one-point PDF  for primordial non-Gaussian initial conditions  as this has a direct application on the calculation of the halo mass function and rare event abundance. This area has seen an exponential expansion starting with the works of \cite{MVJ,LoVerde:2007ri}. However for applications to  e.g., the Cosmic Microwave Background (CMB) an expression for the multi-variate joint PDF is needed.

It is common practice to describe non-Gaussianity as a Taylor expansion of the potential of the form $\Phi = \phi + f_{\rm NL} (\phi^2 -\langle\phi^2\rangle)$ \cite{Verde:1999ij,Komatsu:2001rj}, where $\phi$ is a Gaussian random field.  This can be extended to the cubic power if the quadratic term is  sub-dominant or zero, in which case the free parameter is $g_{\rm NL}$. This is the so-called local type non-Gaussianity because it is a local transformation in real space. More general kinds of non-Gaussianity  beyond the  local type can be described by their bispectrum (although this description is not exhaustive as fields with the same bispectrum  and power spectrum can have different higher-order correlations) e.g.,~\cite{Chen2010Adv} and references therein. 

In addition to the above,  it is the multi-variate joint  PDF of the non-Gaussian primordial field  which is predicted by theory and which has important clues about the physics of the perturbations of de Sitter space; therefore having an exact estimate of it goes beyond the simple experimental merit  of constraining one parameter, as it may yield clues on fundamental physics.

Previous work on estimating the PDF for the CMB  bispectrum or  the non-Gaussianity parameter $f_{\rm NL}$ has focussed on performing Monte Carlo simulations, but these are extremely expensive and may not  fully sample the tails of the distribution (i.e. beyond 3$\sigma$) \cite{Creminelli:2006gc,Elsner:2009md,Smith:2011rm,Smith:2012ta}. Further complications arise form the fact that the PDF will depend at some level on the value of $f_{\rm NL}$ itself.  Furthermore, the general techniques developed here may find useful applications in non-Gaussian late-time fields, such as the galaxy density field in the presence of nonlinear bias or nonlinearity from gravitational instability.

In this paper we circumvent these limitations  and provide an exact solution based on analytical methods. In particular, we provide an exact analytical formula for the one-point PDF in Eq.~(\ref{eq:1pointexact}) that can be used for any computation of the mass function of dark matter halos. 

Furthermore, an exact formula for the joint multi-variate PDF is provided in Eq.~(\ref{eq:pmultiexactM}) and (\ref{eq:70}), which can be numerically integrated. For the CMB we provide a very accurate formula for the PDF (out to more than $7\sigma$ fluctuations in the field for $f_{\rm NL}$ values not ruled out by present data) expressed in terms of the observed bispectra and trispectra. This serves as the basis of a fully Bayesian analysis of $f_{\rm NL}$.   Previous work in this direction can be found in \cite{VielvaSanz:2009,Elsner:2010hb,Elsner:2010gd,Ensslin:2008iu}. The approach, approximations and applicability regimes considered here are fully complementary.

The paper is organised in a  pedagogical way, first, as a warm up, solving the straightforward Gaussian case in \S~3 and then the non-Gaussian one in \S~4, where we also discuss comparison with previous estimators and present an analytic approximation which is valid out to more than $7\sigma$ fluctuations in the field for $f_{\rm NL}$ values not ruled out by present data and applies to non-Gaussianities beyond the local type. We conclude in \S~5 with an application of the formalism to the mass function of dark matter halos and the CMB. The reader interested only in practical applications can find our relevant formulas in Eqs. (\ref{eq:70}, \ref{eq:multivariateedgeworth}, \ref{eq:multivariateedgeworthCMB}).

\section{Set-up}
\label{sec:setup}
Let us define a field $A$, which is the observable field and  can be the temperature of the CMB or the $a_{\ell}^m$ or the  matter overdensity field etc.
The value of the field at any spatial point $i$ is related to the underlying potential $\Phi$  via
\begin{equation}
A({\bf x})=A_i=\int d^3y {\cal M}({\bf x},{\bf y})\Phi({\bf y})\equiv ({\cal M},\Phi)\,
\label{eq:defA} 
\end{equation}
where in the second equality we have introduced a short hand notation.
 For many applications ${\cal M}({\bf x}, {\bf y})={\cal M}(|{\bf x}-{\bf y}|)={\cal M}(r)$.
For example if $A({\bf x})$ is the  (linearly evolved) density field $\delta({\bf x})$ smoothed on a given radius $R$ then we could express this in Fourier space: 
\begin{equation}
 \widetilde{\cal M}(k)\equiv \int d^3y e^{-i{\bf k}\cdot {\bf y}} {\cal M}(|{\bf y}|)=\widetilde{W}_R(k)T(k)g(k)
 \label{eq:defMdelta}
\end{equation}
where $\widetilde{W}_R(k)$ is the Fourier transform  of the assumed filter (typically a top hat for the mass function), $T(k)$ is the transfer function and $g(k)=-2/3(k/H_0)^2\Omega_{m,0}^{-1}$.
The fact that Fourier quantities are complex should not be a problem; one works with a complex field rather than a real one. If the $A_i$  denote the spherical harmonic coefficients of the CMB temperature field, $a_{lm}$, then the equation becomes:
\begin{equation}
a_{lm}=4 \pi (-i)^l \int \frac{d^3 k}{(2\pi)^3}\Phi({\bf k})g_l(k)Y^*_{lm}(\hat{k})  
\label{eq:defalm}
\end{equation}
where $g_l(k)$ denotes the radiation transfer function and $\Phi({\bf k})=\int d^3 x \exp [-i {\bf k}\cdot {\bf x}]\Phi({\bf x})$.  Effectively ${\cal M}$ acquires extra parameters: ${\cal M}^{lm}(k)$.
If we observe a Fourier space quantity then we can relate it to the real space $\Phi$ by
\begin{equation}
A_j=\int d^3y\exp(-i{\bf k}\cdot {\bf y})\Phi({\bf y})
\end{equation}
i.e. ${\cal M}({\bf k,y})=\exp(-i{\bf k}\cdot {\bf y})$, and $j$ labels the wavevector {\bf k}.

Several physically-motivated models for primordial non-Gaussianity are expressed as local transformations of an underlying Gaussian field. For example, the non-Gaussianity of the so-called {\it local} type \cite{Luo:1993yi,Verde:1999ij,Komatsu:2001rj} is given by
\begin{equation}
\Phi = \phi + f_{\rm NL} (\phi^2 -\langle\phi^2\rangle)=\phi + f_{\rm NL} (\phi^2 -\sigma^2_{\phi})\,,
\label{eq:deffnl}
\end{equation}
where $\phi$ is a Gaussian random field and $f_{\rm NL}$ is the non-Gaussianity parameter which is assumed to be small. Here  by {\it small} we mean that this expression is a truncation to linear order in $f_{\rm NL}$ of a physical model  that includes  e.g., the self coupling of the field during inflation. In the following we would like not to make any  further approximations if possible, thus deriving whenever possible  expressions  that are valid  even if $|f_{\rm NL}\sigma_{\phi}|$ is not $\ll 1$.
Other local models include the so called  $g_{\rm NL}$ parameter \cite{Okamoto:2002ik,Enqvist:2005pg,LoVerde:2007ri}:
\begin{equation}
\Phi = \varphi + g_{\rm NL}(\varphi^3-3\varphi\sigma_{\varphi}^2)\equiv \phi+g'_{\rm NL}\phi^3
\label{eq:defgnl2}
\end{equation}
where $\varphi$, $\phi$ are Gaussian random fields.

In both cases the observable field $A$ can be written as a sum of a Gaussian term $A^G$, corresponding to $\phi$  plus a non-Gaussian correction $A^{NG}$, corresponding to the term $\propto \phi^2$ or $\propto \phi^3$. 

The full information about the Gaussian field $\phi$ is given by the Gaussian generating functional ${\cal P[\phi]}$ (see e.g., Eq 11 of Ref.~\cite{MVJ}) which is the basis of several calculations in this paper. For example, the joint probability of $A_1,...A_n$ is given by:  
\begin{equation}
{\cal P}(A_1,A_2,...A_n)  = 
 \int \left [ {\cal D} \phi \right ] {\cal P} ( \phi )\prod_{i=1}^n\delta^D\left[A_i- ({\cal M},\Phi(\phi))\right]
 \label{eq:jointprob}
\end{equation}
where we have explicitly written that the field $\Phi$ is a function of the Gaussian random field $\phi$. 
Here, 
\begin{equation}
{\cal P}(\phi)=\frac{\exp[-1/2(\phi,K_0,\phi)]}{\int D[\phi]\exp[-1/2(\phi,K_0,\phi)]}
\label{eq:gaussianpathintegral}
\end{equation}
and we have used the shorthand notation
\begin{equation}
\int d^3y\, d^3z\,\phi({\bf y})K({\bf y}-{\bf z})\phi({\bf z})\equiv (\phi,K,\phi),
\label{eq:shorthand}
\end{equation}
 and 
 with $K_0$ is defined by the relation to the 2-point correlation function of $\phi$, $\xi^\phi$:
 \begin{equation}
\int d^3y K_0(|{\bf x}-{\bf y}|)\xi^{\phi}(|{\bf y}-{\bf z}|)=\delta(|{\bf x}-{\bf z}|).
\label{eq:defK0}
\end{equation}
Then to compute, for example, the probability distribution of a product of $n$ $A$'s we use

\begin{equation}
{\cal P}(B\equiv A_1\ldots A_n) = \int dA_1 dA_2 ... dA_n {\cal P}(A_1,..., A_n) \delta^D(B-A_1... A_n)\,.
\label{eq:prodprob}
\end{equation}
 Here $n$ is 3 for the three-point function (in Fourier space it is related to the bispectrum), $n=2$ for the  correlation function  (in Fourier space it is related to the power spectrum) of $A$ and   $n=1$ for  the (one-point)  PDF of $A$.

In these types of integrals, the Fourier representation of the  Dirac delta function involves  highly oscillatory  functions and complicates the integrals. On the other hand, observations are often discretized/pixelized, so one possibility is to discretize these integrals  to simplify their analytic calculation, then  if needed take the continuous limit at the end.
Let us pixelize our field $A$ (say $A$ is known at each pixel of a map) and we have $L$ pixels, let us also pixelize the $\Phi$ field for which we have in total  $M$ pixels ($L\le M$).   As we will see this simplifies a lot the spatial integrals and the integrals in ${\cal D}[\phi]$. The integrals in $A_i$ are  however continuous integrals over all possible values that the field can take at position $i$.

\begin{table}
\begin{center}
\begin{tabular}{|c|c|c|}
\hline
Index & range & meaning \\
\hline\hline
$a$:$g$ & $1$:$M$ & number of elements of the full discretized space of $\Phi$, $\phi$\\
$i$:$z$ & $1$:$n$&    number of elements of the discretized A array\\
&& for which we want the joint probability\\
$\nu$ or greek & $n+1$:$M$ & the remaining elements of  the full discretized space\\
\hline
\end{tabular}
\caption{Summary of the index convention.  The number of elements of the observable field $A$ is $L$. Of course $n\le L$, $L\le M$, and $n\le M$. Note that for ${\cal M}$ to be invertible a necessary but not sufficient condition is that the maximum possible size of $A$ to be $M$ i.e. $M=L$.}
\end{center}
\end{table}

Let us define the index convention we will use.
$L$ denotes the size of the observable array $A$, of which we want to compute the  $n$ point distribution $n\le L$,  $M$ denotes the size of the pixelized array $\phi$ (or $\Phi$). Note that $n\le L\le M$. Let us call $M-n=N$.
The $i$ indices  (or latin indices of the last part of the alphabet: $i$--$z$) run from $1$ to $n$ and   $\nu$ (or greek indices)  runs from $n+1$ to $M$. From now on latin indices of the first part of the alphabet ($a$--$h$) runs from 1 to $M$.
Note that if $A$ and $\Phi$ have the same dimensions then ${\cal M}$ becomes a square matrix ($M \times M$) and everything below is  relatively simple.
However this does not need to be, as it is clear in the case of the $a_{lm}$ where the $i$ index runs over the --discrete-- $l,m$ but the index $j$ represent the integral in $k$ which is in principle  a continuous variable (but is discretized in most practical applications).

Discretizing  we obtain for Equations  such as Eq.~(\ref{eq:defK0}) and Eq.~(\ref{eq:defA}):
$\sum_{b} K_{0ab}\xi^{\phi}_{bc}={\cal I}_{ac}$ and $\int d^3y {\cal M}(x_j,{\bf y})  \Phi ({\bf y})  \longrightarrow \sum_{a}{\cal M}_{ja}\Phi_a$ etc.
 Also we can express $D[\phi]=d\phi_i d\phi_\nu$.

The Gaussian integral in the denominator of Eq.~(\ref{eq:gaussianpathintegral}) gives
\begin{equation}
\sqrt{\frac{(2\pi)^M}{\det {\bf K}_0}}\,.
\end{equation} 
The Dirac delta function becomes
$\delta (A_i-\sum_j{\cal M}_{ij}\phi_j)$.

Finally, let us  report here a result for Gaussian integrals with we will use repeatedly below.
The Gaussian integral is 
\begin{equation}
\int d{\varphi} \exp\left[-\frac{1}{2} {\bf \varphi}^T{\bf K} {\bf \varphi}+{\bf B}^T {\bf \varphi}\right]=\sqrt{\frac{(2\pi)^m}{\det{\bf K}}} \exp\left[ \frac{1}{2} {\bf B}^T {\bf K}^{-1} {\bf B}\right]
\label{eq:gaussianintegral} 
\end{equation}
 with ${\bf K}$ a square matrix $m\times m$ and ${\bf B}$ and ${\varphi}$ are vectors of size $m$.

\section{Gaussian  field $\Phi=\phi$}
\label{sec:Gaussian}
To illustrate the approach and familiarize ourselves with some of the expressions, let us start in the simplest case where $f_{\rm NL}=0$, $g_{\rm NL}=0$. We begin by considering the case where ${\cal M}$ is a square, invertible matrix.  We then progressively consider more realistic/complicated cases: ${\cal M}$ not invertible and the  addition of (Gaussian) noise. In what follows, we denote with the over script arrow the vector composed of the value of the field in all the pixels, e.g., $\vec{A} = \{ A_1, ..., A_n \}$.

\subsection{${\cal M}$ invertible, $n=L=M$}
In this case we can do the following change of variables:
$\phi \longrightarrow  {\cal M}\phi$ or $\hat{\phi}_{a}=\sum_{ab} {\cal M}_{ab}\phi_b$ and $\hat{K_0}=\left({\cal M}^{-1}\right)^TK_0{\cal M}^{-1}$ (matrix multiplication); $\prod_ad\phi_a=\frac{\prod_ad\hat{\phi}_a}{\det({\cal M})}$.
We thus obtain:
\begin{eqnarray}
\!\!\!\!\!\!\!\!\!\!{\cal P}(\vec{A})&=&\sqrt{\frac{\det K_0}{(2\pi)^M}}\int \frac{\prod_{i  } d\hat{\phi}_i}{\det({\cal M})} \prod_i\delta^D(A_i-\hat{\phi}_i)\exp\left[-\frac{1}{2} \sum_{ij}\hat{\phi}_i\hat{\phi}_j\widehat{K}_{0\,ij}\right] \label{eq:Pchangevariables} \nonumber \\
&=&\sqrt{\frac{\det \widehat{K}_0}{(2\pi)^M}}\exp\left[-\frac{1}{2} \sum_{ab}A_a A_b\widehat{K}_{0ab}  \right], \label{eq:pofAMeqn}
\end{eqnarray}
which is not surprising given that if the $\phi$ field is a multivariate Gaussian, the $A$ field properties are also those of a multi-variate Gaussian, albeit with a different covariance.

Note that $\widehat{K_0}$ is defined so that
\begin{equation}
\int d^3y \widehat{K_0}(|{\bf x}-{\bf y}|) \xi_A(|{\bf y}-{\bf z}|)=\delta(|{\bf x}-{\bf z}|).
\end{equation}
In other words, $\widehat{K_0}$  is the equivalent quantity of  $K_0$,  not for the  underlying field $\phi$, but  for the {\it observable} field $A$.

\subsection{${\cal M}$ identity matrix or invertible, $n<L$, $L=M$}
\label{sec:Midentity}
Let us start with the simpler, special case where ${\cal M}$ is the identity matrix, for example, when looking at the distribution of the product of $n$ pixel values of an unsmoothed field.
The values taken by the field $\phi$ will be denoted by $\varphi$ (here $\varphi$ plays the role of $A$ but we have changed notation to make clear that one is really seeing the same field).

We start from:
\begin{eqnarray}
&{\cal P}&(\varphi_i,...,\varphi_n)=\sqrt{\frac{\det K_0}{(2\pi)^M}}\int \prod_{i,\nu} d{\phi}_i d{\phi}_{\nu} \prod_i\delta^D(\varphi_i-{\phi}_i)\times   \\
&&\!\!\!\!\!\!\!\exp\left[-\frac{1}{2} \sum_{ij}{\phi}_i{\phi}_j{K_0}_{ij} - \sum_{i,\nu}{\phi}_i{\phi}_{\nu}{K_0}_{i\nu}-\frac{1}{2} \sum_{\mu,\nu}{\phi}_{\mu}{\phi}_{\nu}{K_0}_{\mu\nu} \right]\,. \nonumber
\end{eqnarray}
The integral over $d{\phi}_i$ can be done trivially exploiting the Dirac delta function yielding:
 \begin{eqnarray}
\!\!\!\!\!\!\!\!\!\!&{\cal P}&(\varphi_i,...,\varphi_n)=\sqrt{\frac{\det K_0}{(2\pi)^M}} \int \prod_{\nu} d{\phi}_{\nu}  \times \label{eq:intermediate}  \\ \nonumber
&&\!\!\!\!\!\!\!\!\!\!\!\!\!\!\! \exp\left[-\frac{1}{2} \sum_{ij}\varphi_i\varphi_j{K_0}_{ij} - \sum_{i,\nu}\varphi_i{\phi}_{\nu}{K_0}_{i\nu}-\frac{1}{2} \sum_{\mu,\nu}{\phi}_{\mu}{\phi}_{\nu}{K_0}_{\mu\nu} \right]\,.
\label{eq:intermgaussianintegral}
 \end{eqnarray}
Here we recognize the Gaussian integral of Eq.~(\ref{eq:gaussianintegral})  where
 in particular $B_{\nu}=\sum_i \varphi_i{K_0}_{i\nu}$ and ${\bf K}$ is the matrix made of the elements  ${K_0}_{\mu \nu}$.
 
 Note that the matrix $K_0$ can be split in block matrices  as 
 
 \begin{equation}
 \begin{matrix}
  {\bf K}_0
 \end{matrix}
 =
 \begin{pmatrix}
 K  & {\Bbbk} \\
 {\Bbbk}^T & {\cal K}
 \end{pmatrix}
\end{equation}
where $K$ is made of the  $K_{0_{ij}}$ elements (it is a square sub matrix of size $n\times n$), ${\cal K}$ is made by the $K_{0_{\mu \nu }}$ elements (it is a square matrix of size $N\times N$), and $\Bbbk$ is an $n\times N$ matrix.

Finally we obtain that:
\begin{equation}
{\cal P}(\vec{\varphi})= \sqrt{\frac{\det({\cal G})}{(2\pi)^{n}}}\exp\left[-\frac{1}{2}\sum_{ij=1}^n\varphi_i\varphi_j{\cal G}_{ij}\right]\,.
\label{eq:pofA} 
\end{equation}
Here
\begin{equation}
{\cal G}_{ij}={K_0}_{ij}-\sum_{\mu \nu} {K_0}_{i\nu } {K_0}_{j\mu } {\bf {\cal K}}^{-1}=K-\Bbbk{\cal K}^{-1}\Bbbk^T\,,
\label{eq:defG} 
\end{equation}
and we have used the fact that 
\begin{equation}
\det {\bf K_0}=\det\left(^ {K\Bbbk}_{\Bbbk\,\, {\cal K}}\right)=\det({K}-{\bf \Bbbk} {\cal K}^{-1}{\bf \Bbbk}^T)\det({\bf {\cal K}})
\label{eq:detproperty} 
\end{equation}
implying $\det({K}_0)/ \det({\cal K}) = \det({\cal G})$.

Eq.~(\ref{eq:pofA}),  not surprisingly, is a Gaussian but with a different covariance from that of $\phi$.
It is not difficult to see, using the Woodbury formula
\begin{equation}
(A+UCV)^{-1} = A^{-1}-A^{-1}U(C^{-1}+VA^{-1}U)^{-1}A^{-1}
\end{equation}
(where $A$, $U$, $C$, and $V$ denote generic matrices) that
\begin{equation}
[{\cal G}^{-1}]_{ij} = [K_0^{-1}]_{ij} = \xi_{ij}
\end{equation}
as expected.

In case ${\cal M}$ is not the identity matrix but it is still an invertible matrix, as long as the array $A$ has been pixelized in as many elements as $\phi$, it is straightforward to compute ${\cal P}({\vec A})$ even if $n<M$. 
One just does the substitution: $\phi_i \longrightarrow  \sum _a{\cal M}_{ia}\phi_a$ etc. as above.
In total one gets,
\begin{equation}
{\cal P}(\vec{A})= \sqrt{\frac{\det(\hat{\cal G})}{(2\pi)^{n}}}\exp\left[-\frac{1}{2}\sum_{ij=1}^nA_i A_j\hat{\cal G}_{ij}\right]\,,
\label{eq:pofAfull} 
\end{equation}
where $\hat{K}_0=\left({\cal M}^{-1}\right)^T{K}_0{\cal M}^{-1}$ (matrix multiplication) and $\hat{\cal G}$ is the equivalent of ${\cal G}$ (see Eq.~(\ref{eq:defG})) but  made from $\hat{K}_0$ instead of $K_0$,  
 $[\hat{\cal G}^{-1}]_{ij}=[\hat{K}_0^{-1}]_{ij}=\xi_{A\,ij}$.
That is we have a Gaussian with covariance given by that of the  observed field.

\subsection{${\cal M}$ non invertible, $n<L$, $L<M$}
 This is only a slightly more complicated case for a Gaussian field (we will see  below that for a non-Gaussian field, making ${\cal M}$ non invertible complicates the resulting expressions considerably). 
In this case the observable field $A$ is pixelized in $L$ elements, but the dimension of the field $\phi$ (and the field $\Phi$) is larger, for example $M=  L \times H$ (i.e. there is one (or more) extra dimensions which gets pixelized to have $H$ elements).  To keep in mind a possible relevant application, this is the case of a CMB temperature  map: the  primordial field $\Phi$ is three-dimensional, the observed  temperature fluctuation field  around a given observer is two-dimensional, therefore  ${\cal M}$  is non invertible.

It is easy to see that one can  rewrite  
 Eq. (\ref{eq:jointprob})  considering that if $\phi$ is Gaussian then its projection, e.g., to $a_{lm}$, will also be Gaussian (${\cal M}$ is a linear operator). So we only have to reinterpret Eq. (\ref{eq:Pchangevariables}) in light of Eq. (\ref{eq:defA}) to have $\hat{\phi}=A^G$ and $\widehat{K}_0$ to be the equivalent quantity for the Gaussian $A^G$, in the case of the $a_{lm}$ it will be connected to the $C_{\ell}$.  In this case there is no need to invert ${\cal M}$.
 It is however  instructive  in view of the non-Gaussian case, to derive the same result going through the actual calculations.

We start from 
  \begin{eqnarray}
&{\cal P}&(\vec{A})=\sqrt{\frac{\det K_0}{(2\pi)^M}}\int {\prod_{i,\nu} d{\phi}_id{\phi}_{\nu}} \prod_i\delta^D(A_i-\sum_{ia}{\cal M}_{ia}{\phi}_a)\times  \nonumber \\
&&\exp\left[-\frac{1}{2} \sum_{ij}{\phi}_i{\phi}_j{K_0}_{ij} - \sum_{i,\nu}{\phi}_i{\phi}_{\nu}{K_0}_{i\nu}-\frac{1}{2} \sum_{\mu,\nu}{\phi}_{\mu}{\phi}_{\nu}{K_0}_{\mu\nu} \right]\,,
\label{eq:PgaussMNI} 
\end{eqnarray}
and split ${\cal M}$ into block matrices as:
 \begin{equation}
 {\cal M}=\left({ \aleph\,\,\,\,\,\,{\cal N}}\right),
 \end{equation}
where $\aleph $ is a $n\times n$ sub matrix and ${\cal N}$ is $n\times N$.
We also split the vector $\phi=\left\{\phi_i,\phi_{\nu}\right\}=\left\{\phi',\phi'' \right\}$.
 We start by re-writing  the argument of the Dirac delta as:
 \begin{equation}
 A_i-\sum_j{\cal M}_{ij}\phi_j-\sum_{\nu}{\cal M}_{i\nu}\phi_{\nu}=A_i-(\aleph \phi')_i -({\cal N}\phi'')_i
 \end{equation}
 and make the following change of variables
 \begin{equation}
 A \longrightarrow \aleph^{-1} A\equiv \tilde{A}\,.
 \end{equation}
 $\aleph$ is invertible if the rank of ${\cal M}$ is maximal. Note that there is some freedom in splitting the sum in Eq.(3.14), but  we argue that the rank of  ${\cal M}$ is always  maximal if the information in $A$ is not redundant (e.g.,  in the absence of noise if pixels are not counted more than once).
 This change of variables gives a multiplicative overall factor $|\det \aleph|$  and which  changes the argument of the Dirac delta to
 \begin{equation}
 \tilde{A}_i-\phi_i-(\aleph^{-1}{\cal N}\phi'')_i\equiv \tilde{A}_i-\phi_i-\tilde{\phi}''_i\,.
 \end{equation}
Now we can integrate over the $d\phi_i$ and eliminate the Dirac deltas giving the following substitution $\phi_i\longrightarrow \tilde{A}_i-\tilde{\phi}''_i$.
 We get:
 \begin{eqnarray}
\!\!\!\!\!\!\!\!\!\!\!\!\!\!{\cal P}(\vec{A})&=&
 \sqrt{\frac{\det K_0}{(2\pi)^M}}|\det{\aleph}|\int d\phi''\exp \left[ -\frac{1}{2}\left( \tilde{A}^T K\tilde{A} +2\tilde{A}^T{\cal R}\phi''+\phi''^T{\cal Q}\phi''\right)\right]\\
 \!\!\!\!\!\!\!\!\!\!\!\!\!\!\!\!&=& \sqrt{\frac{\det K_0}{(2\pi)^M}}|\det{\aleph}|\int d\phi''\exp [-\frac{1}{2}\{\tilde{A}, \phi''\}^T H \{\tilde{A}, \phi''\}]
 \end{eqnarray}
 where 
 \begin{equation}
 H=
 \begin{pmatrix} 
 K & {\cal R}\\
 {\cal R}^T & {\cal Q}\\
 \end{pmatrix}
 \end{equation}
 \begin{equation}
 {\cal R}=\Bbbk-K\aleph^{-1}{\cal N}
 \end{equation}
 and 
 \begin{equation}
{\cal Q}={\cal K}+{\cal N}^T\aleph^{-1\,T}K\aleph^{-1}{\cal N}-{\cal N}^T \aleph^{-1\,T}\Bbbk-\Bbbk^T\aleph^{-1}{\cal N},
\end{equation}
which is an $N \times N$ matrix.

We then recognize a Gaussian integral yielding
\begin{equation}
{\cal P}(\vec{A})=
\sqrt{\frac{\det K_0}{(2\pi)^n}}|\det{\aleph}|\frac{1}{\sqrt{|\det {\cal Q}|}}\exp\left[-\frac{1}{2}A^T\tilde{\cal F}A\right]
\label{eq:37}
\end{equation}
where
\begin{equation}
{\cal F}=K-{\cal R}{\cal Q}^{-1}{\cal R}^T
\end{equation}
and therefore, using Woodbury again, $[{\cal F}^{-1}]_{ij}=[H^{-1}]_{ij}$, 
and $\tilde{\cal F}=\aleph^{-1\,T}{\cal F}\aleph^{-1}$.  
Of course, in this simple example we are using the general framework to compute the PDF of another Gaussian field, $\tilde A$, whose covariance is simply ${\cal M}^T\xi {\cal M}$.  It is straightforward using the Woodbury formula to demonstrate that we recover this result.  For the Gaussian case there is therefore no need to invert $\cal{Q}$  or $\aleph$.

Equivalently,  one could change the basis for $\phi$ to $\psi$ so that the matrix ${\cal M}$ is ${\cal M}'=\left(\aleph',0\right)$.
This being a linear operation means that $\psi$ is still Gaussian.
Note that this means: take the rectangular  matrix given by$\left\{\aleph,{\cal N}\right\}$ and SVD decompose it as $U\Sigma V^T$ (or $U\Sigma V^*$ (conjugate transpose)). Then do the following change of variables: $\phi \longrightarrow U\phi=\psi$ and therefore ${\cal M}\longrightarrow U^{-1}{\cal M}={\cal M}'$. The Jacobian of the transformation is therefore $\det U$.

Thus the argument of the Dirac delta becomes $A_i-\aleph'\psi'$ and as before  we make the change of variable $A\longrightarrow \aleph'^{-1}A\equiv \tilde{A}$.
This now simplifies the equations above as ${\cal Q}'={\cal K'}$ and ${\cal R}'=\Bbbk '$. (Of course the Jacobian of the transformation must be  computed; alternatively one makes sure, a posteriori, that the probability is suitably normalized).
We find:
\begin{equation}
{\cal P}(\vec{A})=\sqrt{\frac{\det K_0}{(2\pi)^n}}|\frac{|\det{\aleph}'|}{|\det U|} |\frac{1}{\sqrt{|\det {\cal Q}'|}}\exp\left[-\frac{1}{2}{A}^T\tilde{\cal G}'{A}\right]
\label{eq:40} 
\end{equation}
where $\tilde{\cal G}'= \aleph'^{-1\,T}{\cal G}'\aleph'^{-1}$ and ${\cal G}'=K'-{\cal R}'{\cal Q}'^{-1}{\cal R}'^T$ (where prime denotes the quantity in the $\psi$ basis.

Of course Eqs.~(\ref{eq:37}) and (\ref{eq:40}) are equivalent but depending on the context one or the other might be more suitable for practical implementation.
Note that  $\hat {\cal G}'=\left({\cal M}'^T\xi'{\cal M}'\right)^{-1}$ where we use the prime to indicate the quantity for the $\psi$ field. Since $\xi' =U\xi U^T$ and ${\cal M}'=U^{-1}{\cal M}$ we obtain ${\cal M}'^T\xi'{\cal M}'={\cal M}^T\xi{\cal M}=\xi_A$ and therefore $\tilde{\cal G}'=\hat{\cal G}$,  $[\tilde{\cal G}'^{-1}]_{ij}=\xi_{A\,ij}$.
 
 To interpret $\tilde{\cal G}$ e.g., in Eq.~(\ref{eq:37}) we can readily write down the 1-point PDF ($n=1$), where $\tilde{\cal G}$ is a scalar, $\tilde{\cal G}_{11}$:
 \begin{equation}
{\cal P}(A_1)=\frac{\tilde{\cal G}}{\sqrt{2\pi}}\exp\left[-\frac{1}{2}A_1^2\tilde{\cal G}_{11}\right].
\end{equation}
 If $A$ is the field in real space then $\tilde{\cal G}(R)$ is $1/\sigma_R^2$, the inverse variance of the field, with $R$ indicating the smoothing scale. If $A$ is in Fourier space then $\tilde{\cal G}(k)$ is $1/{\cal P}(k)$.

 \subsection{Adding Gaussian noise}
 \label{sec:addingnoise}

 Having Gaussian  noise in a map means that  the observed field ${\cal A}$ is given by  the superposition of the field $A$ as before and an independent  Gaussian  field $\epsilon$, with covariance $\Sigma_{\epsilon}$, ${\cal A}=A+\epsilon$.  Thus in Eqs.~(\ref{eq:jointprob}, \ref{eq:pofAMeqn}, \ref{eq:PgaussMNI} ) etc. the Dirac delta gets substituted by a Gaussian  with covariance $\Sigma_{\epsilon}$.  It is easy to see that  if $\Sigma_{\epsilon}^{-1}\equiv {\cal E}$  then the second line of Eq. (\ref{eq:pofAMeqn}) remains the same provided that $\widehat{K}_0\longrightarrow {\cal G}$ where ${\cal G}=\left[\widehat{K}_0^{-1}+{\cal E}^{-1}\right]^{-1}$. In doing so we have used the result for the Gaussian integral Eq.~(\ref{eq:gaussianintegral}) and that ${\cal E}-{\cal E}({\cal E}+\widehat{K}_0)^{-1}{\cal E}=\left[\widehat{K}_0^{-1}+{\cal E}^{-1}\right]^{-1}$.
 
Therefore for all  the subsequent cases above one must just make the substitution ${\cal G}\longrightarrow \left[{\cal G}^{-1}+{\cal E}\right]^{-1}$.
Note that we could have obtained this result  more directly as follows: the PDF of the superposition of two independent  processes  ($A$ and $\epsilon$) is the convolution of the individual PDFs. If the two PDFs are Gaussian the resulting PDF will be a Gaussian with inverse covariance   given by the sum of the individual inverse covariances.

\section{Quadratic local non-Gaussianity, $f_{\rm NL}\ne 0$}
 After having used our approach for a Gaussian   field to reproduce known results and familiarize ourselves  with the manipulation of some of the expressions, we go on to tackle the case of a quadratic, local non-Gaussianity. As we will see this  apparently simple, non-linear transformation of a Gaussian field, in some cases --those of most practical interest-- complicates the final expressions quite a lot because it spoils the analyticity of some of the final  steps we were able to use in the Gaussian case.
 Here we will proceed somewhat  in parallel to Sec. \ref{sec:Gaussian}, by first considering the simplest case and then introducing complications progressively.

We begin by  noting that $f_{\rm NL}$ appears only in the argument of the Dirac delta function (see Eq. (\ref{eq:jointprob})). It is much easier to derive the final result in the case where  ${\cal M}$ is the  identity matrix.  While this case might not be of great practical interest, although some  applications to future data are possible and are mentioned below, it is a useful preliminary step to then  tackle more complicated and realistic cases.
\subsection{${\cal M}$ identity matrix}
\label{sec:4.1}

\begin{figure}
 \includegraphics[scale=0.82]{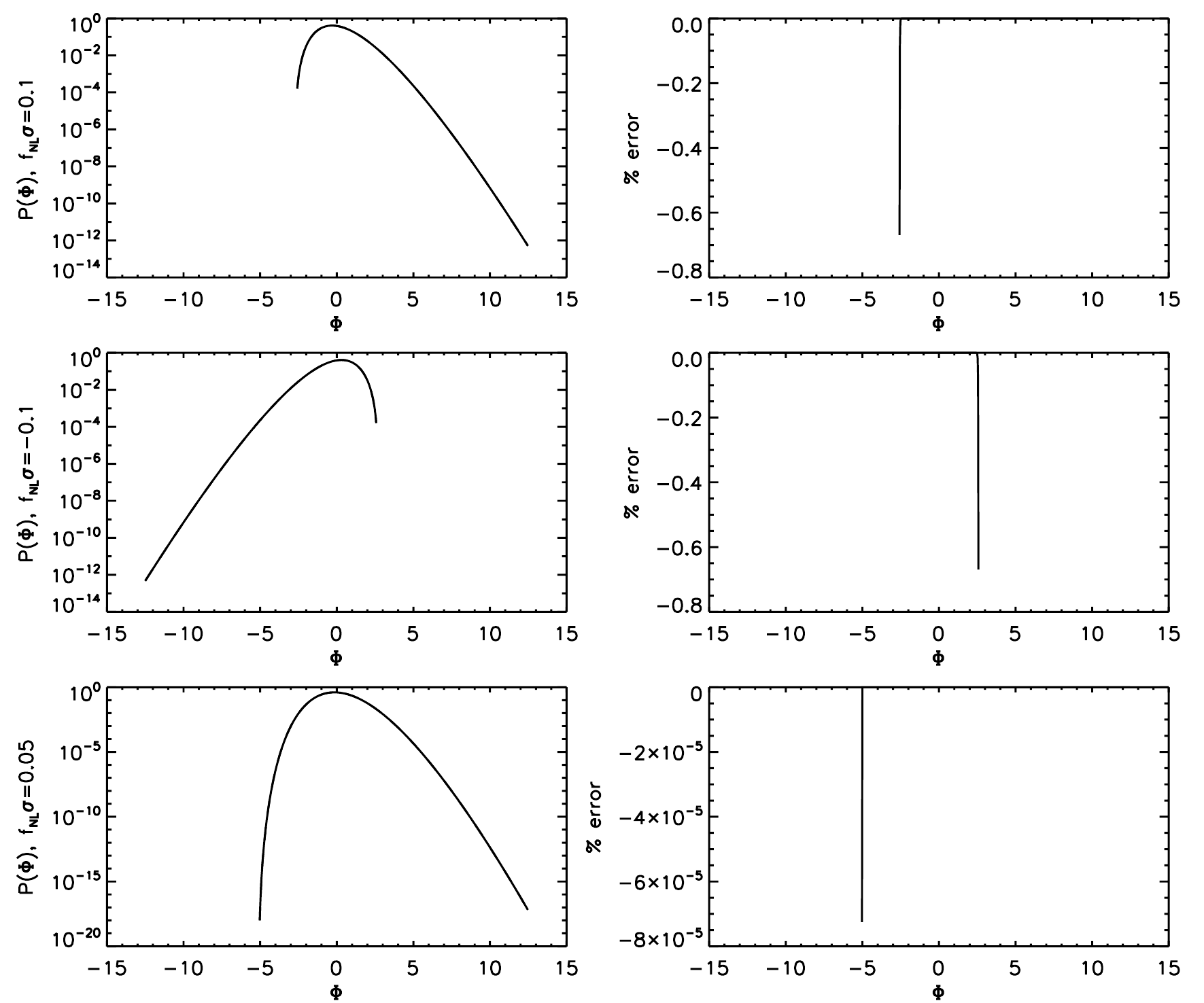}
\caption{
Exact one-point PDF for $\Phi$ (normalized to have unit variance) on the left panel and the effect of neglecting the exponentially suppressed root for ``small" $f_{\rm NL}$ on the right panel. Several choices for $f_{\rm NL}\sigma_{\phi}$ are shown.  Where the distribution is not plotted is because it becomes imaginary (i.e. there is no solution). Neglecting the root $r=1$  is a very good approximation since deviations are small and non-zero only close to the limit where the PDF is not defined. }
\label{fig:neglectroot}
 \end{figure}

 If  ${\cal M}$ is identity matrix, one possible way to proceed to evaluate Eq. (\ref{eq:jointprob}) is to use the property of the Dirac delta:
\begin{equation}
\delta(f(x))=\sum_r\frac{\delta(x-x_r)}{|f'(x_r)|}
\end{equation} 
where $r$ here denotes the roots of the function $f$.  In fact if  ${\cal M}$  is the identity matrix, we have to solve
$A_i+C-\phi-f_{\rm NL}\phi^2=0$, where $C$ is a constant, to make the field $\Phi$ to have zero average ($C=f_{\rm NL}\langle\phi^2\rangle$).
The roots of this are:
\begin{equation}
\frac{-1\pm\sqrt{1+4f_{\rm NL}(A_i+C)}}{2f_{\rm NL}}
\end{equation}
and 
\begin{equation}
f'=1+2f_{\rm NL}\phi\,.
\end{equation}
Our Dirac delta becomes 
\begin{equation}
\prod_i\left[\frac{\delta\left(\phi_i-\frac{-1+\sqrt{1+4f_{\rm NL}(A_i+C)}}{2f_{\rm NL}}\right)}{\sqrt{1+4f_{\rm NL}(A_i+C)}}+\frac{\delta\left(\phi_i-\frac{-1-\sqrt{1+4f_{\rm NL}(A_i+C)}}{2f_{\rm NL}}\right)}{\sqrt{1+4f_{\rm NL}(A_i+C)}}\right]\,.
\end{equation}
In the case $n=1$ there is no product, only the two deltas.
Let us call 
\begin{eqnarray}
f_1(A_i)&=&\frac{-1+\sqrt{1+4f_{\rm NL}(A_i+C)}}{2f_{\rm NL}}\\
f_2(A_i)&=&\frac{-1-\sqrt{1+4f_{\rm NL}(A_i+C)}}{2f_{\rm NL}}\nonumber \,.
\end{eqnarray}
Then Eq.(\ref{eq:jointprob}) becomes:
 \begin{eqnarray}
{\cal P}(\vec{A}|f_{\rm NL})&=&\sqrt{\frac{\det K_0}{(2\pi)^M}} \prod_i\frac{1}{\sqrt{1+4f_{\rm NL}(A_i+C)}}\int \prod_{\nu} d\phi_{\nu} \exp\left[-\frac{1}{2} \sum_{\mu,\nu}{\phi}_{\mu}{\phi}_{\nu}{K_0}_{\mu\nu}  \right]\\ \nonumber
&\times& \sum_{rq} \exp\left[-\frac{1}{2} \sum_{ij}f_r(A_i)f_q(A_j){K_0}_{ij} - \sum_{i,\nu}f_r(A_i)\phi_{\nu}{K_0}_{i\nu}\right]\,. \\ \nonumber
 \end{eqnarray}
As is well known, from the pioneering work of \cite{babich05} to the more recent  \cite{Elsner:2010gd}, for small $f_{\rm nl}$  the root with the minus sign will be exponentially suppressed.  For illustration, Fig.~(\ref{fig:neglectroot}) shows the effect of neglecting the root  on the 1 point PDF (which can be easily computed) for several choices for $f_{\rm NL}\sigma$. For simplicity we have taken $\phi$  to have unit variance; where the distribution is not plotted is because it becomes imaginary (i.e. there is no solution). The figure shows that  neglecting the root $r=2$  is a very good approximation since deviations are small and non-zero only close to the limit where the PDF is not defined. 
Neglecting the $r=2$ root  we obtain:
\begin{eqnarray}
{\cal P}(\vec{A}|f_{\rm NL})&=&\sqrt{\frac{\det K_0}{(2\pi)^M}} \prod_i\frac{1}{\sqrt{1+4f_{\rm NL}(A_i+C)}}\int \prod_{\nu} d\phi_{\nu} \\ \nonumber
&\times& \left. \exp\left[-\frac{1}{2} \sum_{ij}f_1(A_i)f_1(A_j){K_0}_{ij} - \sum_{i,\nu}f_1(A_i)\phi_{\nu}{K_0}_{i\nu}-\frac{1}{2} \sum_{\mu,\nu}{\phi}_{\mu}{\phi}_{\nu}{K_0}_{\mu\nu} \right] \right.\,. \\ 
 \end{eqnarray}

If $n<M$ we can recognize the Gaussian integral again and simplify the above equation:
\begin{equation}
{\cal P}(\vec{A}|f_{\rm NL})=\sqrt{\frac{\det {\cal G}}{(2\pi)^n}}\prod_i\frac{1}{\sqrt{1+4f_{\rm NL}(A_i+C)}}\left\{  \exp\left[-\frac{1}{2}\sum_{ij} f_1(A_i) f_1(A_j) {\cal G}_{ij}\right]
\right\}\,.
\label{eq:55}
\end{equation}

For $n=M$ this is instead:
\begin{equation}
 {\cal P}(\vec{A}| f_{\rm NL})=\sqrt{\frac{\det {K_0}}{(2\pi)^n}}\prod_i\frac{1}{\sqrt{1+4f_{\rm NL}(A_i+C)}}\left\{ \exp\left[-\frac{1}{2}\sum_{ij} f_1(A_i) f_1(A_j) {K_0}_{ij}\right] \right\}\equiv {\cal P}(\Phi)\,.
 \label{eq:59}
\end{equation}

Note that this is the  multivariate distribution  of the $\Phi$ field (${\cal M}$ having been set to the identity).  Note the similarity between this and Eq. 14 in \cite{Elsner:2010gd}. 
Of course, to obtain the  same quantity for ${\cal M}$  not the identity matrix,   one can then always write that 
\begin{equation}
{\cal P}(\vec{A}| f_{\rm NL})=\int {\cal D}[\Phi] {\cal P}(\Phi)\delta^D(A-\int M \Phi)\,.
\label{eq:60}
\end{equation}

This then must be integrated numerically via Monte-Carlo methods,  as for example  explored in \cite{Elsner:2010gd,Elsner:2010hb}. While Ref. \cite{Elsner:2010gd, Elsner:2010hb} proceeds by computing this (or a related) expression directly with a Monte Carlo approach,  here we attempt to carry on analytically.
If ${\cal M}$ is the identity matrix,  the  one-point function  can be obtained by setting, $i, j \equiv 1$ as for the Dirac delta of Eq.~(\ref{eq:55}).
One then recognizes the Gaussian integral in $\phi$ obtaining
 
\begin{equation}
\boxed{{\cal P}(A_1| f_{\rm NL})=\sqrt{\frac{\det({\cal G})}{2\pi}}\frac{1}{\sqrt{1+4f_{\rm NL}(A_1+C)}} \left( \exp\left[-\frac{1}{2}f_1^2(A_1){\cal G}_{11}\right] + \exp\left[-\frac{1}{2}f_2^2(A_1){\cal G}_{11}\right] \right)\,.}
\label{eq:1pointexact}
\end{equation}

\subsection{Approximations}
\label{sec:approximations}
 We now take a detour (subsections \ref{sec:approximations} and \ref {sec:estimators}). We first  explore  possible approximations of this exact PDF expression, their performance and  range of validity. For simplicity and clarity we do so in the case where  ${\cal M}$ is the identity matrix, i.e., where we can ignore the transfer function. Therefore the results will be valid only qualitatively and not strictly quantitatively in the (practically interesting case) when there is a transfer function, especially if it   makes ${\cal M}$ not invertible. Nevertheless the performance of the approximations considered will be directly  useful in Sec.\ref{sec:4.6approximations}. In addition they could be useful to find approximations to Eq. \ref{eq:70}.

It is possible to further simply  our expressions  in Sec. \ref{sec:4.1} by expanding to first (or higher) order in $f_{\rm NL}$. For example  expanding at first order  the square root we obtain  $\left(1+2 f_{\rm NL}\sum_i(A_i+C)\right)$  both in Eqs. (\ref{eq:55}) and (\ref{eq:59}),   in the same spirit then $f_1(A_i)\longrightarrow A_i+C-A_i^2f_{\rm NL}$.
\begin{figure}
 \includegraphics[scale=0.82]{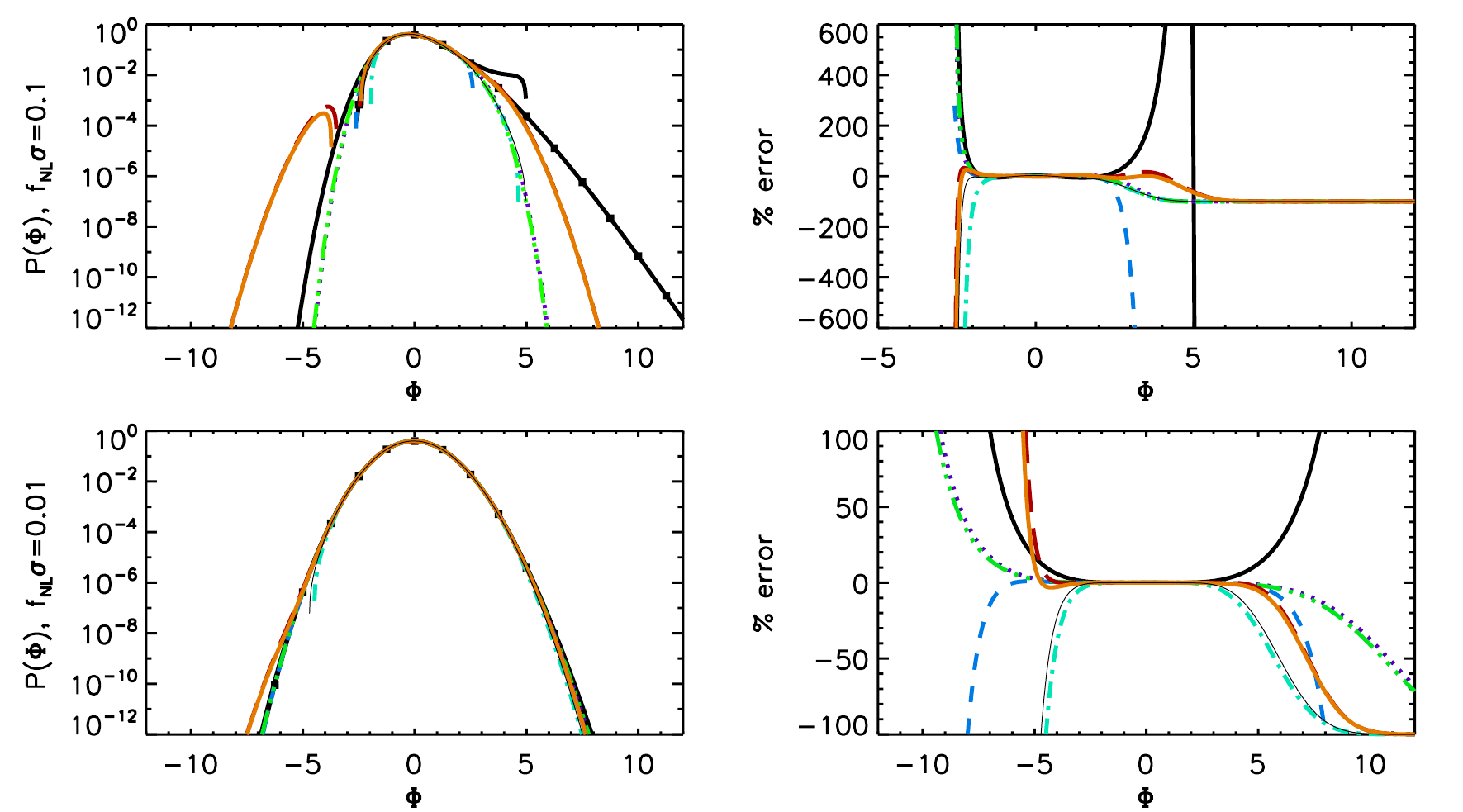}
\caption{Left: the one-point PDF (the underlying Gaussian field $\phi$ is normalized to have unit variance ($\sigma=1$)). Solid  line with diamonds: exact. Solid (black) line: keeping only terms proportional to  $f_{\rm NL}\sigma$ both in the argument of the exponential and in the factor in front.  Solid thin (black) line: First-order expansion. Dotted line (purple):  keeping terms proportional to  $f_{\rm NL}\sigma^2$ both in the argument of the exponential and in the factor in front. Short dashed line (blue):as above but expanding the terms $\propto f_{\rm NL}\sigma^2$ in the argument of the exponential as $\exp(x)=(1+x)$. Dot-dashed (light blue):  expression  where starting from an expansion that keeps terms proportional to  $f_{\rm NL}\sigma^2$ both in the argument of the exponential and in the factor in front, only terms $\propto \Phi$ and $\Phi^2$ are kept in the argument of the exponential, the rest is expanded as $\exp(x)=1+x$.
Solid thick (red): second-order Edgeworth expansion.
 Long dashed (red):  Expansion to second order in $f_{\rm NL}\sigma$. Dot--dot--dashed line (green): approximation of Babich (2005)  Right: \% error in the approximation (same linestyle as in the  left panel). See text for more details.}
\label{fig:expand}
\end{figure}
Figure~(\ref{fig:expand}) shows how  different approximations perform in the case of the one-point PDF (where ${\cal M}$ is the identity matrix).  We  illustrate the performance of the different approximations by using the one-point PDF and taking $\phi$ to have unit variance ($\sigma=1$). Note that  due to this normalization the 
real expansion parameter is $f_{\rm NL} \sigma$. For clarity below we will use  the symbol $f_{\rm nl}$ to indicate the non-Gaussianity parameter for a field with unit variance, thus  $f_{\rm nl}=f_{\rm NL}\sigma$.

The solid (black) line shows the first-order expansion of both the square root and the argument of the exponential:
\begin{equation}
{\cal P}^{1st}(\Phi|f_{\rm nl})=\frac{1}{\sqrt{2\pi}} (1-2f_{\rm nl}\Phi)\exp\left[-\frac{1}{2}\Phi^2+f_{\rm nl}(\Phi^3-\Phi)\right]\,.
\label{eq:expanfirstorder}
\end{equation}

Note that the Babich (2005) \cite{babich05} (B05 from now on) approximation of the PDF (see their Eq. 20)  would correspond to approximating $(1-2f_{\rm nl}\Phi)\sim \exp(-2f_{\rm nl}\Phi)$.
On the other hand the first-order Edgeworth expansion (see e.g. \cite{LoVerde:2007ri} their Eq.~(4.10))  corresponds to further expanding the non-Gaussian part in the exponential by $\exp[f_{\rm nl}(\Phi^3-\Phi)]\sim 1+f_{\rm nl}(\Phi^3-\Phi)$ (thin solid (black) lines).
The popular non-Gaussianity estimator \cite{Komatsu:2003iq,babich05,Creminelli:2006gc}  where the term $(\Phi^3-3\Phi) $ is considered stems from an  expansion of the PDF (more on this below).

In order to make the path integrals in what follows analytic it is worth to note that  an integrand made by the product of a polynomial  with an exponential of a quadratic expression is integrable analytically but an exponential of an expression of power higher than quadratic is not. However expanding the $f_{\rm nl}\Phi^3$ term in the exponent as $(1+f_{\rm nl} \Phi^3)$ performs very badly as Fig.~(\ref{fig:expand}) shows (dot-dashed).

Fig.~(\ref{fig:expand}) shows that the first-order  expansion in $f_{\rm nl}$ does not provide a good approximation to the PDF, and indeed it is well known for example that a first-order Edgeworth expansion can give negative values for the PDF (see discussion in \cite{LoVerde:2007ri}). Going to second order  helps.
Expanding to second order both  the square-root and the argument of the exponential yields the dotted line in the figure. If the second-order part of the exponential is also expanded as $\exp(x)=1+x$ we obtain the short dashed line.

Expanding  the full expression to second order in $f_{\rm nl}$ (thus  expressing the PDF as a Gaussian multiplied by a correction) we obtain something similar (but not quite identical) to the  second-order Edgeworth expansion:
\begin{equation}
{\cal P}^{\rm 2nd}(\Phi | f_{\rm nl})=\frac{1}{\sqrt{2\pi}} \exp\left[-\frac{1}{2}\Phi^2\right]\left[1+f_{\rm nl}(\Phi^3-3\Phi)+\frac{f_{\rm nl}^2}{2}\left(\Phi^6 -11\Phi^4 +23\Phi^2-5\right)\right]
\label{eq:expandsecondorder}
\end{equation}
where the second-order Edgeworth expansion would be:
\begin{equation}
{\cal P}^{\rm 2nd EW}(\Phi |  f_{\rm nl})=\frac{1}{\sqrt{2\pi}} \exp\left[-\frac{1}{2}\Phi^2\right]\left[1+f_{\rm nl}(\Phi^3-3\Phi)+\frac{f_{\rm nl}^2}{2}\left(\Phi^6 -13\Phi^4 +33\Phi^2-9\right)\right]\,.
\label{eq:expandsecondorderEdgeworth}
\end{equation}
The fact that Eq.~(\ref{eq:expandsecondorder}) and Eq.~(\ref{eq:expandsecondorderEdgeworth}) are different at second order should not be too surprising. In fact  while  Eq.~(\ref{eq:expandsecondorder}) is a polynomial expansion around $f_{\rm nl}=0$ using polynomials in their {\it natural} form, Eq.~(\ref{eq:expandsecondorderEdgeworth}) is an expansion in Hermite polynomials, which, for the problem at hand,  are an orthogonal  basis. It is well known that expansions in orthogonal polynomials have  smaller truncation errors, much better convergence and extrapolation properties than other polynomial expansions. Therefore a truncated   Edgeworth expansion should be preferred to a  Taylor expansion  truncated at the same order. Not unexpectedly, our numerical investigations have shown that the difference between  Eq.~(\ref{eq:expandsecondorder}) and  Eq.~(\ref{eq:expandsecondorderEdgeworth}), at least for  relatively small $f_{\rm nl}$, is small.

Note that the B05 \cite{babich05} second-order expansion would correspond to 
\begin{equation}
{\cal P}^{\rm 2nd Babich}(\Phi | f_{\rm nl})=\frac{1}{\sqrt{2\pi}} \exp\left[-\frac{1}{2}\Phi^2+f_{\rm nl}(\Phi^3-3\Phi)+\frac{f_{\rm nl}^2}{2}\left(-5\Phi^4 +14\Phi^2-5\right)\right]\,.
\end{equation}

\begin{figure}
\begin{center}
 \includegraphics[scale=0.5]{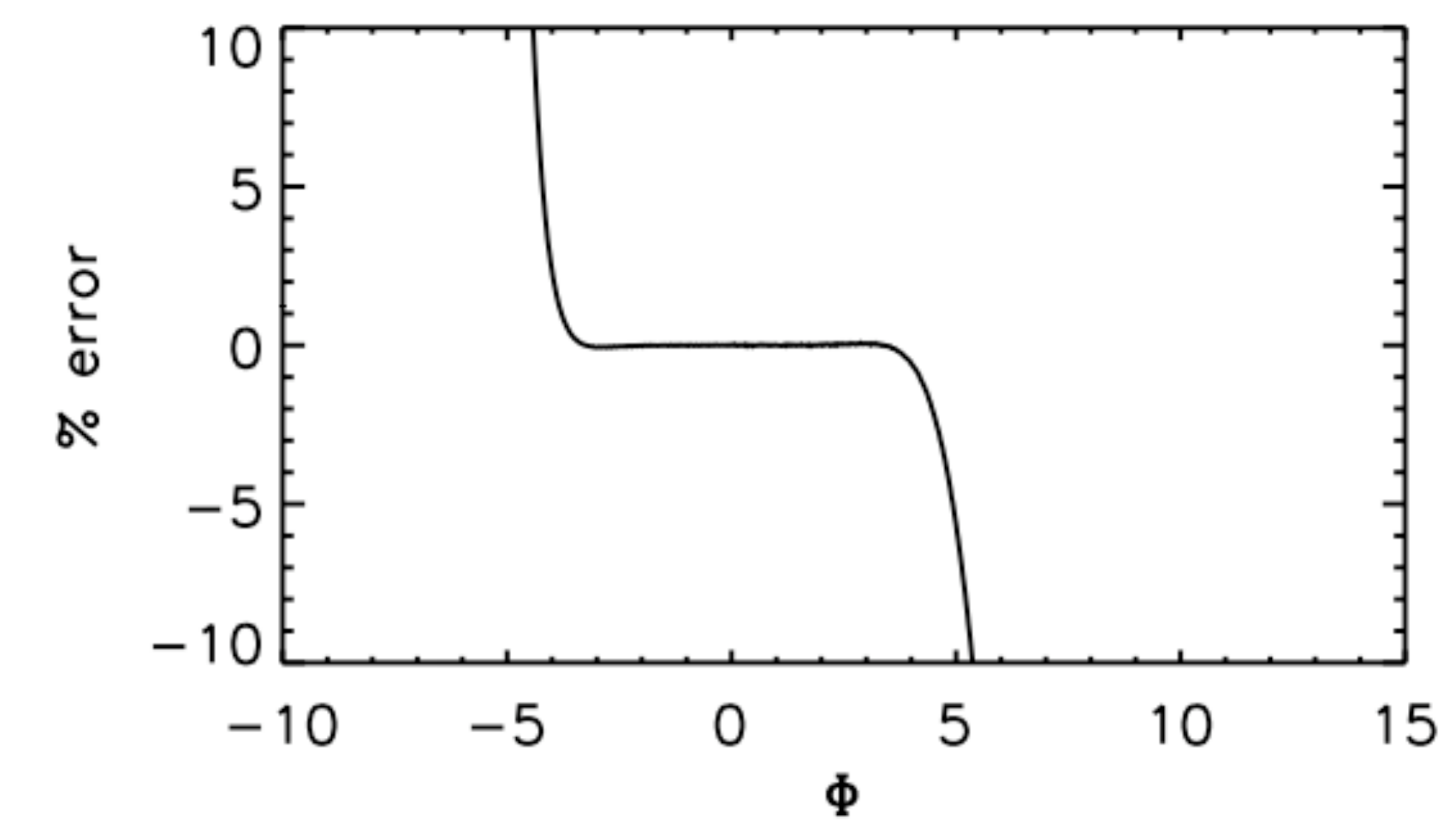}
 \end{center}
\caption{Hybrid approach: the exponential part preserves the Gaussian structure as it  has only terms  $\propto \phi$ and $\propto \phi^2$. Again $\phi$ is normalized to have unit variance.}
\label{fig:approxgaussian}
\end{figure}

 One ``hybrid" approach is to look for simplicity of integration,  first expand to second order both the square root and the argument of the exponential, then leave in the argument of the  exponential only terms $\propto \phi^2$ and $\propto f_{\rm nl}\phi$ the rest expand as $\exp(x)=1+x+x^2$. This is shown in Fig.~(\ref{fig:approxgaussian}).

Having different approximations used in the literature and the exact expression allows us to quantify their performance. In table (\ref{tab:pdfapprox}) we show the range of $\Phi/\sigma$ where the different approximations work at better than 20\%.
We conclude that a second-order  expansion (Edgeworth or plain second order or hybrid) reproduces well the PDF  except in the extreme tails ( $>5\sigma$)  for relatively high $f_{\rm nl}$ values and in addition offers the possibility of performing  many of the relevant   path integrals analytically. In what follows we will therefore use this approximation when needed.

Note that in the form of B05 the log of the PDF  to second order is written as a Gaussian part plus  a first-order term given by $I_1\times f_{\rm NL}$  and a second-order term  $I_2\times f_{\rm NL}^2$.  $\langle I_1\rangle=6 f_{\rm NL}$ and  $\langle  I_2 \rangle=6$. If we write the second-order expansion or the  second-order Edgeworth expansion in the B05 form  we obtain the same $I_1$ expression but different expressions for $I_2$. Note however that $\langle I_2\rangle $ is always equal to 6.

While so far we have interpreted the PDF as a probability of $\Phi$ for a given $f_{\rm NL}$, we can  reinterpret it, assuming  a uniform prior on $f_{\rm NL}$, as the PDF for $f_{\rm NL}$ given $\Phi$.
 Doing so, as we will show in details below, will uncover some properties of this PDF which we believe were missed so far. These stem from the fact that in ${\cal P}(\Phi|f_{\rm NL})$, $f_{\rm NL}$ is fixed and it is therefore physically motivated to chose it to be small. Conversely,  when considering ${\cal P}(f_{\rm NL}|\Phi)$,  the $f_{\rm NL}$ values are not bounded and can in principle range from $-\infty$ to $\infty$. For large $|f_{\rm NL}|$ (or combinations of relatively small $|f_{\rm NL}|$ but large $|\Phi|$),  ${\cal P}(f_{\rm NL}|\Phi)$ has discontinuities which prevent any expansion in $f_{\rm NL}$. It is only when working in a regime away from the discontinuities that  expansions in $f_{\rm NL}$ are meaningful and moments of the distribution are well-defined so the the central limit theorem applies.

\begin{table}
\begin{center}
\label{tab:pdfapprox}
\begin{tabular}{ccccc}
\hline
$f_{\rm NL}\sigma$ & 1st order Babich(2005) & 2nd order  & second-order Edgeworth  & Hybrid\\
\hline
0.1 & 3 & 4 & 5  &  4\\
0.01 & 4 & 6 & 6  &  6 \\
0.001 & 7 & 10 & 10 & 8 \\
\hline
\end{tabular}
\end{center}

\caption{Number of sigmas where the different approximations fit the exact PDF to better than 20\%. Recall that  for simplicity throughout this section we have taken the field to have unit variance.}
\end{table}

\subsection{Estimators vs. full PDF}
\label{sec:estimators}
\begin{figure}
\begin{center}
 \includegraphics[scale=0.7]{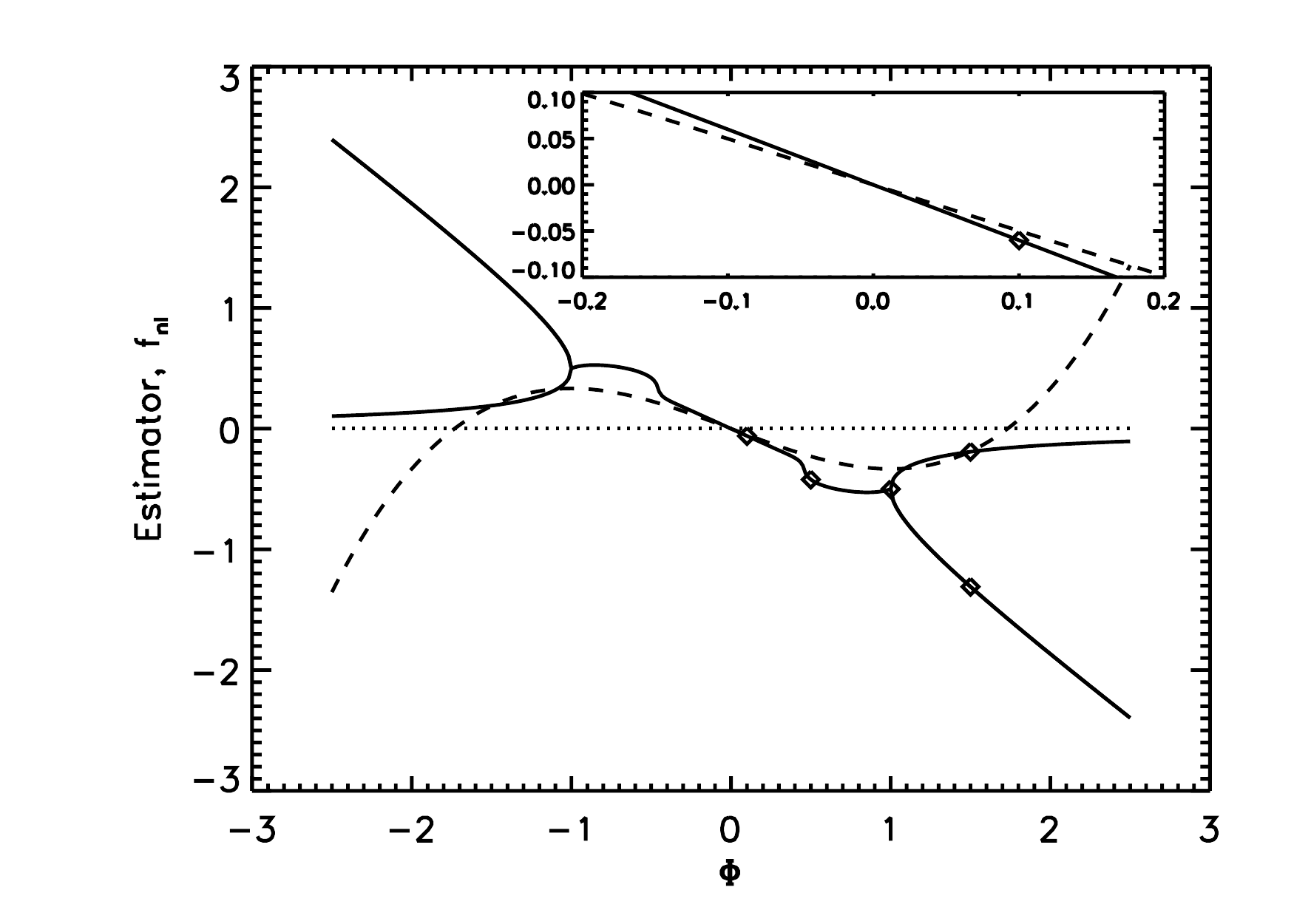}
 \includegraphics[scale=0.7]{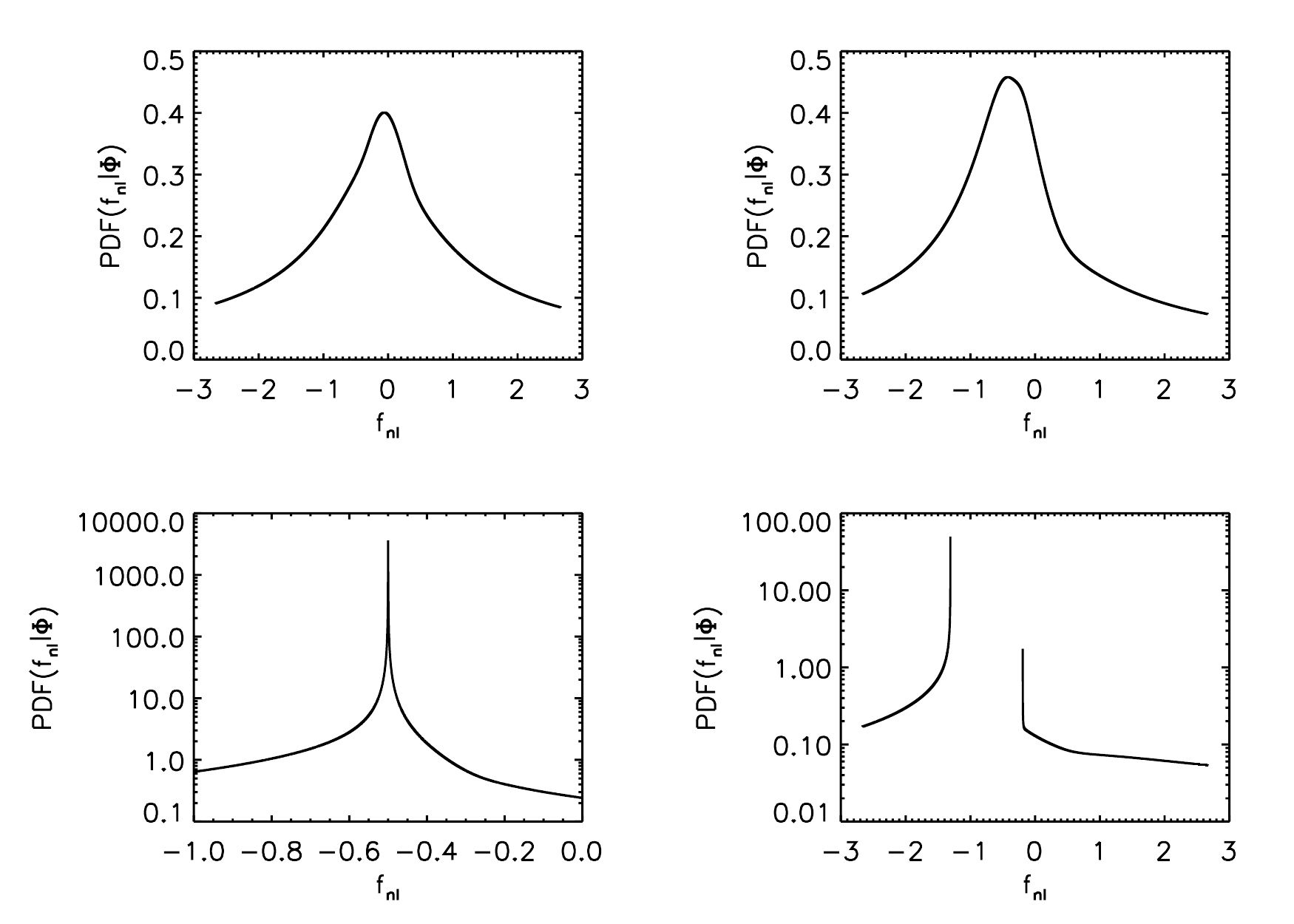}
 \end{center}
\caption{Top panel. Solid line: the numerically-determined  maximum likelihood  as a function of $f_{\rm nl}(=f_{\rm NL}\sigma)$ for several values of $\Phi$. Here the underlying Gaussian field $\phi$ is taken to have unit variance ($\sigma=1$). For  $|\Phi|\ge 1$ the PDF diverges making a maximum likelihood estimator not well-defined. Dashed line: the Babich (2005) estimator. The dotted line is for reference at $f_{\rm nl}=0$. The inset shows a zoom in in the region around $f_{\rm nl}=0$. The Bottom 4 panels show slices through the $f_{\rm nl}$ likelihood for fixed values of $\Phi$ ($0.1$, $0.5$, $1$, $1.5$ from top left to bottom right), which maximum correspond to the locations of the diamonds. Note that the likelihood for $f_{\rm nl}$ is always highly non-Gaussian. For $|\Phi|>1$ there are two points where the PDF diverges that are separated by a region where PDF is zero.  For $|\Phi|\ge 1$ the derivatives of the likelihood are non continuous. }
\label{fig:mle}
\end{figure}

We continue our detour to discuss the choice of an estimator for $f_{\rm NL}$ given the exact and approximated expressions of the PDF in the simplified one-point case discussed above.  We use this simplified case for clarity, as we will see below, it still captures all the interesting physics, features and limitations of known estimators. As we will see below, this exercise is  useful to clarify some common misconceptions.
While all statistical  information is enclosed in the full PDF,  if the PDF is complicated or hard to compute it is often useful to work with estimators. An estimator represents a massive reduction of information: in this case for example  the  full PDF is substituted by a single  number (the estimated $f_{\rm NL}$ value).
It is worth recalling that the maximum likelihood estimator  (MLE) is often desirable because, if it exists, under mild conditions, it is asymptotically the best unbiased estimator. To derive a MLE one interprets the PDF (probability of the data given the parameters) as the likelihood for the parameters given a set of data, and finds its maximum. This conditions give a combination of the data ($\Phi$ in our case at hand) which is a MLE of $f_{\rm NL}$ (or $f_{\rm nl}$) i.e. $\widehat{f}_{\rm NL}$ (or $\widehat{f}_{nl}$) . The drawback of a MLE is that in many problems the full PDF is not known or if it is known, its maximum cannot be found analytically. For our exact formulation of the PDF, it is clear that the maximum must be found numerically. In addition, in some regimes the PDF diverges making a MLE not well-defined. Fig.~(\ref{fig:mle}) --solid line, top panel-- shows the locus of the maximum of the likelihood as a function of $\Phi$ for the simple one-point PDF investigated above (i.e., in this sub-section we keep considering $\phi$ to have unit variance). The dotted line is for guidance at $f_{\rm NL}=0$. The dashed line corresponds to the estimator of B05. In correspondence of the diamond symbols in the bottom panels we show the shape of the likelihood as a function of $f_{\rm nl}$  for fixed $\Phi$ values ($0.1$, $0.5$, $1$, $1.5$). The likelihood is well-behaved as long as $\Phi$ is small. The location of the ``cusp" in the curve corresponds to $|\Phi|=1$ and for $\Phi=1+\epsilon$ (where $\epsilon$ is a small number) the PDF diverges, the  discriminant becomes negative (and the PDF is zero) splitting the divergence through the middle. For $|\Phi|>1$ there are two points where the PDF diverges, separated by a region where the discriminant is negative and the PDF is zero.

Furthermore, the support of the PDF depends on the parameter, since it is zero for $\Phi<-f_{\rm nl}-1/(4f_{\rm nl})$, and this makes the Cram\' er-Rao bound invalid \cite{Tenorio:2006}.

These conditions make the MLE not well-defined.   Note that the B05 estimator is not a MLE except for $f_{\rm nl}=0$. Its error is shown to satisfy the Cram\' er-Rao bound, but that  just corresponds to an expansion of the PDF i.e. it approximates the PDF as a function of $f_{\rm nl}$ as a Gaussian distribution. The plots in Fig.~(\ref{fig:expand}) show that this is not a good approximation: even  when $\Phi$ is very small  the likelihood has broad wings. While it is true that the central limit theorem should make the PDF  close to Gaussian for practical applications where many independent measurements are considered, the full PDF shape carries much more information than its mean and the variance, as discussed at length in \cite{Kamionkowski:2010me}. 
Since we have the exact expression for the PDF we can compute the possible bias of the estimator by computing  (numerically) the expectation value of it.

Numerically we find that the estimator of B05  is unbiased even for large values of $f_{\rm nl}$. In particular the bias of the estimator scales like $f_{\rm nl}^2$ as follows:
\begin{equation}
{\rm bias\,\, of\,\,\, estimator} = 0.001 \left ( \frac{f_{\rm nl}}{0.1} \right )^2\,.
\end{equation}
Given that the maximum likelihood  estimator is not well-defined and that the full likelihood is known, and that the PDF carries more information than just the mean and variance of an estimator, it is much more useful   (although not necessarily simpler) to work directly with the  full PDF. This is what we set up to do here.

The above discussion is valid for a single pixel. In any practical application  many independent measurements will be combined, and a full analysis would include in-field correlations (e.g. \cite{Marinucci:2006}).   For illustration, we consider first a full posterior analysis for $f_{\rm NL}$ based on the complete PDF in $n$ pixels.

Using Bayes' theorem, the probability of the non-Gaussianity parameter having a value $f_{\rm nl}$ is $
{\cal P}(f_{\rm nl}|\Phi) \propto {\cal P}(\Phi|f_{\rm nl})$, for a uniform prior on $f_{\rm nl}$.  For $i=1,\ldots,n$ pixels, assumed independent, the probability of $f_{\rm nl}$ given a set of pixel values $\{\Phi_i\}\equiv\vec{\Phi}$ (normalized so that the underlying Gaussian field $\phi$ has unit variance for the case at hand) is
\[
{\cal P}(f_{\rm nl}|\vec{\Phi}) = \prod_i {\cal P}(f_{\rm nl}|\Phi_i)  \propto \prod_i {\cal P}(\Phi_i|f_{\rm nl})
\]
so
\[
\ln {\cal P}(f_{\rm nl}|\vec{\Phi}) = \sum_i \ln {\cal P}(\Phi_i|f_{\rm nl}) + const\,.
\]
If the true $f_{\rm nl}$ is $f^{\rm true}_{nl}$, then the probability of a pixel having a value between $\Phi$ and $\Phi+d\Phi$ is ${\cal P}(\Phi|f^{\rm true}_{nl})d\Phi$. So the sum over the set of pixels becomes an integral over $\Phi$, with a weight given by ${\cal P}(\Phi|f^{\rm true}_{nl})$ (and a normalisation given by the number of pixels):
\[
\ln {\cal P}(f_{\rm nl}|\vec{\Phi}) = n\int d\Phi {\cal P}(\Phi|f^{\rm true}_{nl})  \ln {\cal P}(\Phi|f_{\rm nl}) + const.
\]
Note that the for the purpose of this illustration the number of pixels enters only as a normalization factor; in fact  the sum is evaluated as an integral and the limits of integration are (formally) from $-\infty$ to $\infty$. In practice, given a finite number of pixels,  the chances of  sampling the tails of the distribution are small and thus effectively the limits of integration should depend on the number of pixels.

\begin{figure}
\begin{center}
\includegraphics[scale=0.43]{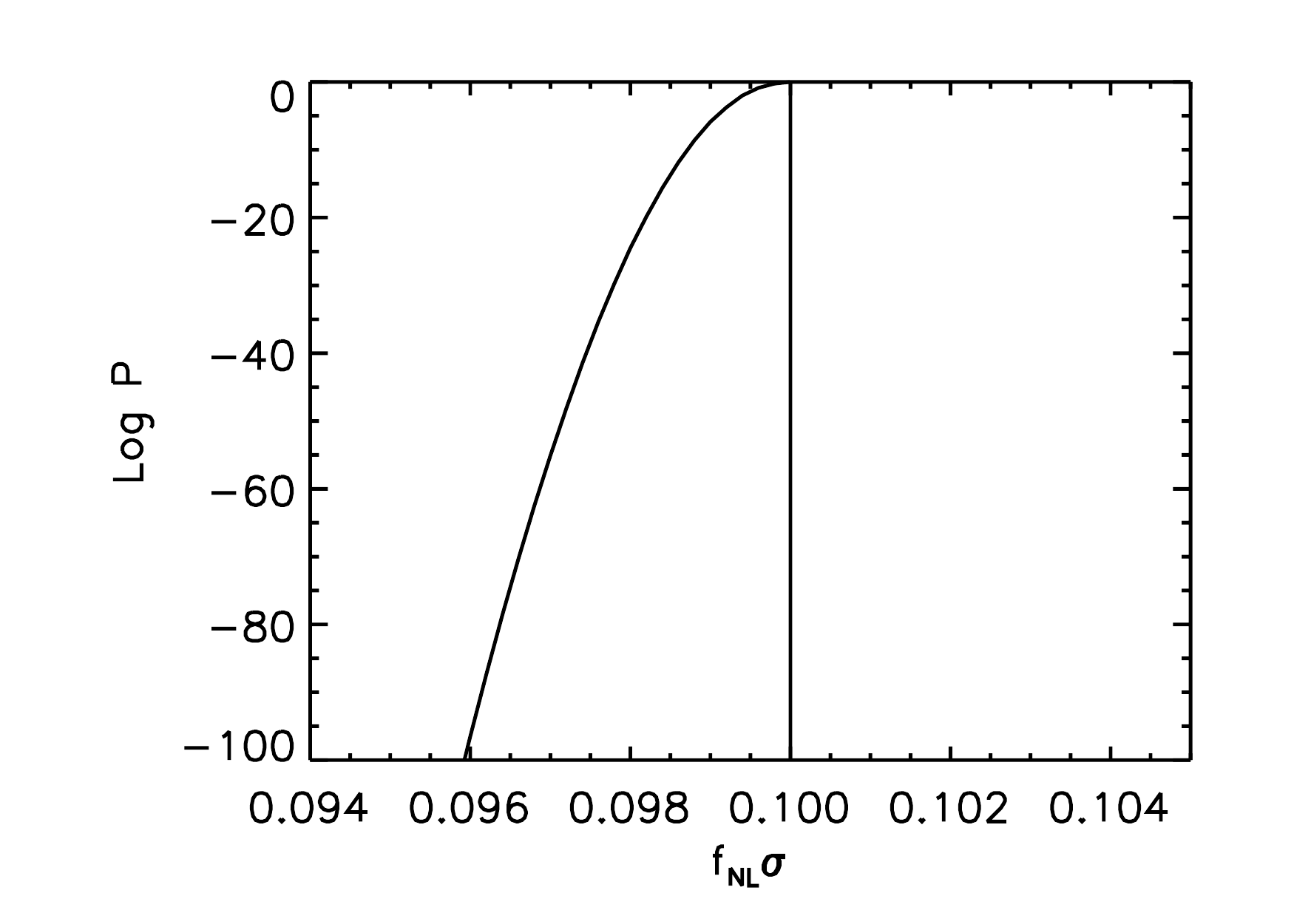}
\includegraphics[scale=0.43]{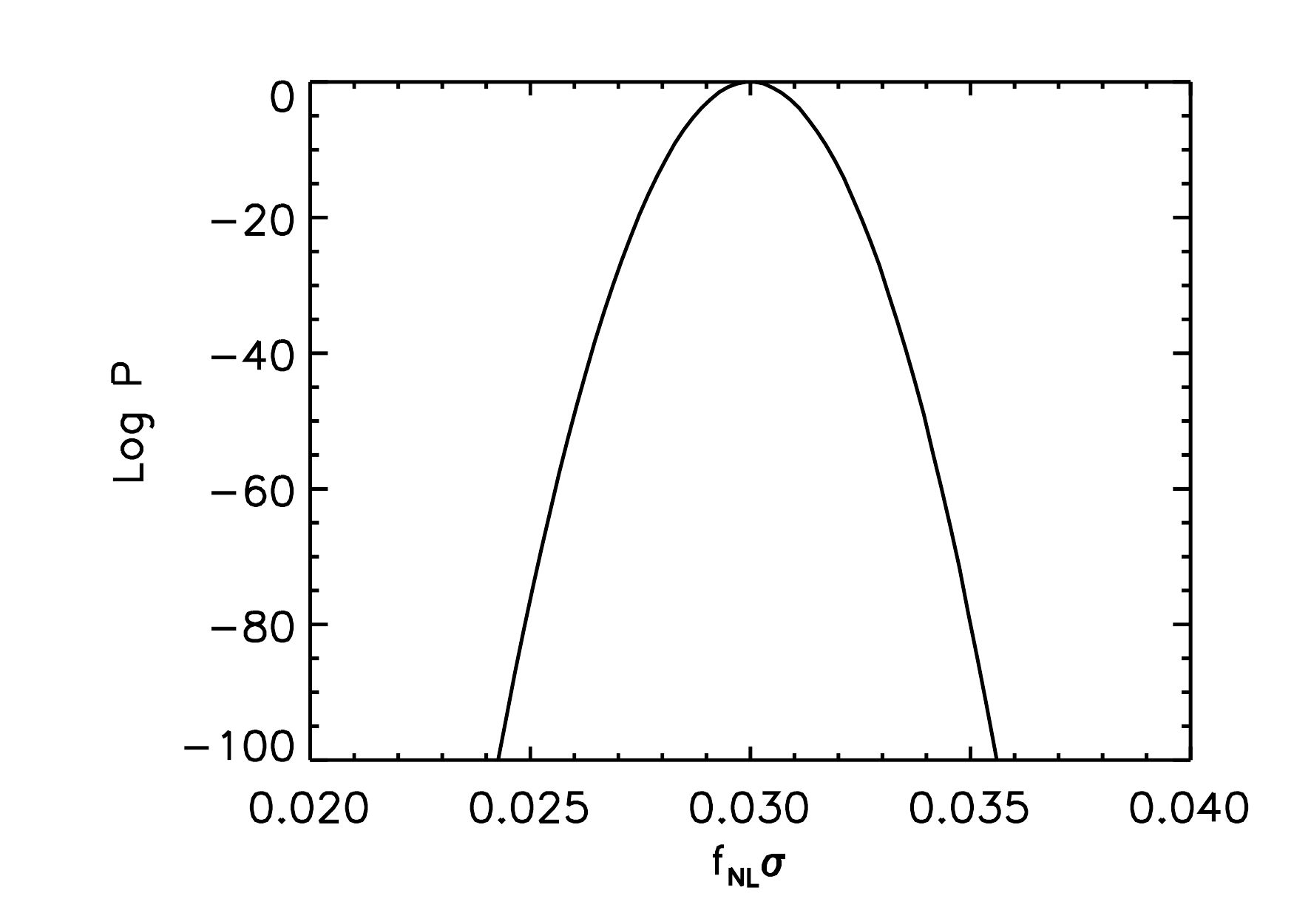}
\caption{{\it Left panel}: Log(Posterior) for $f_{\rm nl} (=f_{\rm NL}\sigma)$, with true $f^{\rm true}_{nl}$=0.1 and $10^6$ pixels but ignoring sampling issues and noise. $f_{\rm nl}>f^{\rm true}_{nl}$ is excluded, as the true PDF is non-zero for some values of $\Phi$ where the trial PDF is zero. {\it Right panel:}  Log(Posterior) for $f_{\rm nl}(=f_{\rm NL}\sigma)$ given $10^6$ pixels, and  $f_{\rm nl}^{\rm true}=0.03$. A Gaussian noise of rms $0.01$ added (recall that the underlying Gaussian field $\phi$ is normalized to have unit variance).  The noise makes all values of $\Phi$ reachable in principle, and so the posterior is always non-zero.  This removes the strict cutoff apparent in the left panel.}
\label{fig:Post1}
\end{center}
\end{figure}

The above equation implies that ${\cal P}(f_{\rm nl}|\vec{\Phi})$ is zero for $ f_{\rm nl}>f^{\rm true}_{nl}$, since there is an excluded region which depends on $f_{\rm nl}$: ${\cal P}(\Phi|f_{\rm nl})=0$ for $\Phi<\Phi_{\rm min}$ where $\Phi_{\rm min}=-f_{\rm nl}-1/(4f_{\rm nl})$ (Note that this limit is very small for small $f_{\rm nl}$, e.g. for $f_{\rm nl}=0.01$, $\Phi>-25.01$).  Hence,  if $f_{\rm nl}>f^{\rm true}_{nl}$, there is a region of $\Phi$ where ${\cal P}(\Phi|f^{\rm true}_{nl})\ne 0$ and $\ln {\cal P}(\Phi|f_{\rm nl}) \rightarrow -\infty$ and the integral diverges. This is borne out by doing the integration (see left panel of Fig.~(\ref{fig:Post1})) (Note that this argument applies to small $f_{\rm nl}$, in which case $\Phi_{\rm min}$  increases with $f_{\rm nl}$). 
 This sharp discontinuity will also prevent any truncated expansion of the PDF being a good approximation.

However, in a practical case, the chances of reaching the `excluded' region which is forbidden for a small (positive) $f_{\rm nl}$ are basically vanishingly small, so the {\em cliff} in the posterior is not going to happen in practice, because of sampling, or it could also happen because of noise.  For example, if we introduce a small Gaussian noise into the measurements, then the probability of reaching the otherwise forbidden region of $\Phi$ is no longer zero, and this allows values of $f_{\rm nl}>f^{\rm true}_{nl}$ to have non-zero probability.  The right panel of Fig.~(\ref{fig:Post1}), shows the posterior for local non-Gaussianity with a Gaussian noise added, with rms only 0.01; the zero probabilities have gone away, and the posterior becomes quite Gaussian.  This behaviour is quite robust for small smoothing; for large smoothing the posterior gets wider.


The cutoff disappears also if we now consider that for a sample of finite size the tails of the distribution are not well sampled. For example for $f_{\rm nl}=0.05(0.1)$, the region $\Phi<-5.05(-2.6)$  is not sampled if the sample is  smaller than  about $1.7\times 10^6$ ($80$).
The disappearance of the {\it cliff} makes  truncated expansions in $f_{\rm NL}$ of the PDF viable. Note however that a  truncation to first order  will yield to a ${\cal P}^{1st}(f_{\rm NL})$ that does not have a maximum, which therefore cannot be a good approximation. At the minimum to have a maximum the approximated PDF must be truncated at second order or higher in $f_{\rm NL}$. 

\begin{figure}
\begin{center}
 \includegraphics[scale=0.4]{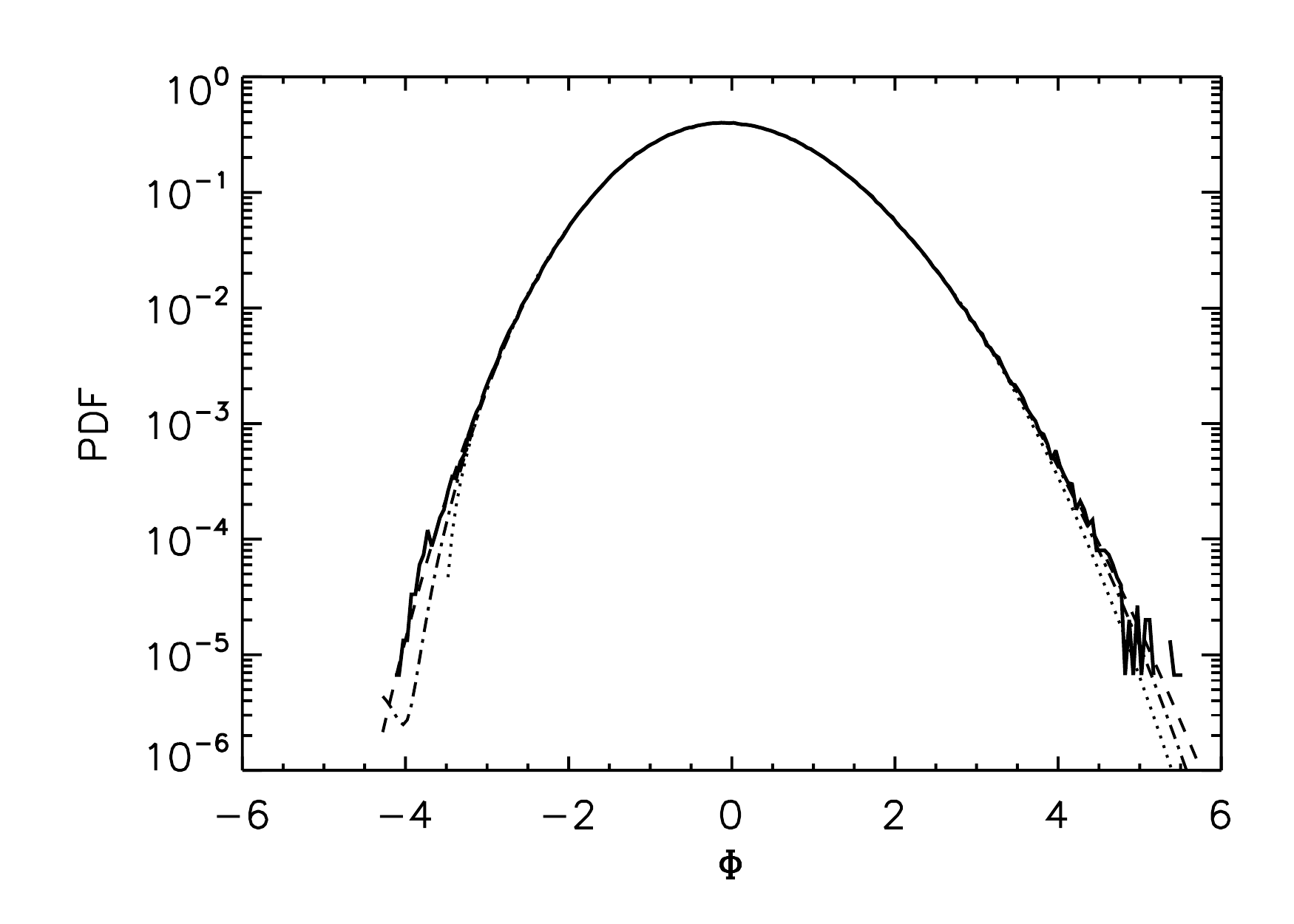}
 \includegraphics[scale=0.4]{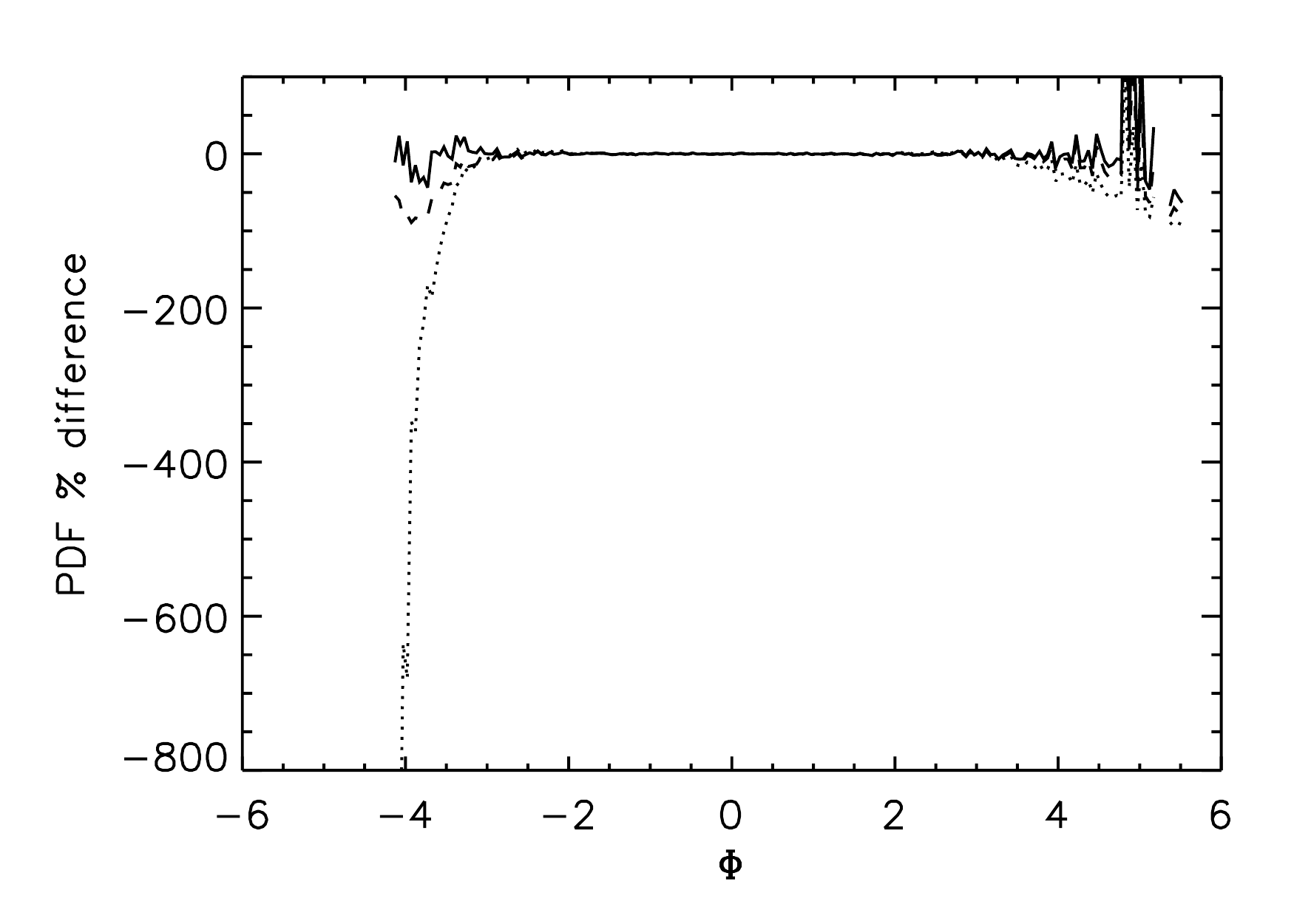}
\end{center}
\caption{Left  panel: PDF simulated (3 million $\Phi$'s, thick solid line), analytical exact expression (dashed line), first order (dotted) and second order (dot-dashed). Right panel: \% difference  compared with simulated PDF, analytical exact expression (solid), first order (dotted) and second order (dashed). Here  the underlying Gaussian field $\phi$ has unit variance and $f_{\rm nl}^{\rm true}=0.03$.}
\label{fig:distrib}
\end{figure}

We next explore the performance of the estimator, the exact PDF and  its approximations in the regime where  sampling is finite.  To do so we generated simulations ({\it realizations}) of uni-variate non-Gaussian fields for given values of $f_{\rm nl}$ (which we refer to as $f_{\rm nl}^{\rm true}$) and  proceed to recover (measure) it from {\it samples} of the field (which we call $f_{\rm nl}^m$, where $m$ stands for {\it measured}). Before proceeding we verified that the  distribution of the realization matches to analytic exact PDF for all $f_{\rm nl}$ values. This is shown in Fig.~(\ref{fig:distrib}) for $f_{\rm nl}^{\rm true}=0.03$.

\begin{figure}
\begin{center}
 \includegraphics[scale=0.41]{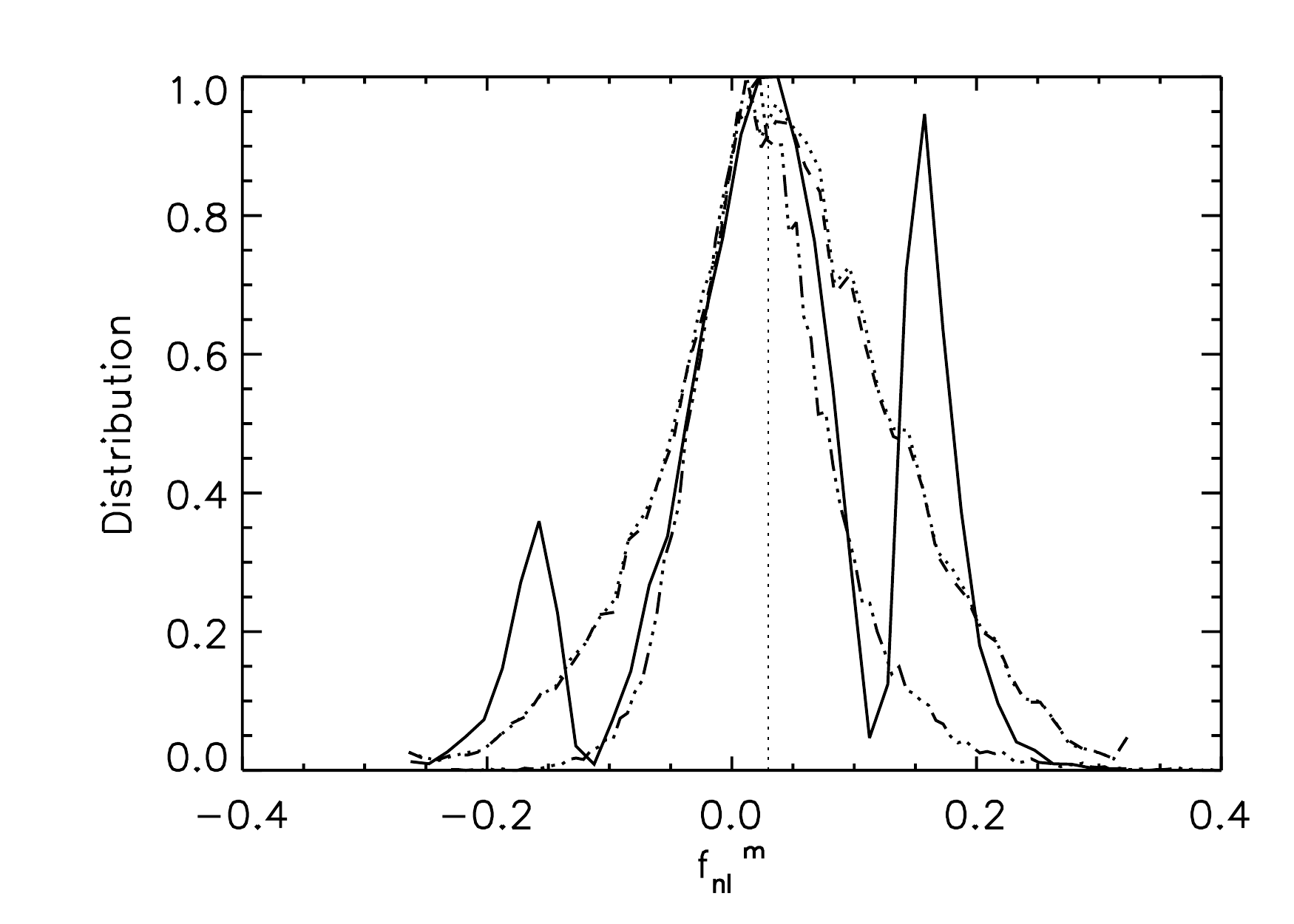}
 \includegraphics[scale=0.41]{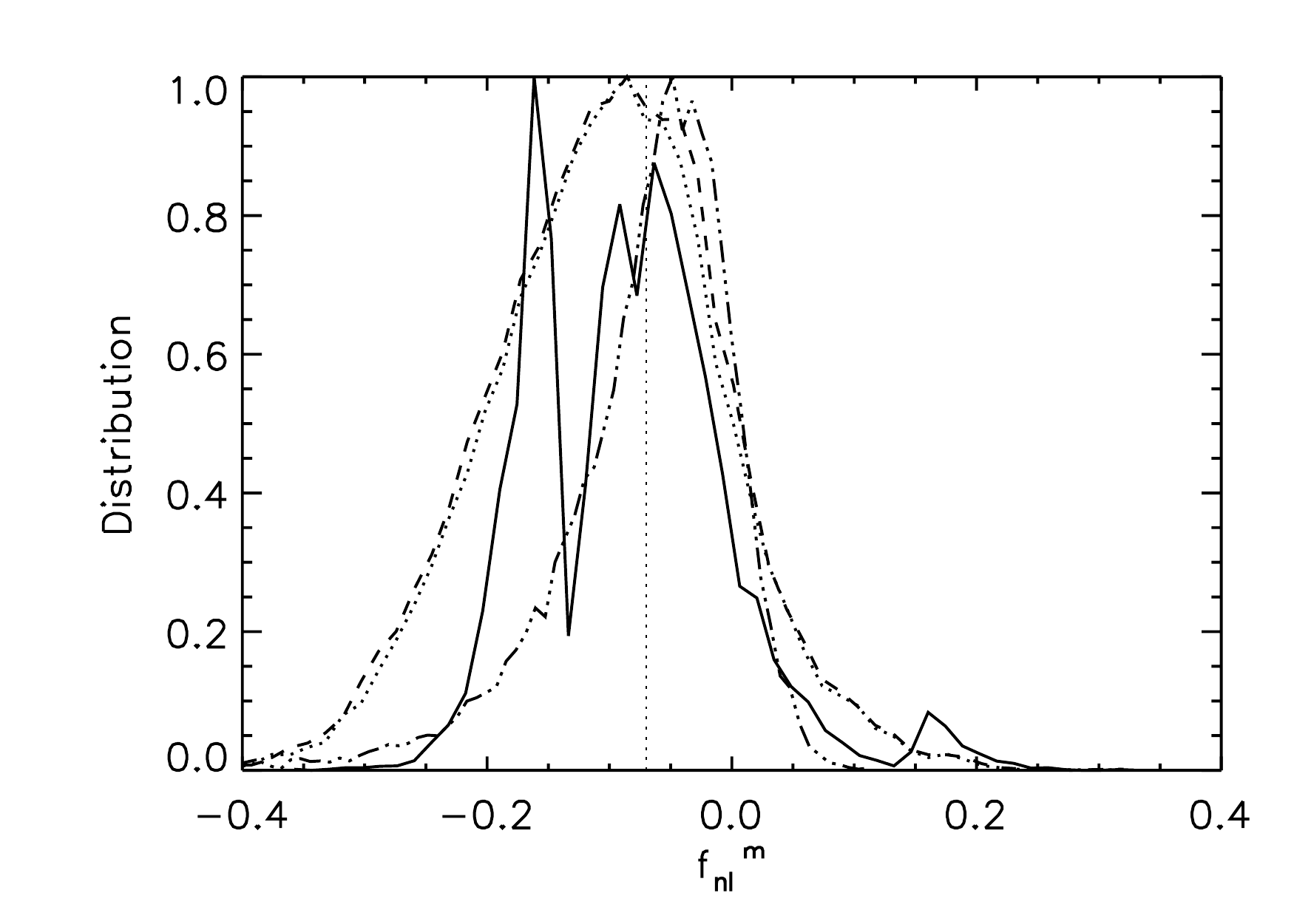}
\end{center}
\caption{Distribution of $f_{\rm nl}^m$ for the exact PDF (solid line), first order (dotted) second order (dashed), B05 estimator (Dot-dot-dot-dashed) in the small sample size limit. The thin vertical dotted line shown the location of $f_{\rm nl}^{\rm true}$ ($0.03$ on the left panel and $-0.07$ on the right panel).}
\label{fig:SSS}
\end{figure}

We concentrate here on  two regimes:  one where there are few samples (only $50$ $\Phi$s; {\it small sample size})   and one where there are many more samples ($1500$ $\Phi$s; {\it large sample size}).  For simplicity we assume that the pixels (samples) are uncorrelated. 
Fig.~(\ref{fig:SSS}) shows the distribution of the $f^m_{nl}$ obtained using the exact PDF (solid line), the first and second-order PDF (dashed and dotted, almost indistinguishable) and the B05 estimator (Dot-dot-dot-dashed). To be explicit, since the pixels are uncorrelated we can multiply the individual likelihoods ${\cal P}(f_{\rm nl}|\Phi_1...\Phi_n)=\prod_i {\cal P}( f_{\rm nl}|\Phi_i)$ and $f^m_{nl}$ is the value that maximizes such product.

Clearly the distribution of   $f^m_{nl}$ is non-Gaussian. In particular we note that if $f_{\rm nl}^{\rm true}$ is small ($0.03$ in this example) there is a non-negligible probability  that the $f_{\rm nl}$ of the sample is $\sim 0.15$ if the sample is small. The shape of the PDF for small sample size reflects the  behaviour seen in Fig. (\ref{fig:mle}) bottom right panel.

\begin{figure}
\begin{center}
 \includegraphics[scale=0.42]{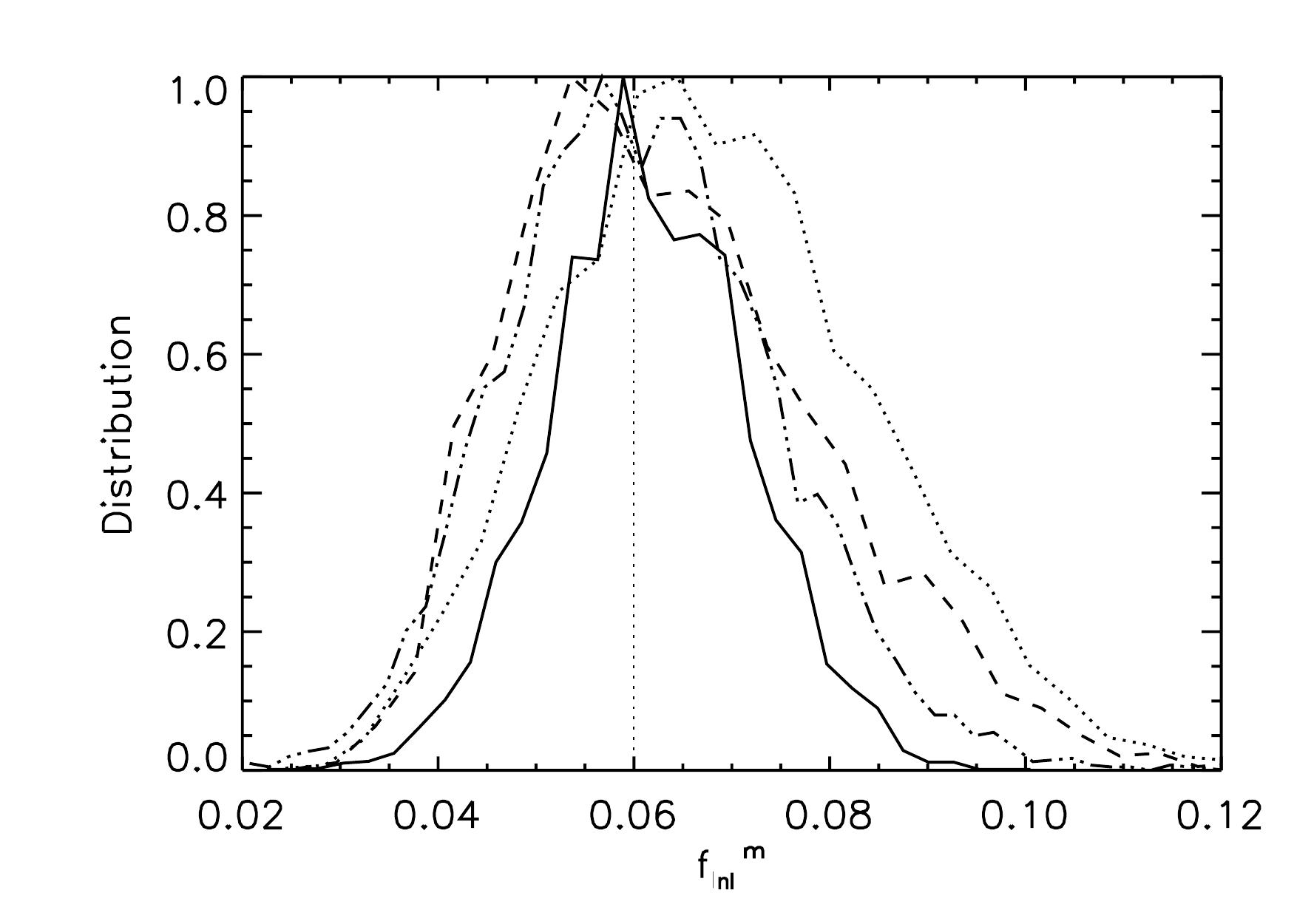}
  \includegraphics[scale=0.42]{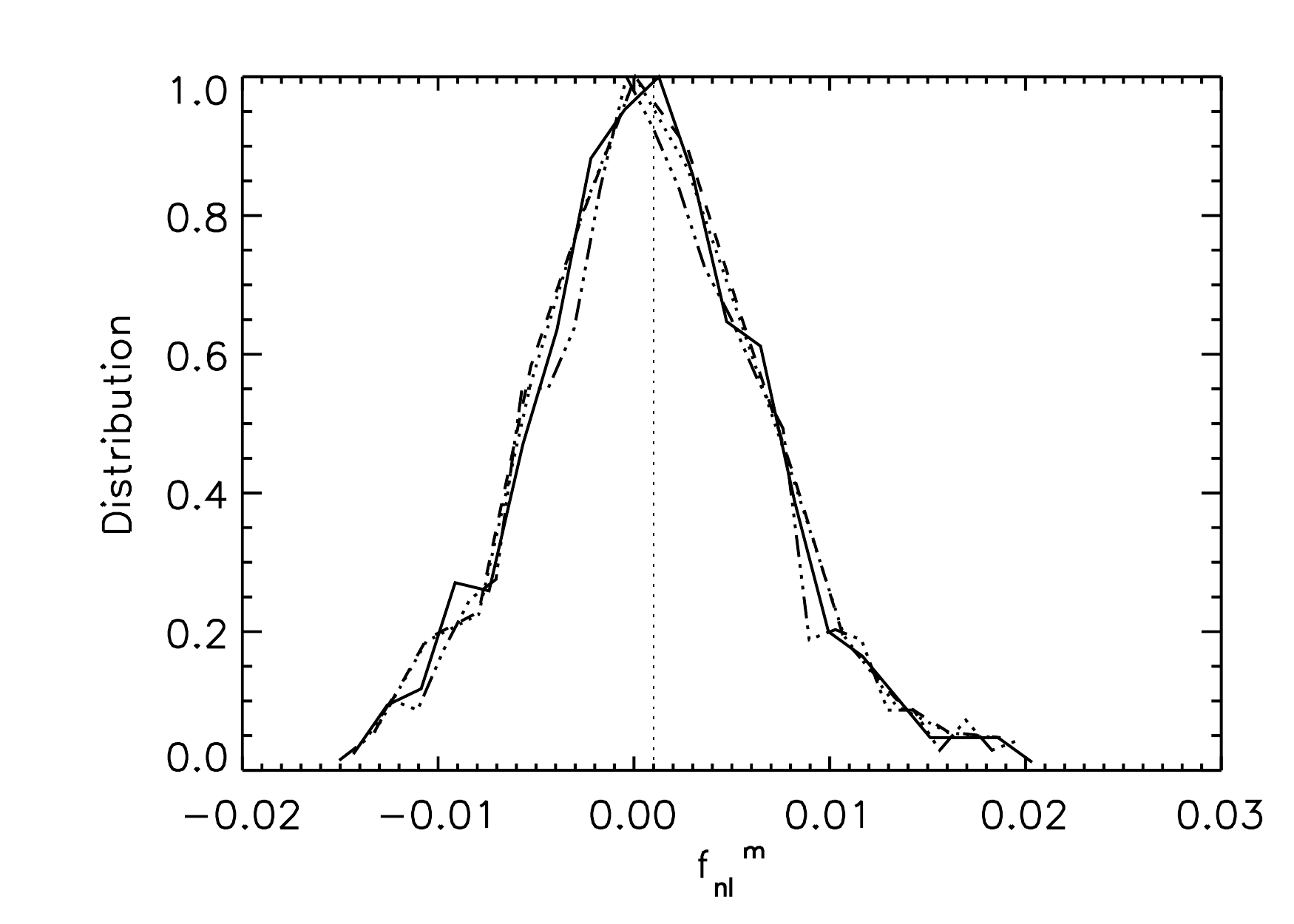}
 \end{center}
\caption{Distribution of $f_{\rm nl}^m$ for the exact PDF (solid line), first order (dotted) second order (dashed), B05 estimator (Dot-dot-dot-dashed) in the large sample size limit. The thin vertical dotted line shows the location of $f_{\rm nl}^{\rm true}$. Note that all approximations  overestimate the width of the true distribution for large $f_{\rm nl}^{\rm true}$. As $f_{\rm nl}^{\rm true}\longrightarrow 0$ the PDF, its approximation and estimator distributions converge.}
\label{fig:LSS}
\end{figure}

This behaviour disappears when the sample size is increased as shown in Fig.(\ref{fig:LSS}). Note that all approximations and the estimator have a variance that is larger than the one given by the exact PDF when $f_{\rm nl}^{\rm true}$ is large. As $f_{\rm nl}^{\rm true}\longrightarrow 0$ the PDF, its approximation and estimator distributions converge. The resulting distribution remains however non-Gaussian. This demonstrates the importance of working with a PDF (exact  if  $f_{\rm nl}^{\rm true}$ is large, exact or approximated if  $f_{\rm nl}^{\rm true}$ is small): the confidence intervals for $\gg 1 \sigma$ or $\gg 68\%$ levels, cannot be estimated from the {\it rms}. In the absence of a PDF, many simulations of the experiment would be required to estimate these confidence intervals via Monte Carlo  simulation. On the other hand knowing the PDF allows one to  compute  any desired confidence interval. Although in this section we have worked in the case where ${\cal M}$ is the identity matrix, we expect that these findings hold qualitatively in the general case. 

Our detour  ends here; we now go back at deriving analytic expressions for the full PDF for increasingly complicated (realistic) cases.
 
\subsection{If ${\cal M}$ is not the identity matrix}
\label{sec:minvertible}
If ${\cal M}$ is not the identity matrix but it is an invertible square matrix    then
 we need to go back to the expression for ${\cal P}$ as:
 \begin{eqnarray}
{\cal P}(\vec{A} |  f_{\rm NL})&=&\sqrt{\frac{\det K_0}{(2\pi)^M}}\int {\prod_{i,\nu} d{\phi}_id{\phi}_{\nu}} \prod_i\delta^D(A_i-\sum_{a}{\cal M}_{ia}{\Phi}_a)\times  \nonumber \\
&&\exp\left[-\frac{1}{2} \sum_{ij}{\phi}_i{\phi}_j{K_0}_{ij} - \sum_{i,\nu}{\phi}_i{\phi}_{\nu}{K_0}_{i\nu}-\frac{1}{2} \sum_{\mu,\nu}{\phi}_{\mu}{\phi}_{\nu}{K_0}_{\mu\nu} \right]\,.
\label{eq:Pfnl} 
\end{eqnarray}

If $n=M$ then one could do 

 \begin{equation}
A_b\longrightarrow \sum_a ({\cal M}^{-1})_{ba} A_a\equiv \tilde{A}_b\,,
\end{equation}
which also introduces a factor of $1/\det{\cal M}^{-1}=\det{\cal M}$ in front of the RHS.
This complicates the Dirac delta function but leaves us with a Gaussian integral.

We obtain
\begin{eqnarray}
{\cal P}(\vec{A} |  f_{\rm NL})&=&\sqrt{\frac{\det{K_0}}{(2\pi)^M}}\det{\cal M}\prod_a\frac{1}{\sqrt{1+4f_{\rm NL}(\tilde{A}_a+C)}}\times  \nonumber \\
&&\sum_{rq}\left\{  \exp\left[-\frac{1}{2}\sum_{ab} f_r(\tilde{A}_a) f_q(\tilde{A}_b) { K_0}_{ab}\right]\right\}\,;
\label{eq:pmultiexactM}
\end{eqnarray}
if we keep only one root for small $f_{\rm}$, $r=q=1$. 

For a possible practical application ${\cal M}$ is a square invertible matrix;  for example when $A$ corresponds to the  matter overdensity field $\delta$ be it in real or Fourier space. If $n<M$ this route cannot be used (see below for the approach to use) but one could always write :
\begin{equation}
{\cal P}(\vec{A} |  f_{\rm NL})=\int{\cal D}[{\cal A}]{\cal P}({\cal A})\delta^D(A-{\cal A})\,.
\label{eq:massfn101}
\end{equation} 
 So for example  for the one-point PDF one should do: ${\cal P}(A_1)=\int{\cal D}[{\cal A}]{\cal P}({\cal A})\delta^D(A_1-{\cal A})$. By applying the Dirac delta one  reduces the dimensionality of the path integral  by one. The remaining integral might still have to be performed numerically.

\subsection{${\cal M}$ not invertible,  $n<M$}
\label{sec:4.5}
 To obtain directly the probability of ${\cal P}(\vec{A})$ if $n<M$ and ${\cal M}$ is not invertible let us go back to Eq.~(\ref{eq:Pfnl}) and  
as before let us now split the Dirac delta function so its argument is given by the elements of:
\begin{equation}
A-\aleph \Phi'-{\cal N}\Phi''=A-\aleph \phi'-\aleph f_{\rm NL}\phi'^2-{\cal N}\phi''-f_{\rm NL}{\cal N}\phi''^2+f_{\rm NL}\sigma^2_{\phi}{\rm c}
\end{equation}
where the elements of the matrix ${\rm c}$ are given by ${\rm c}_i=\sum_{a}{\cal M}_{ia}$.
From now on let us redefine $c=f_{\rm NL}\sigma^2_{\phi}{\rm c}$.

Now note that to impose the Dirac delta function one must first integrate in $\phi'$ as such  it is convenient  first to change variables  so that the argument of the Dirac delta function is multiplied by $\aleph ^{-1}$, which yields a $|\det \aleph|$ in front of the equation and then it is useful to define  :
\begin{equation}
\tilde{A}(\phi''|f_{\rm NL})=\aleph^{-1}A-\aleph^{-1}{\cal N}\phi''-f_{\rm NL}\aleph^{-1}{\cal N}\phi''^2+\aleph^{-1}c\,.
\label{eq:67}
\end{equation}
with this the roots of the quadratic equation are:
\begin{equation}
\phi_{r,i}=\frac{-1\pm\sqrt{1+4f_{\rm NL}\tilde{A}_i(\phi'' | f_{\rm NL})}}{2f_{\rm NL}},
\end{equation}
and the Dirac delta function becomes
\begin{equation}
\frac{\delta^D(\phi_i-\phi_{r=1})+\delta^D(\phi_i-\phi_{r=2})}{\sqrt{1+4f_{\rm NL}\tilde{A}_i}}.
\end{equation}
Thus we arrive at
\begin{equation}
\addtolength{\fboxsep}{5pt}
\boxed{
\begin{gathered}
{\cal P}(\vec{A} | f_{\rm NL}) = \sqrt{\frac{\det K_0}{(2\pi)^M}} \det \aleph \!\!\int d\phi''\prod_i\left(\sqrt{1+4f_{\rm NL}\tilde{A}_i(\phi''|f_{\rm NL})}\right)^{-1}\!\!\!\!\!\!\!\exp\!\left[-\frac{1}{2}\sum_{\nu\mu}\phi_{\mu}\phi_{\nu}{K_0}_{\mu \nu}\right]\!\!\times  \\ 
 \sum_{r=1,2}\exp\left[-\frac{1}{2}\left(\sum_{ij}\phi_{r,i}{K_0}_{ij}\phi_{r,j} +2 \sum_{i\nu}\phi_{r,i}{K_0}_{i \nu}\phi''_\nu\right)\right].
\end{gathered}
}
\label{eq:70}
\end{equation}
Again one can safely ignore the exponentially suppressed root.
Note that even if only an integral in $\phi''$ is left,  there are complications to perform this analytically:   $\phi''$ appears not only in front of the exponential as an argument of the function $\tilde{A}$ (under the square root),  but also in the exponential both under square root  in  the $\phi_r$ and explicitly.

This equation  is one of the main results of this paper as it can be interpreted as the probability of $f_{\rm NL}$ given the observable vector $A$. The expression is exact but the path integral must be performed numerically for different $f_{\rm NL}$ values. Note that the dimensionality of the integration has been reduced by a factor $L$ compared to e.g., Eq.~(\ref{eq:60})-- i.e. for a  full sky CMB map the dimensionality of the integral has been reduced by a factor ${\cal O}(10^6)$. The  integrand  is close to  a Gaussian  so the  numerical integration should converge easily but detailed implementation will depend on the particular application one is interested in. We  will return on this in Sec.~(\ref{sec:CMBappl}).

\subsection{Approximations}
\label{sec:4.6approximations}
In order to progress analytically  approximations must be made. 
 Taylor-like expansions   around $f_{\rm NL}=0$ could  be considered  and they could be chosen so that the integral (Eq.\ref{eq:70}) becomes analytic, in the same spirit as  Sec. (\ref{sec:approximations}).
Sec.~(\ref{sec:approximations}) and (\ref{sec:estimators}) indicate that a second-order (Edgeworth) expansion is a good approximation for the full PDF if $f_{\rm NL}$ is small and the sample size is reasonably large. 
While we have quantified this numerically for the single pixel and  Poisson, uncorrelated pixels case, we expect that qualitatively the same will hold in the more general, multi-variate case. 
 While for the  one-point PDF where ${\cal M}$ is the identity matrix it is simple  to expand directly the PDF, as done in Sec.~(\ref{sec:approximations}), it  becomes, still doable, but unnecessarily complicated if we want to work with Eq.~(\ref{eq:70}). In fact the work leading up to Eq. ~(\ref{eq:70}) and the route we chose, are necessary if one wants to find a closed form for the exact PDF.  If one instead  wanted  {\it a priori}  to obtain an approximated expression, truncated at a given order, the simplest route to follow would have been different. In fact it is well know that a general expression for any (well behaved) PDF can be obtained from its generating functional by expanding the PDF in terms of all its (2nd and higher-order) correlations.

This expansion  can be written as  (see e.g.,  \cite{MLB86} , the integrand of their  Eq. 9, and \cite{GrinsteinWise86}):
\begin{equation}
{\cal P}(\vec{A}|f_{\rm NL})=\frac{\det {\bf K}^{1/2}}{(2\pi)^{n/2}}\exp\left[\sum_{i=3}^\infty (-1)^i\sum_{[r_i]=1}^{n}\frac{\xi^{(i)}_{[r_i]}}{i!}\prod_{j=1}^{i}\frac{\partial}{\partial A_{r_j}}\right]\exp\left[-\frac{1}{2} ({\vec A}, {\bf K}, {\vec A})\right]
\label{eq:MLB86}
\end{equation}
where ${\bf K}$ denotes the inverse covariance matrix for $\vec{A}$ and $[r_i]$ is a compact notation to account for all relevant combinatorials  for example for $i=3$ the second sum runs over all triplets that can be combined out of the $A_1...A_n$. Here $\xi^{(i)}$ denotes the connected correlation function (or corresponding polyspectra) of order $i$.
This expression is exact, but it involves an infinite sum which must be truncated. In fact this expression  simply says that any  non-Gaussian distribution is specified by {\it all} its higher order correlations (only a Gaussian distribution is completely specified by its covariance). 
 In other words Eqs.~(\ref{eq:70}) and (\ref{eq:MLB86}) are mathematically equivalent for local non-Gaussianity  once the $\xi^n$ are the ones of the  field $A$. For a practical application however, when the exact PDF is needed, Eq.~(\ref{eq:70}) is preferable to Eq.~(\ref{eq:MLB86}) because  of the infinite series in Eq.~(\ref{eq:MLB86}) which would require evaluation of an infinite sequence of correlation functions for $A$.
Conversely,  if for some applications an approximated expansion of the PDF is what is needed, it is overwhelmingly easier to work with Eq.~(\ref{eq:MLB86}), truncated at the appropriate order.

As an exercise it is not too algebraically cumbersome   to see  that at first order in $f_{\rm NL}$ Eq.~(\ref{eq:70}) and Eq.~(\ref{eq:MLB86}) are truly equivalent.  The univariate case for unit variance field was presented in Sec.~(\ref{sec:approximations}). For the multi-variate case  with a generic covariance, the most cumbersome part is to keep track of the  various (symmetrized) permutations  that go in $\xi^{(3)}$ for the local non-Gaussian case.
Since in the simpler case analyzed in  Sec.~(\ref{sec:approximations}) and (\ref{sec:estimators})  we could compute the exact PDF, its first and second-order expansions, and compare the performance of the exact PDF with its approximations, we conclude that a second-order expansion  of  Eq.~(\ref{eq:MLB86}) is suitable for many interesting  practical applications ($f_{\rm NL}$ in agreement with current constraints i.e., Ref. \cite{planckNG}, sample size not too small). Here therefore we derive an explicit expression for the multi-variate  non-Gaussian PDF to second-order using Eq.~(\ref{eq:MLB86}).

We expand to second order in $f_{\rm NL}$ by first  truncating the first sum to $i=4$ -- i.e., the trispectrum-- and then expanding the exponential in Taylor series to second order.

It is easy to see that if $A=\Phi$, $n=1$ and $\Phi$ has unit variance, we recover Eq.~(\ref{eq:expandsecondorderEdgeworth}). In doing so it is useful to keep in mind that for local non-Gaussianity $\xi^{(3)}=6f_{\rm NL}$ and $\xi^{(4)}=24f_{\rm NL}^2$.

In the general case by performing the derivatives we obtain:
\begin{equation}
\boxed{
\begin{gathered}
{\cal P}(\vec{A}|f_{\rm NL}) = \frac{\sqrt{\det {\bf K}}}{(2\pi)^{n/2}}\exp\!\left[-\frac{1}{2} ({\vec A}, {\bf K}, {\vec A})\right]\!\times\!\!
\left\{ 1+\frac{1}{6} \sum_{{\rm all }\, l}  \xi^{(3)}_{l_1l_2l_3}\!\left[({\bf K}{\vec A})_{l_1}({\bf K}{\vec A})_{l_2}({\bf K}{\vec A})_{l_3}\!-3 {\bf K}_{l_1l_2}({\bf K}{\vec A})_{l_3}\right] +\right. \\
\left. \frac{1}{24}  \sum_{{\rm all}\,l}\xi^{(4)}_{l_1l_2l_3l_4}\left[3 K_{l_1l_2}K_{l_3l_4}-6 {\bf K}_{l_1l_2} ({\bf K}{\vec A})_{l_3} ({\bf K}{\vec A})_{l_4}+ ({\bf K}{\vec A})_{l_1} ({\bf K}{\vec A})_{l_2}({\bf K}{\vec A})_{l_3} ({\bf K}{\vec A})_{l_4}\right] \right.+ \\
\left. \frac{1}{72}\sum_{l_1,...,l_6}\xi^{(3)}_{l_1l_2l_3}\xi^{(3)}_{l_4l_5l_6} \left[({\bf K}{\vec A})_{l_1} ({\bf K}{\vec A})_{l_2}({\bf K}{\vec A})_{l_3} ({\bf K}{\vec A})_{l_4} ({\bf K}{\vec A})_{l_5} ({\bf K}{\vec A})_{l_6}-15 {\bf K}_{l_1l_2}{\bf K}_{l_3,l_4}{\bf K}_{l_5l_6} \right.\right. \\
\left.\left.-15 {\bf K}_{l_1l_2}({\bf K}{\vec A})_{l_3} ({\bf K}{\vec A})_{l_4}({\bf K}{\vec A})_{l_5} ({\bf K}{\vec A})_{l_6}  +45 {\bf K}_{l_1l_2}{\bf K}_{l_3l_4}({\bf K}{\vec A})_{l_5} ({\bf K}{\vec A})_{l_6} \right] \right\}
\label{eq:multivariateedgeworth}
\end{gathered}}
\end{equation}
where $l_i$ run from $1$ to $n$.
Here we recognize that in this approximation the PDF is given by a  suitable combination of bispectrum and trispectrum.  We will return to this point below in the application to the CMB in Sec.~(\ref{sec:CMBappl}).
The great advantage of  this approach is that the resulting expression is valid for   non-Gaussianity  beyond the local  type. While we have tested quantitatively the performance of this approximation  for the local case, we expect that it will hold  also e.g., for equilateral, orthogonal or flattened type.
 Of course, Eq. (\ref{eq:multivariateedgeworth}) is strictly second-order in $f_{\rm NL}$ for non-Gaussianity of the type where $\Phi^{NG} \sim f_{\rm NL}\phi *\phi$ ($*$ denoting convolution or multiplication). Should  other types of non-Gaussianity be considered,  then the expression  would not be necessarily second-order in the non-Gaussianity parameter. For example for the $g_{\rm NL}$ type where $f_{\rm NL}=0$, the terms with $\xi^{(3)}$ would drop out and the resulting expression would be first-order in $g_{\rm NL}$.
 
Given that in a practical application  the $A_i$ are measured (given) quantities,  the PDF  can be interpreted as  the  probability distribution for $f_{\rm NL}$. Within a given model  (like for example the local model)   where the bispectrum and trispectrum coefficients are related (one being $\propto f_{\rm NL}$ and the other $\propto f_{\rm NL}^2$) the PDF can be used to constrain  the parameter  (i.e. find the maximum likelihood and the  confidence intervals). In a more model-independent way, one could imagine finding joint constraints on the coefficients of the bispectrum and trispectrum and then comparing those with theory predictions.  
In light of these considerations and the fact that for $f_{\rm NL}$ values allowed by present data and for not too small sample sizes  this expression  gives and  an extremely good approximation  for the exact PDF, Eq.~(\ref{eq:multivariateedgeworth})  represents one of the main results of this paper.  For a CMB application  of Eq.~(\ref{eq:multivariateedgeworth}), see Sec.~(\ref{sec:CMBappl}).

\subsection{Adding Gaussian (or near-Gaussian) noise}
\label{sec:addinggaussianoiseindirac}
As before (and using the same convention as of Sec.~(\ref{sec:addingnoise})) we can derive the result in two ways. 
The direct way is to recall that ${\cal P}({\cal A})={\cal P}(A)\star {\cal P}(\epsilon)$ which will be further explored below (Sec. \ref{sec:addingnoise2}) in the context of measuring  $f_{\rm nl}$ from a CMB map. In fact, although  the approach is general,  the implementation  depends on the details of the concrete application.  This approach might be best if the noise has  complicated pixel-to-pixel correlations. The other way is to substitute the Dirac delta function by a Gaussian  with width given by the noise covariance $\Sigma_{\epsilon}$ or by the noise distribution if it is non-Gaussian. Here we  illustrate the approach in the Gaussian noise case.

In this case our starting equation becomes Eq.~(\ref{eq:jointprob}) but where the product of the Dirac delta function is instead $1/(2\pi)^{n/2}\exp[-1/2(A-{\cal M}\Phi,K_\epsilon,A-{\cal M}\Phi)]$ where $K_\epsilon$ denotes the inverse covariance of the noise and  ${\cal M}\Phi$ should due interpreted as matrix times a vector. Unfortunately the resulting path integral cannot be done exactly analytically but can be performed numerically. However even in the signal-only regime the last path integral has to be performed numerically (see discussion around Eq.~(\ref{eq:70})) thus correctly including the effect of noise  does not represent an additional computational burden.

On the other hand, if the noise contribution is sub-dominant compared to the signal (and yet non-negligible) a saddle point approximation  performs very well: Eq.~(\ref{eq:70}) then becomes the  first term in a saddle point approximation.
 Note that Ref.\cite{Elsner:2010gd} already noted that the presence of Gaussian  noise  has the effect of changing  the Dirac delta functions for Gaussians. They approach the problem by  implementing numerically the expressions corresponding with our Eq.~(\ref{eq:jointprob})  and Eq.~(\ref{eq:60}).

\section {Application I: measuring $f_{\rm nl}$ from a CMB map} 
\label{sec:CMBappl}
\subsection{Signal-only regime}
It is possible to  consider Eq.~(\ref{eq:70}) and interpret it as the probability of $f_{\rm NL}$ given the observable vector $\vec{A}$ (which could be $a_{\ell m} $ or  $T$ etc.). This expression is exact. One would need to still do a path integral numerically but only for a grid of  values for $f_{\rm NL}$  i.e. only a  finite, reduced number of times, to find the maximum likelihood and its confidence levels. The path integral  can be computed using standard lattice QCD techniques. The  integrand  is close to  a Gaussian  so integration should converge easily. This makes possible to increase the size of the lattice, which sample $\phi''$ in Eq.~(\ref{eq:70}). In QCD applications the size of the lattice can be as large as billions, comparable with the size we estimate would be needed for a CMB application.   Alternatively Monte-Carlo techniques such as those explored in \cite{Elsner:2010hb} can be employed. We leave the numerical implementation for the evaluation of Eq.~(\ref{eq:70}) for future work, but note that {\it a)} with  respect to the formulation of Eq.~(4.11) the dimensionality of the integral has been reduced by a factor ${\cal O} (10^6 )$  and {\it b)} given  a set of data (e.g., a CMB map) the integral for several $f_{\rm NL}$ values must be computed only once to obtain the best-fit value  and confidence intervals.

This approach of using the PDF does not have the problem of the bispectrum that is that  the number of triplets one can make grows much faster than the number of pixels in a map and therefore different bispectrum estimates are not independent (and thus the central limit theorem needs not to apply) \cite{YadavWandelt2010}. One possible limitation is that the expression is exact only in the signal-dominated regime (i.e. no noise). This can be overcome as we discuss below.

\subsection{Adding  noise}
\label{sec:addingnoise2}

As done in Sec.~(\ref{sec:addingnoise}) consider that  the observable field  is ${\cal A}$ and is given by the superposition of two independent random processes, the signal  $A$  and the noise $\epsilon$, (${\cal A}_i=A_i+\epsilon_i$). 
For the sum (superposition) of two independent random processes the probability distribution is the convolution of the two probabilities. Doing the convolution however might not be possible analytically. 
One thing to keep in mind is that if the noise is uncorrelated  one need not do a $N_{\rm pix}$-dimensional convolution, but one can do $N_{\rm pix}$ 1-dimensional convolutions. On the other hand, what we expect in the presence of noise  is that the probability distribution of $f_{\rm NL}$ is broadened by the noise:
\begin{equation}
{\cal P}_{\cal A}(f_{\rm NL}|{\cal A})={\cal P}_{\cal A}(f_{\rm NL}|{\cal A})\star {\cal P}_{\epsilon}(f_{\rm NL}|\epsilon)
\end{equation}
i.e. a convolution between the result of the signal-only regime with  the distribution for the noise-only regime.  Given that the noise has zero mean, $P_{\epsilon}(f_{\rm NL}|\epsilon)$ must only depend on the noise covariance $\Sigma_{\epsilon}$. The advantage of this approach  is that one is left to do only a 1 dimensional convolution.

While $P_{\epsilon}(f_{\rm NL}|\epsilon)$ has not been computed, if the noise is Gaussian or close to Gaussian, we know that it must be very well approximated by the distribution for the KSW estimator  \cite{Komatsu:2003iq} applied to a purely Gaussian map (as it saturates the Cram\' er-Rao bound \cite{Kamionkowski:2010me}). This distribution has been shown to be very well approximated by a Gaussian \cite{Smith:2011rm} --since for a large  number of pixels the central limit theorem holds-- with variance $\sigma^2_{f_{\rm NL}}= 1/(6\sigma_{\epsilon}^2N_{\rm pix})$. 
In summary, since the noise is not correlated with the signal,  in the presence of noise the signal-only PDF, Eq.~(\ref{eq:70}), must be convolved with the noise-only PDF for $f_{\rm NL}$.  If the noise is Gaussian, the noise-only PDF has a closed analytical expression; if not, it would need to be computed on a case-by-case basis. Having $P_{\cal{A}}(f_{\rm NL}|{\cal A})$ already computed on a grid of $f_{\rm NL}$ values (and $P_{\epsilon}(f_{\rm NL}|\epsilon)$ as an analytic expression or computed on  the same grid values for $f_{\rm NL}$) makes the convolution a discrete sum. Of course this is useful if the noise-only PDF is less broad  than the signal-only PDF (i.e., significantly non zero  over a smaller range of $f_{\rm NL}$ values than the signal-only PDF ). If not, then one would be working in the noise-dominated regime where little useful information on the primordial $f_{\rm NL}$ can be extracted, or, alternatively, where the signal PDF is only a correction   to the noise-only PDF. Thus rough approximations of the signal PDF  can be  safely used in this regime.
 
Another equivalent approach is  the one outlined in sec. (\ref{sec:addinggaussianoiseindirac}): substitute the Dirac delta function in Eq.(\ref{eq:Pfnl})  by the noise distribution.  Then Monte-Carlo like approaches as adopted e.g. in \cite{Elsner:2010gd} can be used to evaluate the multi-dimensional integral. Depending on the nature of the noise  one or the other will be best suited.

\subsection{Second-order PDF expansion}
Eq.(\ref{eq:multivariateedgeworth}) as it is can be interpreted as the PDF given the temperature pixels in a CMB map (i.e. $A=T$ or $A=\Delta T/T$, $n$ the number of pixels, $l$ the pixel index and $w^i$ the $i$-point correlation function). On the other hand it is useful to re-write it explicitly   as a function of the spherical harmonic coefficients $a_{\ell m}$ also because it will make explicit the relation of our findings with previous work in the literature.
In the following we denote by $a$ the vector of $a_{\ell}^m$ and by $n$ the number of harmonic coefficients considered. We obtain:
\begin{equation}
\boxed{
\begin{gathered}
{\cal P}(a | f_{\rm NL}) = \frac{(\det C^{-1})^{1/2}}{(2\pi)^{n/2}}\exp\left[-\frac{1}{2} \sum_{\ell \ell' m m'} a^{*m}_{\ell}(C^{-1})_{\ell m \ell' m'}a_{\ell'}^{m'})\right]\times\\
\left\{ 1+\frac{1}{6} \sum_{{\rm all }\, \ell_i m_j}  \langle a_{\ell_1}^{m_1} a_{\ell_2}^{m_2} a_{\ell_3}^{m_3}\rangle\left[(C^{-1} a)_{\ell_1}^{m_1}(C^{-1}a)_{\ell_2}^{m_2}(C^{-1}a)_{\ell_3}^{m_3}-3 (C^{-1})_{l_1,l_2}^{m_1 m_2}(C^{-1}a)_{l_3}^{m_3}\right] +\right. \\
\frac{1}{24}  \sum_{{\rm all}\,\ell m}\langle  a_{\ell_1}^{m_1} a_{\ell_2}^{m_2} a_{\ell_3}^{m_3} a_{\ell_4}^{m_4}\rangle\left[3 (C^{-1})_{\ell_1\ell_2}^{m_1 m_2} (C^{-1})_{\ell_3,\ell_4}^{m_3 m_4}\right.\\
\left.-6 (C^{-1})_{\ell_1,\ell_2}^{m_1 m_2} (C^{-1}a)_{\ell_3}^{m_3} (C^{-1}a)_{\ell_4}^{m_4}+ (C^{-1}a)_{\ell_1}^{m_1}(C^{-1}a)_{\ell_2}^{m_2}(C^{-1}a)_{\ell_3}^{m_3}(C^{-1}a)_{\ell_4}^{m_4} \right]+ \\
\left. \frac{1}{72}\sum_{l_1,...,l_6}\langle a_{\ell_1}^{m_1}a_{\ell_2}^{m_2}a_{\ell_3}^{m_3}\rangle \langle a_{\ell_4}^{m_4}a_{\ell_5}^{m_5}a_{\ell_6}^{m_6}\rangle  \left[(C^{-1}a)_{\ell_1}^{m_1} (C^{-1}a)_{\ell_2}^{m_2}(C^{-1}a)_{\ell_3}^{m_3} (C^{-1}a)_{\ell_4}^{m_4} (C^{-1}a)_{\ell_5}^{m_5} (C^{-1}a)_{\ell_6}^{m_6}\right.\right.\\
-15 (C^{-1})_{\ell_1\ell_2}^{m_1 m_2}(C^{-1}a)_{\ell_3}^{m_3} (C^{-1}a)_{\ell_4}^{m_4}(C^{-1}a)_{\ell_5}^{m_5} (C^{-1}a)_{\ell_6}^{m_6} \\
\left.\left.-15 (C^{-1})_{\ell_1\ell_2}^{m_1 m_2}(C^{-1})_{\ell_3\ell_4}^{m_3 m_4}(C^{-1})_{\ell_5\ell_6}^{m_5 m_6} 
 +45 (C^{-1})_{\ell_1\ell_2}^{m_1 m_2}(C^{-1})_{\ell_3\ell_4}^{m_3 m_4}(C^{-1}a)_{\ell_5}^{m_5} (C^{-1}a)_{\ell_6}^{m_6} \right] \right\}
\end{gathered}}
\label{eq:multivariateedgeworthCMB}
\end{equation}

 Note that in the second line of Eq. (\ref{eq:multivariateedgeworthCMB}) we recognize the  standard formulation for approximating the PDF \cite{babich05}  which is the starting point to derive the standard KSW estimator \cite{Komatsu:2003iq}. In fact  we recognize  Eq. (5) of \cite{Komatsu1003.609}.  
In the next two lines we recognize Eq. (32), but more specifically Eq. (158), of \cite{Regan1004.2915}.

The last three lines are ``new" terms that arise from expanding the exponential to second order in $f_{\rm NL}$ thus involving a term proportional to  the bispectrum squared. 
 
 We argue here that interpreting this as a likelihood for $f_{\rm NL}$  enables one to combine optimally bispectrum and trispectrum measurements  and obtain both best-fit value and  confidence intervals for the non-Gaussianity parameter. As discussed above this expression is valid for non-Gaussianities more general than the local form  as long as departures from Gaussianity are small as  quantified in Sec.~(\ref{sec:estimators}). This expression is strictly second order in $f_{\rm NL}$ for the type of non-Gaussianity where $\Phi^{NG} \sim f_{\rm NL}\phi *\phi$ ($*$ denoting convolution or multiplication) which includes local, equilateral, flattened and orthogonal shapes. for other types of non-Gaussianity, this expression will not necessarily be second order in the non-Gaussianity parameters. Within a given non-Gaussian model where the bispectrum and trispectrum amplitudes are specified by a single parameter, this PDF can be used to constrain such a parameter. Moreover, in a  more model-independent approach one could use the above PDF to find joint constraints on the amplitude of the bispectrum and trispectrum for comparison with theory. Eq.~(\ref{eq:multivariateedgeworthCMB})  is one of the main results of this paper. 
 Moreover note that  this approximation to the PDF, being quadratic in $f_{\rm NL}$,  always has a maximum. In the standard formulation, where the  PDF expansion  is truncated to first  order in $f_{\rm NL}$, the PDF depends linearly on $f_{\rm NL}$ and therefore at face value has no maximum.  
{\color{black} Implicitly, the quadratic term is assumed to have a  constant coefficient (the average of the true coefficient in the case of B05), and the estimator normalised later to be unbiased.  With the complete PDF,
it is easy to  write down the maximum after expansion to second order.  We rewrite Eq. (\ref{eq:multivariateedgeworthCMB}) is a compact form 
 \begin{equation}
 {\cal P}(a | f_{\rm NL})=G*[1+f_{\rm NL}b*w_1+f_{\rm NL}^2(t*w_2+h*w_3)]
 \end{equation} 
 where $G$ denotes the Gaussian factor, $b*w_1$ corresponds to the part on the second line of  Eq.~(\ref{eq:multivariateedgeworthCMB}) that is proportional to $f_{\rm NL}$ evaluated for $f_{\rm NL}=1$, and similarly the terms $t*w_2$ and $h*w_3$ are defined by that equation.  Then the formal maximum would be  
 \begin{equation}
 \hat{f}_{\rm NL}= -\frac{b*w_1}{2(t*w_2+h*w_3)}
 \end{equation}
and its variance  $1/(t*w_2+h*w_3)$.}
\section{$g_{\rm NL}$ case}

We can extend our formalism to the so-called $g_{\rm NL}$ model, in which 
\begin{equation}
\Phi = \phi + g_{\rm NL} (\phi^3 -3 \langle\phi^2\rangle \phi) \equiv \phi + g'_{\rm NL} \phi^3 \,,
\label{eq:defgnl}
\end{equation}
which corresponds to the case when $f_{\rm NL}$ is zero.  Having worked out in pedagogical detail the Gaussian case and the local $f_{\rm NL}$ one, here we will only sketch the derivation. In fact the steps are closely similar, only the resulting expressions (given the increasing number of roots involved) become quickly long and cumbersome. If $\cal M$ is the identity matrix, we obtain

\begin{eqnarray}
{\cal P}(\vec{A})&=&\sqrt{\frac{\det K_0}{(2\pi)^M}} \prod_i\frac{1}{|1+3 g_{\rm NL} \phi_r^2|}\int \prod_{\nu} d\phi_{\nu} \exp\left[-\frac{1}{2} \sum_{\mu,\nu}{\phi}_{\mu}{\phi}_{\nu}{K_{0\mu\nu} } \right]\\ \nonumber
&\times& 
 \exp\left[-\frac{1}{2} \sum_{ij}f_r(A_i)f_r(A_j){K_{0ij}} - \sum_{i,\nu}f_r(A_i)\phi_{\nu}K_{0i\nu}\right] \\ \nonumber
\label{eq:gnl1}
 \end{eqnarray}
where $\phi_r$ is the real root of the third order equation $A - \phi - g_{\rm NL} \phi^3 = 0$.  The integral in $\phi$ can be performed analytically  as done before. 
 If ${\cal M}$ is invertible but not the identity matrix, then $A\longrightarrow {\cal M}^{-1}A$ and a factor of $\det {\cal M}$ should be included in front of Eq.~(\ref{eq:gnl1}). In the general case where ${\cal M}$ is  not invertible,  the technique presented in Sec.~(\ref{sec:4.5})  can be similarly applied here to obtain a closed expression.   The resulting expression  can be obtained straightforwardly, but it is long and cumbersome and therefore we do not write it explicitly here.

\section{Application II: Mass function of dark matter halos}
Note that in the case  where we need  the PDF of the linearly extrapolated $\delta^L$,  ${\cal P}(\delta^L)$, ${\cal M}$ is invertible and we have obtained above an exact analytic expression: 
\begin{equation}
{\cal P}(\delta^L| f_{\rm NL})=\frac{1}{(2\pi)^{M/2}}\frac{1}{\sigma_R}\frac{1}{\sqrt{1+4f_{\rm NL}(\tilde{\delta^L}+f_{\rm NL}\sigma^2_{\phi,R})}}
\sum_{r=1,2}\left\{  \exp\left[-\frac{1}{2}\frac{ f_r^2(\tilde{\delta^L})}{\sigma^2_{\phi,R}} \right]\right\}\,.
\label{eq:mfexact}
\end{equation}
Here $\delta^L$ is the linearly extrapolated  overdensity field  smoothed on scale $R$, corresponding to the halo mass one is interested in; $\sigma_R$ denotes the RMS of $\delta^L$ (smoothed on scale $R$);  $\sigma_{\phi,R}$ is the corresponding quantity for the primordial field $\phi$. The sum is over the two roots (recall that the exponentially suppressed root can be safely neglected). $\tilde{\delta}={\cal M}^{-1}\delta$.  Finally $f_{1,2}(\delta)=1/(2f_{\rm NL})\left(-1 \pm \sqrt{1+4f_{\rm NL}\delta}\right)$.
While in principle the inversion of ${\cal M}$ here requires solving the Poisson equation, recall that  in our approach the fields are pixelized and therefore techniques developed for setting the initial conditions for  dark matter cosmological N-body simulations   can be directly applied here.
 ${\cal P}(\delta^L)$ is the basic ingredient to compute the mass function. In the old-fashioned Press-Schechter approach one only had to integrate the one-point PDF for $\delta^L$ above the threshold for collapse $\delta_c$ (and introduce the fudge factor of 2). In this framework our approach gives an {\it exact} expression  for the Press-Schechter mass function, which should not have the problem of negative PDF tails (as in the Edgeworth expansion, see  \cite{LoVerde:2007ri,Maggiore:2009rv}) or being inaccurate for not-so-rare peaks \cite{MVJ}. The standard Press-Schechter approach fails in many respects but the one-point PDF remains the key ingredient for more sophisticated formulations of the mass function \cite{Maggiore:2009rv,D'Amico:2010ta,Paranjape:2012jt,Achitouv:2012ux}. Ref.~\cite{Ferraro:2012bd} showed the  equivalence of  simple barrier crossing to more modern and sophisticated approaches and ref.~\cite{Paranjape:2012jt}  showed how to compute the (Gaussian and non-Gaussian) mass functions starting from the 1-point PDF and using a suitably-defined threshold.

\subsection{Application III: three-dimensional matter overdensity  (e.g. 21cm surveys)}

In the case of surveys that map  directly the three-dimensional  $\delta$ field, the size of the ${\cal M}$ matrix is  $M=n$ as in Sec.~(\ref{sec:minvertible}), which simplifies the computations over the CMB case.  Although one still needs to compute the transfer function because non-Gaussianity is  in the $\phi$ field, the equations simplify significantly,  as the last path integral can be performed analytically, see Eq.~(\ref{eq:pmultiexactM}).

\section{Conclusions}

We have presented a new formalism to  write down a closed expression for the exact multi-variate joint probability distribution function  (PDF) of a non-Gaussian field of the local type. We provide  an exact formula for such  PDF for the non-Gaussianity parameter  $f_{\rm NL}$  in Eq.~(\ref{eq:70}).  
This exact formulation has several applications ranging from the  halo mass function to the analysis of CMB maps. In its more generic formulation (and this for applications that rely on the generic formulation such as application to CMB maps) the final integration must be performed numerically.   The dimensionality of the integral is high but it has been reduced by a factor ${\cal O} (10^6)$ compared to the initial and intermediate  formulation.  While we estimate that the numerical calculation is doable it remains to be seen how computationally intensive it might be in practice for a realistic (Planck-size) CMB map. This is left for future work.  However in special cases (e.g., as the fundamental step for the derivation of the halo  mass function) the resulting expression is fully analytic. We present the exact analytic formula for the 1-point non-Gaussian PDF in Eq.~(\ref{eq:1pointexact}) which provides the key ingredient for an exact formulation of  the mass function of dark halos; a formula to do so is provided in Eq.~(\ref{eq:mfexact}). 

These results serve as the basis for a fully Bayesian analysis of the non-Gaussianity parameter which has several advantages compared to the presently-used, state-of-the-art frequentists estimators. Most notably the fact that the PDF encloses the full  information on the parameter of interest and therefore prohibitively expensive Monte Carlo estimation of the errors is not needed.
While the analytic derivation is done in the signal-only regime,  we also discuss how to include (analytically) the effect of instrumental noise 
 and show that   Gaussian noise  can be included at little extra computational expense.

Having a full, exact PDF expression and  formulating the problem in a Bayesian fashion, enabled us to uncover some features of the PDF for local non-Gaussianities that were previously ignored. Notably the fact that the PDF is neither smooth nor continuous, making any approximation based on expansions on the $f_{\rm NL}$ parameter   of the PDF  sometimes  unsuitable (and applicability of the central limit theorem limited). We however identify a regime where a second-order expansion is a good approximation, this happens for a combination of $f_{\rm NL}$ not too large (i.e., not ruled out by current measurements) and a sample size which is not too small (few tens or equivalently for a full sky  CMB analysis ($\ell <\ell_{\rm max}$ where $\ell_{\max} \sim $ few or larger).  That is, for a full sky CMB analysis that explores a wide $\ell$ range, the second-order expansion is a good approximation, however if one wanted to explore localized features or fine scale-dependence, the full PDF behaviour might have to be considered.
 We present explicit  expressions for this  second-order expansion approximation both in the general case, Eq.~(\ref{eq:multivariateedgeworth}), and for  the  specific application to a CMB map, Eq.~(\ref{eq:multivariateedgeworthCMB}).  In such expansions we can recognize previously known terms (such as the KSW estimator of Ref.~\cite{Komatsu:2003iq}, and the optimal trispectrum  estimator of Ref.~\cite{Regan1004.2915}) but also a new term $\propto f_{\rm NL}^2$. This formulation enables one to combine optimally bispectrum and trispectrum measurements and obtain both best-fit value and confidence intervals for the non-Gaussianity parameter. As an added bonus, this truncated expression is valid for non-Gaussianity beyond the local type. While we have tested quantitatively the performance  and regime of validity of this approximation for the local case, we expect that qualitatively it will hold also e.g. for equilateral, orthogonal or flattened type and in general  for non-Gaussianity where $\Phi^{NG} \sim f_{\rm NL}\phi *\phi$ ($*$ denoting convolution or multiplication).
 Within a given model (like for example the local model) where the bispectrum and trispectrum coefficients are related,  the PDF can be used to constrain the parameter (i.e. find the maximum likelihood and the confidence intervals). In a more model-independent way, one could imagine finding joint constraints on the coefficients of the bispectrum and trispectrum and then comparing those with theory predictions.
 Our formulas may also find application in large-scale galaxy surveys in searching for non-Gaussian features,  compute the confidence intervals and in general do statistical inference  also comparing and combining  different data sets.

\acknowledgments{We thank Miguel  Angel Vazquez-Mozo, A. Sabio Vera  and Benjamin Wandelt for discussions and  Michele Liguori  for comments on the draft. LV acknowledges support from  FP7-IDEAS-Phys.LSS 240117. LV and RJ are supported by FPA2011-29678- C02-02 and acknowledge hospitality of Galileo Galilei Institute for Theoretical Physics  during  some initial stages of  this work. We thank Imperial College London for travel support which facilitated this collaboration. S.M. acknowledges partial financial support from ASI/INAF Agreement I/072/09/0 for the Planck LFI Activity of Phase E2 and PRIN 2009 project ``La Ricerca di non-Gaussianit\`a Primordiale".}

\end{document}